\documentclass[aps,prd,reprint,superscriptaddress,amsmath,amssymb,floatfix,nofootinbib,raggedbottom]{revtex4-2}
\usepackage[T1]{fontenc}
\usepackage[utf8]{inputenc}
\usepackage{lmodern}
\usepackage{graphicx,bm,microtype}
\usepackage{booktabs}
\usepackage{xcolor}
\usepackage{hyperref}
\hypersetup{colorlinks=true,linkcolor=blue,citecolor=black,urlcolor=blue,
  citebordercolor={0 1 0},
  pdftitle={Correlation Harvesting in 2+1 Dimensions}}

\makeatletter
\newsavebox{\CH@citebox}
\AtBeginDocument{%
  \let\CH@natlinkstart\hyper@natlinkstart
  \let\CH@natlinkend\hyper@natlinkend
  \def\hyper@natlinkstart#1{%
    \begingroup
    \begin{lrbox}{\CH@citebox}%
    \CH@natlinkstart{#1}%
  }%
  \def\hyper@natlinkend{%
    \CH@natlinkend
    \end{lrbox}%
    \leavevmode
    \begingroup
      \setlength{\fboxsep}{0pt}%
      \setlength{\fboxrule}{0.4pt}%
      \rlap{\kern-\fboxrule
        \smash{\fcolorbox{green}{white}{\phantom{\usebox{\CH@citebox}}}}}%
      \usebox{\CH@citebox}%
    \endgroup
    \endgroup
  }%
}
\makeatother

\newcommand{\trajectorypanel}[2]{%
 \begin{minipage}[t]{0.235\textwidth}
 \centering
 \IfFileExists{#1}{%
  \includegraphics[width=\linewidth,height=2.6cm,keepaspectratio]{#1}%
 }{%
  \fbox{\parbox[c][2.6cm][c]{0.90\linewidth}{\centering\small Insert figure}}%
 }\\[0.2em]
 {\footnotesize #2}
 \end{minipage}%
}

\begin{document}
\title{Unruh-like effects and correlation harvesting between particle detectors in uniform motion in \((2+1)\)-dimensional Minkowski spacetime}

\author{Siwei Han}
\affiliation{School of Physics, Xi'an Jiaotong University, Xi'an 710049, Shaanxi, China}

\author{Zeinab Imani}
\affiliation{Department of Physics and Astronomy, University of Waterloo, 200 University Ave W, Waterloo, Ontario N2L 3G1, Canada}

\author{Jun Feng}
\email[]{j.feng@xjtu.edu.cn}
\affiliation{School of Physics, Xi'an Jiaotong University, Xi'an 710049, Shaanxi, China}
\affiliation{MOE Key Laboratory for Non-equilibrium Synthesis and Modulation of Condensed Matter, Xi'an 710049, Shaanxi, China}

\author{Robert B. Mann}
\email[]{rbmann@uwaterloo.ca}
\affiliation{Department of Physics and Astronomy, University of Waterloo, 200 University Ave W, Waterloo, Ontario N2L 3G1, Canada}
\affiliation{Perimeter Institute for Theoretical Physics, 31 Caroline St. N., Waterloo, Ontario N2L 2Y5, Canada}
\affiliation{Department of Applied Mathematics, University of Waterloo, 200 University Ave W, Waterloo, Ontario N2L 3G1, Canada}
\affiliation{Institute for Quantum Computing, University of Waterloo, 200 University Ave W, Waterloo, Ontario N2L 3G1, Canada}

\begin{abstract}
We study Gaussian-switched Unruh--DeWitt detectors coupled to a massless scalar field in the $(2+1)$-dimensional Minkowski vacuum and following four classes of uniformly accelerated trajectories: linear, catenary, cusped, and circular motion. We first determine how the single-detector transition probability and effective temperature depend on acceleration and the torsion-to-acceleration ratio $\bar b$, whose continuous variation connects these four trajectory classes. We find that both quantities generally increase with acceleration, while their dependence on $\bar b$ can become nonmonotonic depending on the energy gap. We also examine a correlation-harvesting protocol by two detectors following horizontally displaced copies of the same trajectory. Increasing acceleration or separation generally suppresses the harvested concurrence and quantum mutual information, but the extent of suppression at fixed acceleration depends strongly on the trajectory class. For each of the four trajectory classes, we identify regimes of genuine entanglement harvesting in which vacuum correlations alone suffice to entangle the detectors, while the communication contribution alone cannot overcome local noise.
\end{abstract}

\maketitle
\raggedbottom
\setlength{\emergencystretch}{1em}

\makeatletter
\def\raggedcolumn@skip{%
  \vskip\z@\raggedcolumn@sw{\@plus.0001fil}{}\relax
}
\makeatother

\section{Introduction}
\label{sec:introduction}

The particle content of a quantum field is an observer-dependent notion. For example, even in the Poincar\'e-invariant Minkowski vacuum, a uniformly linearly accelerated observer experiences a thermal response~\cite{Unruh1984}. This celebrated Unruh effect can be described operationally using an Unruh--DeWitt (UDW) detector, a localized quantum system coupled to the field along its worldline~\cite{Unruh1976,DeWitt1979}. In the long-interaction limit, the excitation and de-excitation rates of a detector undergoing uniform linear acceleration satisfy detailed balance at the Unruh temperature,
\begin{equation}
 T_{\mathrm U}=\frac{\hbar a}{2\pi c k_{\mathrm B}},
 \label{eq:unruh-temperature}
\end{equation}
where $a$ is the proper acceleration, $\hbar$ is the reduced Planck constant, $c$ is the speed of light, and $k_{\mathrm B}$ is Boltzmann's constant. Related phenomena in curved spacetime include Hawking radiation from black holes~\cite{Hawking1975} and particle creation in an expanding universe~\cite{Parker1969}. The Unruh effect offers a setting in flat spacetime for investigating the relation between vacuum fluctuations, particle detection, and relativistic motion~\cite{Crispino2008}.

Despite its theoretical importance, a direct experimental verification of the standard Unruh effect remains an outstanding challenge, since equation~\eqref{eq:unruh-temperature} requires an acceleration of approximately $2.4\times10^{20}\,\mathrm{m\,s^{-2}}$ for a temperature of merely $1\,\mathrm K$. This difficulty has motivated alternative routes to observing acceleration-dependent vacuum fluctuations. One possibility is to use analogue spacetime systems, such as condensed-matter systems whose excitations simulate relativistic quantum fields with a propagation speed much lower than the speed of light~\cite{Retzker2008,Gooding2020}. Another practical choice is to replace linear acceleration motion with circular motion, which also possesses constant acceleration and can be maintained within a bounded spatial region for a long time, making it easier to realize in a laboratory. Circular acceleration has been investigated in connection with electron spin polarization in storage rings~\cite{BellLeinaas1983,Unruh1998}, and a cavity-based proposal uses a rotating atom to enhance acceleration-dependent changes in spontaneous emission~\cite{Lochan2020}.

However, linear and circular motion are not the only trajectories with constant proper acceleration. Letaw classified a special set of timelike worldlines in $(3+1)$-dimensional Minkowski spacetime using three constant Frenet--Serret invariants: the magnitude of the proper acceleration and the first and second torsions~\cite{Letaw1981}. This classification yields the inertial class and five accelerated classes: linear, catenary, cusped, circular, and helical motion. Moreover, these classes comprise all possible stationary worldlines in $(3+1)$-dimensional Minkowski spacetime, namely worldlines along which the invariant separation between two events depends only on their proper-time difference. As a result, a detector following any of these trajectories has a stationary transition rate in the long-interaction limit. Linear, catenary, cusped, and circular motion, all of which have vanishing second torsion, can be realized in two spatial dimensions, whereas the helical class, whose second torsion is nonzero, requires a third spatial direction. In fact, the four planar classes form a continuous family. At fixed acceleration $a>0$, varying the torsion-to-acceleration ratio $\bar b=b/a$, where $b\geq0$ is the magnitude of the first torsion, connects linear motion at $\bar b=0$ to catenary motion for $0<\bar b<1$, cusped motion at $\bar b=1$, and circular motion for $\bar b>1$. In particular, catenary motion, also called drifted Rindler motion, can be smoothly deformed into circular motion through the cusped motion~\cite{Good2020,ParryLouko2025}.

To understand the acceleration effects associated with these trajectories, considerable work has been carried out examining single-detector transition probabilities and the effective temperature defined by the ratio of excitation to de-excitation responses~\cite{JuarezAubryMoustos2019,Good2020,Biermann2020,ParryLouko2025}. The effective temperature defined through this detailed balance condition can be compared directly with the Unruh temperature in Eq.~\eqref{eq:unruh-temperature}, even when the field correlations along the detector's trajectory do not satisfy the Kubo--Martin--Schwinger (KMS) condition~\cite{Kubo1957,MartinSchwinger1959}. Unlike the Unruh temperature, which depends only on the acceleration, the resulting effective temperature generally depends on the energy gap of the detector as well as the trajectory. Good \textit{et al.} compared the effective temperatures of detectors following all five classes of uniformly accelerated trajectories in $(3+1)$-dimensional Minkowski spacetime~\cite{Good2020}. In $(2+1)$ dimensions, Biermann \textit{et al.} found a strong suppression of the circular temperature at small energy gaps~\cite{Biermann2020}. Parry and Louko subsequently showed that the vanishing small-gap temperatures of circular and cusped motion are exceptional to these trajectories in $(2+1)$ dimensions, and traced this behavior to the slow decay of the Wightman function along the worldline~\cite{ParryLouko2025}.

On the other hand, while a protocol called entanglement harvesting has been investigated extensively for linearly accelerated detectors, comparable studies of the other accelerated trajectories described above remain relatively limited~\cite{Salton2015,Liu2022Acceleration}. This two-detector setting concerns correlations between distinct worldlines, which cannot be inferred from the individual transition probabilities alone. Such questions fall within relativistic quantum information (RQI), where relativistic motion, spacetime geometry, and causality enter the study of quantum-information processes~\cite{PeresTerno2004,AlsingFuentes2012}. The quantum-field vacuum contains entanglement between spacelike-separated regions~\cite{SummersWerner1985}, and two initially uncorrelated detectors can acquire some of this entanglement through local interactions with the field, a process known as entanglement harvesting~\cite{Valentini1991,Reznik2003,ReznikRetzkerSilman2005}. More generally, correlation harvesting also concerns the total correlations quantified by quantum mutual information~\cite{PozasKerstjens2015} beyond quantum entanglement. These protocols have been used to investigate cosmological expansion~\cite{Martín-Martínez2026}, black-hole backgrounds~\cite{HendersonBlackHole2018}, and gravitational waves~\cite{XuAhmadSmith2020}. Experimental work has also detected spacelike vacuum-field correlations~\cite{Settembrini2022} and separated correlations induced by vacuum fluctuations from those induced by source radiation~\cite{Herter2026}, while a recent proposal describes entanglement harvesting by bound impurities in a condensate~\cite{Perche2026}.

For accelerated detectors in $(3+1)$-dimensional Minkowski spacetime, entanglement harvesting can be enhanced or suppressed depending on the energy gap and the arrangement of the worldlines~\cite{Salton2015,Liu2022Acceleration}. Circular trajectories introduce additional dependence on orbital parameters and relative orientation~\cite{ZhangYu2020}. Quantum mutual information can also distinguish uniformly accelerated detectors in vacuum from inertial detectors in a thermal bath even when their individual responses agree~\cite{Naeem2023}. Bozanic \textit{et al.} compared linear, catenary, cusped, and circular motion in stationary and nonstationary configurations~\cite{Bozanic2023}. Their comparison showed how the choice of configuration affects the acquired correlations. They also identified stationary linear and circular configurations that allow genuine entanglement harvesting even when the detectors are in causal contact. Here, genuine harvesting refers to the extraction of preexisting field entanglement, distinguished from entanglement assisted by communication between the detectors~\cite{ETjoa2018,Zambianco2024}. Separating these two contributions makes it possible to assess harvesting when causal contact is present. Further examples of genuine harvesting during causal contact have been obtained for derivative-coupled detectors~\cite{TeixidoBonfill2024}.

In this paper, we investigate uniformly accelerating UDW detectors coupled to a real massless scalar field in the $(2+1)$-dimensional Minkowski vacuum, using Gaussian switching to study both single-detector responses and two-detector correlation harvesting. Our choice of spacetime dimension is motivated by proposed experiments that realize effective quantum fields in two spatial dimensions. For example, Gooding \textit{et al.} proposed a localized
laser beam moving through a quasi-two-dimensional
Bose--Einstein condensate as a detector of the circular
Unruh effect~\cite{Gooding2020}. Similarly, Bunney
\textit{et al.} proposed using a moving laser beam to probe
third sound, the surface modes of a thin superfluid-helium
film~\cite{Bunney2024}. Both platforms support an effective
$(2+1)$-dimensional relativistic field in the relevant
long-wavelength regime. More recently, Gooding
\textit{et al.} proposed a nondestructive heterodyne
readout scheme for quasi-two-dimensional Bose--Einstein
condensates, allowing unequal-time density correlations
to be extracted along stationary probe
trajectories~\cite{Gooding2026}. Moreover, two-dimensional condensates have already been used experimentally to simulate quantum-field dynamics in curved spacetime~\cite{Viermann2022}. For these planar platforms, a $(2+1)$-dimensional analysis is therefore a more practical choice than a $(3+1)$-dimensional treatment for connecting theoretical detector predictions with experiments.

We study both single-detector responses and two-detector correlation harvesting protocol for UDW detectors. A common Gaussian switching function allows the four planar accelerated classes to be compared at finite interaction duration. For two detectors, we consider horizontally displaced trajectories: two copies of the same worldline related by a constant spatial translation within the plane of motion. Each local Wightman function evaluated through individual worldlines is stationary, but the nonlocal Wightman function evaluated through different worldlines is generally nonstationary. This simple coplanar configuration lets us compare the effects of acceleration, torsion, and separation on local excitation and acquired correlations.

For a single detector, the finite interaction distinguishes our questions from those based on stationary transition rates. Parry and Louko emphasized that a finite duration may change the small-gap behavior inferred from a detector operating for arbitrarily long times~\cite{ParryLouko2025}. Finite-time detector responses have also been studied more generally~\cite{SriramkumarPadmanabhan1996,Fewster2016}, and Barman \textit{et al.} have already considered Gaussian-switched circular detectors in $(2+1)$-dimensional vacuum and thermal states, including collective transitions of initially entangled pairs~\cite{Barman2024}. Recent work on circular motion has further examined simultaneous small-gap and long-interaction limits with suitably scaled switching families~\cite{ParryWaiting2025,ParryFewsterLouko2026}. Here we keep a fixed Gaussian switching function and study the dependence on acceleration, gap, and torsion across the full planar family. We combine numerical calculations with acceleration expansions and an analysis of monotonicity in the torsion-to-acceleration ratio $\bar b$.

The single-detector results show that the leading terms of the transition probability and the effective temperature at the small-acceleration limit are the same for all four trajectories. Torsion first affects the probability at fourth order at the small-acceleration limit, and this torsion-dependent correction may change sign when choosing different values of energy gap. At large-acceleration limit, the normalized excitation probabilities approach a common limit for linear, catenary, and cusped motion, while circular motion has a smaller, gap-dependent limit. The effective temperatures nevertheless exhibit three different behaviors: gap-independent linear growth for linear and catenary motion, slower growth for the cusped motion, and a finite limit for circular motion. At fixed acceleration and energy gap, the excitation probability and effective temperature have the same increasing and decreasing intervals as functions of $\bar b$. Moreover, both decrease throughout the range of $\bar{b}$ at sufficiently small energy gaps for every fixed nonzero acceleration. We determine the critical energy gaps at which this monotonicity first fails, either immediately away from linear motion when $\bar{b}=0$ or at a positive $\bar{b}$. Therefore, transforming from linear motion to circular motion at a fixed acceleration can result in a non-monotonic change in the response of the detector. 

For two detectors, we compute both harvested quantum mutual information and entanglement (using concurrence to measure the latter) at the same accelerations and separations. Their different dependence on the local and nonlocal terms allows quantum mutual information to remain nonzero in regions where concurrence vanishes. We also trace the boundary of positive concurrence and find that the representative circular configuration supports entanglement over a wider separation range than the other three configurations at moderate and large accelerations. Finally, we separate the anticommutator and commutator contributions of the Wightman function to distinguish genuine entanglement harvesting from the quantum vacuum from communication-assisted entanglement. For all four trajectory classes, the horizontally displaced configurations contain separation intervals where genuine concurrence is positive while communication-assisted concurrence vanishes. These intervals become substantially wider at the larger energy gap in the representative settings studied here.

Our paper is organized as follows. Section~\ref{sec:protocol} introduces the detector model and correlation measures. Section~\ref{sec:trajectories} presents the stationary trajectories and their horizontally displaced configurations. Section~\ref{sec:transition-temperature} studies the single-detector excitation probability and effective temperature, including their acceleration limits and torsion dependence. Section~\ref{Concurrence and quantum mutual information between uniformly accelerating detectors} examines the two-detector correlations, the boundary of positive concurrence, and genuine entanglement harvesting. The derivations of the acceleration expansions and monotonicity results are given in Appendices~\ref{app:acceleration-asymptotics} and \ref{app:torsion-monotonicity}, respectively. Throughout this work, we use natural units $\hbar=c=k_{\mathrm B}=1$ and the metric convention $\eta_{\mu\nu}=\operatorname{diag}(-1,1,1)$.

\section{Correlation harvesting protocol with Unruh--DeWitt detectors in \texorpdfstring{$(2+1)$}{(2+1)}-dimensional Minkowski spacetime}
\label{sec:protocol}

\subsection{Detector model and reduced density matrix}
\label{subsec:model}

We study correlation harvesting between two pointlike Unruh--DeWitt detectors, labeled $A$ and $B$, coupled locally to a real massless scalar field in the Minkowski vacuum~\cite{Reznik2003,PozasKerstjens2015,Naeem2023}. The detectors are identical two-level systems with ground states $|g_j\rangle$, excited states $|e_j\rangle$, and energy gap $\Omega>0$, where $j\in\{A,B\}$. Each detector follows a prescribed timelike trajectory $x_j^\mu(\tau_j)$ parameterized by its proper time $\tau_j$. In the interaction picture, the Hamiltonian generating evolution with respect to $\tau_j$ is
\begin{equation}
 H_j^{(\tau_j)}(\tau_j)=\lambda\chi(\tau_j)\hat\mu_j(\tau_j)
 \otimes\hat\phi\bigl(x_j(\tau_j)\bigr),
 \label{eq:proper-hamiltonian}
\end{equation}
where $\lambda$ is the coupling constant and $\chi$ is a real switching function. The detector's monopole moment is
\begin{equation}
 \hat\mu_j(\tau_j)=e^{i\Omega\tau_j}|e_j\rangle\langle g_j|
 +e^{-i\Omega\tau_j}|g_j\rangle\langle e_j|.
 \label{eq:monopole}
\end{equation}
The field operator in Eq.~\eqref{eq:proper-hamiltonian} is evaluated at the detector's position, so the interaction is local along its trajectory.

To describe both interactions using the common Minkowski time $t$, we write the total interaction Hamiltonian as
\begin{equation}
 H_I(t)=\sum_{j=A,B}\frac{d\tau_j}{dt}
 H_j^{(\tau_j)}\bigl(\tau_j(t)\bigr).
 \label{eq:common-hamiltonian}
\end{equation}
The corresponding time-evolution operator is
\begin{equation}
 U_I=\mathcal T_t\exp\!\left[-i\int_{\mathbb R}dt\,H_I(t)\right],
 \label{eq:evolution}
\end{equation}
where $\mathcal T_t$ orders the interactions with respect to $t$. In perturbation theory, $U_I=\mathbb{I}+U_I^{(1)}+U_I^{(2)}+\mathcal O(\lambda^3)$, with
\begin{subequations}
\label{eq:dyson}
\begin{align}
 U_I^{(1)}&=-i\int_{\mathbb R}dt\,H_I(t),\label{eq:dyson-first}\\
 U_I^{(2)}&=-\int_{\mathbb R}dt\int_{-\infty}^{t}dt'\,
 H_I(t)H_I(t').\label{eq:dyson-second}
\end{align}
\end{subequations}
We assume that both detectors are initially in their ground states and uncorrelated with the field. The initial state of the full system is therefore
\begin{equation}
 \rho_0=|g_Ag_B\rangle\langle g_Ag_B|\otimes|0\rangle\langle0|.
 \label{eq:initial}
\end{equation}
After the interaction, tracing out the field gives the reduced detector state $\rho_{AB}=\operatorname{Tr}_{\phi}(U_I\rho_0U_I^\dagger)$. Terms of odd order in $\lambda$ vanish in this reduced state because odd vacuum correlation functions of the free field vanish~\cite{PozasKerstjens2015}.

The second-order contributions are determined by the vacuum Wightman function $W(x,x'):=\langle0|\hat\phi(x)\hat\phi(x')|0\rangle$. For a massless scalar field in $(2+1)$-dimensional Minkowski spacetime, it is~\cite{HodgkinsonLouko2012,Henderson2019,Biermann2020}
\begin{equation}
 W(x,x')=\frac{1}{4\pi}\lim_{\epsilon\to0^+}
 \frac{1}{\sqrt{-(t-t'-i\epsilon)^2+
 |\boldsymbol{x}-\boldsymbol{x}'|^2}},
 \label{eq:wightman}
\end{equation}
where $\boldsymbol{x}=(x,y)$ and the principal branch of the square root is understood. The limit is taken in the sense of distributions; the $i\epsilon$ prescription fixes the phase for timelike separations.

In the basis $\{|g_Ag_B\rangle,|e_Ag_B\rangle,|g_Ae_B\rangle,|e_Ae_B\rangle\}$, the final reduced density matrix is~\cite{PozasKerstjens2015}
\begin{equation}
 \setlength{\arraycolsep}{3pt}
 \rho_{AB}=\begin{pmatrix}
 1-\mathcal L_{AA}-\mathcal L_{BB}&0&0&\mathcal M^*\\
 0&\mathcal L_{AA}&\mathcal L_{AB}^*&0\\
 0&\mathcal L_{AB}&\mathcal L_{BB}&0\\
 \mathcal M&0&0&0
 \end{pmatrix}+\mathcal O(\lambda^4),
 \label{eq:rho}
\end{equation}
where
\begin{subequations}
\label{eq:coefficients}
\begin{align}
 \mathcal L_{ij}={}&\lambda^2\int_{\mathbb R}d\tau_i
 \int_{\mathbb R}d\tau_j'\,
 \chi(\tau_i)\chi(\tau_j')e^{-i\Omega(\tau_i-\tau_j')}
 \nonumber\\[-0.3em]
 &\hspace{3em}\times W\bigl(x_i(\tau_i),x_j(\tau_j')\bigr),
 \label{eq:Lij}\\
 \mathcal M={}&-\lambda^2\int_{\mathbb R}d\tau_A
 \int_{\mathbb R}d\tau_B\,
 \chi(\tau_A)\chi(\tau_B)e^{i\Omega(\tau_A+\tau_B)}
 \nonumber\\[-0.3em]
 &\quad\times\Bigl[
 \Theta(t_A-t_B)W\bigl(x_A(\tau_A),x_B(\tau_B)\bigr)
 \nonumber\\[-0.3em]
 &\qquad+\Theta(t_B-t_A)W\bigl(x_B(\tau_B),x_A(\tau_A)\bigr)
 \Bigr].
 \label{eq:M}
\end{align}
\end{subequations}
Here $t_j=t_j(\tau_j)$ and $\Theta$ is the Heaviside step function. The diagonal elements $\mathcal L_{jj}$ are the excitation probabilities to leading order. The off-diagonal elements $\mathcal M$ and $\mathcal L_{AB}$ determine the leading contributions to entanglement and quantum mutual information, respectively. Our convention identifies $\mathcal M$ with the $|e_Ae_B\rangle\langle g_Ag_B|$ matrix element. The double-excitation probability first appears at fourth order, so Eq.~\eqref{eq:rho} is a perturbative expression for the detector state.

Throughout the following calculations, we use identical Gaussian switching functions,
\begin{equation}
 \chi(\tau_j)=\exp\!\left(-\frac{\tau_j^2}{2\sigma^2}\right),
 \label{eq:switching}
\end{equation}
where $\sigma>0$ sets the characteristic interaction timescale. The switching functions are centered at $t=0$ for the trajectories considered below. In $(2+1)$ dimensions, $\hat\phi$ and $\lambda$ have dimensions of $\mathrm{length}^{-1/2}$, and $\lambda^2\sigma$ is dimensionless. We work in the weak-coupling regime $\lambda^2\sigma\ll1$, with the trajectory and switching parameters held fixed.

\subsection{Measures of detector correlations}
\label{subsec:measures}

We use the concurrence~\cite{Wootters1998} as an entanglement measure for the final detector state. For a two-qubit state, define the spin-flipped matrix $\widetilde\rho_{AB}=(\sigma_y\otimes\sigma_y)\rho_{AB}^*(\sigma_y\otimes\sigma_y)$, where $\sigma_y$ is the Pauli matrix and the complex conjugate is taken in the detector basis. If $r_1\ge r_2\ge r_3\ge r_4$ are the square roots of the eigenvalues of $\rho_{AB}\widetilde\rho_{AB}$, the concurrence is
\begin{equation}
 C_{AB}=\max\{0,r_1-r_2-r_3-r_4\}.
 \label{eq:concurrence-definition}
\end{equation}
For the perturbative detector state~\eqref{eq:rho}, its leading contribution is
\begin{equation}
 C_{AB}=2\max\!\left\{0,|\mathcal M|-
 \sqrt{\mathcal L_{AA}\mathcal L_{BB}}\right\}
 +\mathcal O(\lambda^4).
 \label{eq:concurrence}
\end{equation}
Thus the detectors become entangled at this order when $|\mathcal M|$ exceeds the local noise term $\sqrt{\mathcal L_{AA}\mathcal L_{BB}}$.

Quantum mutual information measures both classical and quantum correlations~\cite{Groisman2005} and is defined by
\begin{equation}
 I_{AB}=S(\rho_A)+S(\rho_B)-S(\rho_{AB}),
 \label{eq:mutual-definition}
\end{equation}
where $S(\rho)=-\operatorname{Tr}(\rho\ln\rho)$ is the von Neumann entropy and $\rho_A$, $\rho_B$ are the reduced states of the individual detectors. At leading order in the coupling, one obtains~\cite{PozasKerstjens2015,Naeem2023}
\begin{align}
 I_{AB}={}&\mathcal L_+\ln\mathcal L_+
 +\mathcal L_-\ln\mathcal L_-
 \nonumber\\
 &-\mathcal L_{AA}\ln\mathcal L_{AA}
 -\mathcal L_{BB}\ln\mathcal L_{BB},
 \label{eq:mutual}
\end{align}
with
\begin{align}
 \mathcal L_\pm=\frac12\Bigl[&\mathcal L_{AA}+\mathcal L_{BB}
 \nonumber\\[-0.3em]
 &\pm\sqrt{(\mathcal L_{AA}-\mathcal L_{BB})^2
 +4|\mathcal L_{AB}|^2}\Bigr].
 \label{eq:Lpm}
\end{align}
We use natural logarithms and the convention $0\ln0=0$. Unlike concurrence, quantum mutual information can remain nonzero for a separable detector state.

\section{Unruh--DeWitt detectors in uniform motion in \texorpdfstring{$(2+1)$}{(2+1)}-dimensional Minkowski spacetime}
\label{sec:trajectories}

\subsection{Single-detector trajectories}
\label{subsec:single}

The response of an accelerated detector depends on its trajectory through the Wightman function evaluated along the worldline. For stationary motion, this function depends only on the proper-time difference, and is therefore invariant under a simultaneous translation of its two proper-time arguments. In the long-interaction limit, the corresponding transition rate is independent of proper time. Linear acceleration is one example of stationary motion, but a constant proper acceleration magnitude can also occur along trajectories with a changing acceleration direction.

Letaw classified stationary worldlines by requiring the invariant interval between any two points to depend only on their proper-time separation~\cite{Letaw1981}. These trajectories have constant Frenet--Serret invariants and are orbits of Minkowski Killing vector fields that are timelike along the orbit. The curvature gives the magnitude of the proper acceleration, while the torsions describe the rotation of an orthonormal frame along the worldline. In $(2+1)$ dimensions there are two independent invariants: the curvature $a>0$ and the torsion, whose magnitude we denote by $b\ge0$~\cite{ParryLouko2025}.\footnote{The general helical stationary family in $(3+1)$ dimensions has nonzero second torsion. A timelike worldline in $(2+1)$ dimensions has a tangent and only two independent normal directions, so there is no independent second torsion. The stationary trajectories allowed in $(2+1)$ dimensions are therefore contained in the four noninertial classes considered here and the inertial case~\cite{Letaw1981,ParryLouko2025}.}

\begin{figure*}[tp]
 \centering
 \trajectorypanel{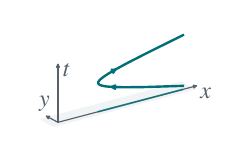}{(a) Linear}\hfill
 \trajectorypanel{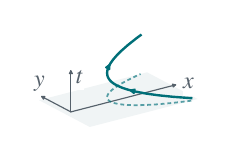}{(b) Catenary}\hfill
 \trajectorypanel{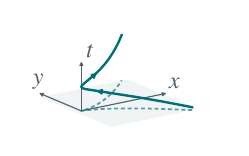}{(c) Cusped}\hfill
 \trajectorypanel{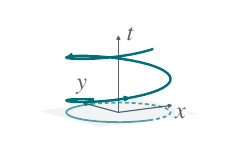}{(d) Circular}
 \caption{Representative stationary worldlines in $(2+1)$-dimensional
 Minkowski spacetime, given by Eqs.~\eqref{eq:linear-trajectory},
 \eqref{eq:catenary-trajectory}, \eqref{eq:cusped-trajectory}, and
 \eqref{eq:circular-trajectory}. Arrows indicate increasing proper time;
 faint dashed curves show spatial projections onto a constant-time
 reference plane.}
 \label{fig:trajectories}
\end{figure*}

We consider worldlines with constant $a$ and $b$ and introduce their dimensionless ratio,
\begin{equation}
 \bar b:=\frac{b}{a}.
 \label{eq:torsion-ratio}
\end{equation}
The four noninertial classes are linear ($\bar b=0$), catenary ($0<\bar b<1$), cusped ($\bar b=1$), and circular ($\bar b>1$). Their geometries are illustrated in Fig.~\ref{fig:trajectories}.

In Minkowski coordinates $x^\mu(\tau)=(t(\tau),x(\tau),y(\tau))$, let
$\Delta\tau=\tau-\tau'$ denote the proper-time difference between two
events on the same worldline. The squared invariant interval entering the
Wightman function is
\begin{equation}
 \Delta s^2(\Delta\tau)=-(t-t')^2+
 |\boldsymbol{x}-\boldsymbol{x}'|^2.
 \label{eq:self-interval-definition}
\end{equation}

For linear acceleration ($\bar{b}=0$), the worldline is
\begin{equation}
 x^\mu_{\mathrm{lin}}(\tau)=
 \left(\frac{\sinh(a\tau)}{a},
 \frac{\cosh(a\tau)}{a},0\right).
 \label{eq:linear-trajectory}
\end{equation}
Combining the time and $x$ differences using the hyperbolic subtraction
identity gives
\begin{equation}
 \Delta s^2_{\mathrm{lin}}(\Delta\tau)
 =-\frac{4}{a^2}\sinh^2\!\left(\frac{a\Delta\tau}{2}\right).
 \label{eq:self-interval-linear}
\end{equation}

For $0<b<a$, the catenary worldline is
\begin{equation}
 \begin{split}
 x^\mu_{\mathrm{cat}}(\tau)=\Bigl(&
 \frac{a\sinh(\sqrt{a^2-b^2}\,\tau)}{a^2-b^2},\,
 \\
 &\frac{a\cosh(\sqrt{a^2-b^2}\,\tau)}{a^2-b^2},\,
 \frac{b\tau}{\sqrt{a^2-b^2}}\Bigr).
 \end{split}
 \label{eq:catenary-trajectory}
\end{equation}
The $t$ and $x$ coordinates retain the same hyperbolic form. And
now the squared invariant interval is
\begin{equation}
 \begin{split}
 \Delta s^2_{\mathrm{cat}}(\Delta\tau)={}&
 \frac{b^2}{a^2-b^2}(\Delta\tau)^2\\
 &-\frac{4a^2}{(a^2-b^2)^2}
 \sinh^2\!\left(\frac{\sqrt{a^2-b^2}\,\Delta\tau}{2}\right).
 \end{split}
 \label{eq:self-interval-catenary}
\end{equation}
Both the worldline and the interval reduce to the linear expressions
at $\bar{b}=0$.

At $b=a$, the catenary parametrization becomes singular. The cusped
worldline at this value can be represented by
\begin{equation}
 x^\mu_{\mathrm{cus}}(\tau)=
 \left(\tau+\frac{a^2\tau^3}{6},
 \frac{a\tau^2}{2},\frac{a^2\tau^3}{6}\right).
 \label{eq:cusped-trajectory}
\end{equation}
The squared invariant interval is then
\begin{equation}
 \Delta s^2_{\mathrm{cus}}(\Delta\tau)=-(\Delta\tau)^2
 -\frac{a^2(\Delta\tau)^4}{12}.
 \label{eq:self-interval-cusped}
\end{equation}

For circular motion ($b>a$), let $R$ be the orbital radius, $\omega$
the angular velocity with respect to Minkowski time, and $\gamma$ the
Lorentz factor. The trajectory is
\begin{equation}
 x^\mu_{\mathrm{cir}}(\tau)=
 \bigl(\gamma\tau,R\cos(\omega\gamma\tau),
 R\sin(\omega\gamma\tau)\bigr),
 \label{eq:circular-trajectory}
\end{equation}
where
\begin{equation}
 \begin{gathered}
 R=\frac{a}{b^2-a^2},\qquad
 \omega=\frac{b^2-a^2}{b},\\[0.3em]
 \gamma=\frac{b}{\sqrt{b^2-a^2}}.
 \end{gathered}
 \label{eq:circular-parameters}
\end{equation}
The orbital speed is $R\omega=a/b=1/\bar b<1$, and
$\omega\gamma=\sqrt{b^2-a^2}$. Thus
\begin{equation}
 \begin{split}
 \Delta s^2_{\mathrm{cir}}(\Delta\tau)={}&-\gamma^2(\Delta\tau)^2\\
 &+4R^2\sin^2\!\left(\frac{\omega\gamma\Delta\tau}{2}\right).
 \end{split}
 \label{eq:self-interval-circular}
\end{equation}
All four worldlines satisfy $\dot x^\mu\dot x_\mu=-1$ and
$\ddot x^\mu\ddot x_\mu=a^2$.

Although the coordinate expressions differ, the four intervals can
be described by a single function of $\bar b$. To see this, rewrite
the catenary interval~\eqref{eq:self-interval-catenary} with $b=a\bar b$:
\begin{equation}
 \begin{split}
 \Delta s^2(\Delta\tau)={}&\frac{\bar b^2}{1-\bar b^2}(\Delta\tau)^2\\
 &-\frac{4}{a^2(1-\bar b^2)^2}
 \sinh^2\!\left(\frac{a\sqrt{1-\bar b^2}\,\Delta\tau}{2}\right).
 \end{split}
 \label{eq:self-interval}
\end{equation}
This expression holds for $\bar b\ne1$ and gives
Eq.~\eqref{eq:self-interval-linear} at $\bar b=0$. For $\bar b>1$,
the argument of the hyperbolic sine is imaginary. Using
$\sinh(ix)=i\sin x$ and Eq.~\eqref{eq:circular-parameters} then gives
the circular interval~\eqref{eq:self-interval-circular}.
As $\bar b\to1$, the quadratic term of the $\sinh^2$ expansion combines
with the first term of Eq.~\eqref{eq:self-interval} to give
$-(\Delta\tau)^2$. The quartic term gives $-a^2(\Delta\tau)^4/12$,
recovering the cusped interval~\eqref{eq:self-interval-cusped}.
The interval is therefore continuous in $\bar b$ at fixed $a$ and
$\Delta\tau$. In all cases, it is timelike for distinct points on the
worldline and satisfies
$\Delta s^2(\Delta\tau)=-(\Delta\tau)^2+\mathcal O((\Delta\tau)^4)$
near coincidence.

For stationary motion, we write $W(x(\tau),x(\tau'))=W_{\bar b}(\tau-\tau')$. Integrating Eq.~\eqref{eq:Lij} over the average proper time with the Gaussian switching functions gives
\begin{equation}
 \mathcal L_{jj}=\lambda^2\sigma\sqrt{\pi}
 \int_{\mathbb R}du\,
 e^{-u^2/(4\sigma^2)}e^{-i\Omega u}W_{\bar b}(u),
 \label{eq:local-response}
\end{equation}
where $u=\tau-\tau'$.

Since all four trajectory classes have an identical expansion of $\Delta s^2(u)=-u^2+\mathcal O(u^4)$ near coincidence, the corresponding short-distance expansion of the Wightman function is
\begin{equation}
 \frac{1}{4\pi\sqrt{\Delta s^2(u-i\epsilon)}}
 =-\frac{i}{4\pi(u-i\epsilon)}
 +\mathcal O(u-i\epsilon).
\end{equation}
Adding and subtracting the singular term before taking
the limit, we write
\begin{equation}
 \begin{split}
 W_{\bar b}(u)={}&-\frac{i}{4\pi(u-i0)}
 \\
 &+\frac{1}{4\pi}\lim_{\epsilon\to0^+}
 \left[
 \frac{1}{\sqrt{\Delta s^2(u-i\epsilon)}}
 +\frac{i}{u-i\epsilon}
 \right].
 \end{split}
 \label{eq:stationary-W-subtraction}
\end{equation}
Here $i0$ denotes the distributional $\epsilon\to0^+$ limit. The short-distance expansion shows that the expression
in brackets is $\mathcal O(u-i\epsilon)$ near coincidence,
so the subtraction removes its singularity.
For real $u\ne0$, the prescribed square-root branch gives
\begin{equation}
 \lim_{\epsilon\to0^+}
 \frac{1}{\sqrt{\Delta s^2(u-i\epsilon)}}
 =-i\,\frac{\operatorname{sgn}(u)}
 {\sqrt{-\Delta s^2(u)}}.
 \label{eq:stationary-W-noncoincident}
\end{equation}
The regular remainder therefore satisfies
\begin{equation}
 \begin{split}
 &\lim_{\epsilon\to0^+}
 \left[
 \frac{1}{\sqrt{\Delta s^2(u-i\epsilon)}}
 +\frac{i}{u-i\epsilon}
 \right]
 \\
 &\qquad=-i\left[
 \frac{\operatorname{sgn}(u)}{\sqrt{-\Delta s^2(u)}}
 -\frac1u
 \right],
 \end{split}
 \label{eq:stationary-W-regular-limit}
\end{equation}
with the right-hand side extended continuously to $u=0$.
Substituting this limit back into the decomposition gives
\begin{equation}
 W_{\bar b}(u)={}-\frac{i}{4\pi(u-i0)}
 -\frac{i}{4\pi}\left[
 \frac{\operatorname{sgn}(u)}{\sqrt{-\Delta s^2(u)}}
 -\frac1u\right].
\end{equation}
The expression in brackets is smooth at coincidence and vanishes there, because $\operatorname{sgn}(u)/\sqrt{-\Delta s^2(u)}=1/u+\mathcal O(u)$. Applying the Sokhotski--Plemelj identity, $(u-i0)^{-1}=\operatorname{PV}(1/u)+i\pi\delta(u)$, and combining the principal-value term with the regular remainder, we obtain
\begin{equation}
 W_{\bar b}(u)=\frac14\delta(u)
 -\frac{i}{4\pi}\operatorname{PV}\!\left[
 \frac{\operatorname{sgn}(u)}{\sqrt{-\Delta s^2(u)}}\right],
 \label{eq:stationary-W}
\end{equation}
where $\operatorname{PV}$ denotes the Cauchy principal value.

Substituting Eq.~\eqref{eq:stationary-W} into Eq.~\eqref{eq:local-response}, the delta-function term contributes $\lambda^2\sigma\sqrt{\pi}/4$. In the remaining term, the Gaussian factor is even and the geometric kernel is odd, so the cosine part of $e^{-i\Omega u}$ integrates to zero. The sine part gives an even integrand, yielding
\begin{align}
 \mathcal L_{jj}={}&\frac{\lambda^2\sigma\sqrt{\pi}}{4}
 \nonumber\\
 &-\frac{\lambda^2\sigma}{2\sqrt{\pi}}
 \int_0^\infty du\,
 \frac{e^{-u^2/(4\sigma^2)}\sin(\Omega u)}
 {\sqrt{-\Delta s^2(u)}}.
 \label{eq:response-real}
\end{align}
For fixed finite $\bar b$, taking $a\to0$ gives $\Delta s^2(u)\to-u^2$ and hence
\begin{equation}
 \left.\mathcal L_{jj}\right|_{a=0}
 =\frac{\lambda^2\sigma\sqrt{\pi}}{4}
 \operatorname{erfc}(\Omega\sigma).
 \label{eq:inertial}
\end{equation}
This agrees with the Gaussian-switched inertial-detector result in Eq.~(C.2) of~\cite{Henderson2019}, where $\operatorname{erfc}(x)=1-\operatorname{erf}(x)$. Its nonzero value at finite $\sigma$ is due to the finite interaction time, a switching effect also present for inertial detectors in $(3+1)$ dimensions~\cite{SriramkumarPadmanabhan1996}.

\subsection{Horizontally displaced trajectories}
\label{subsec:horizontal}

\begin{figure*}[tp]
 \centering
 \trajectorypanel{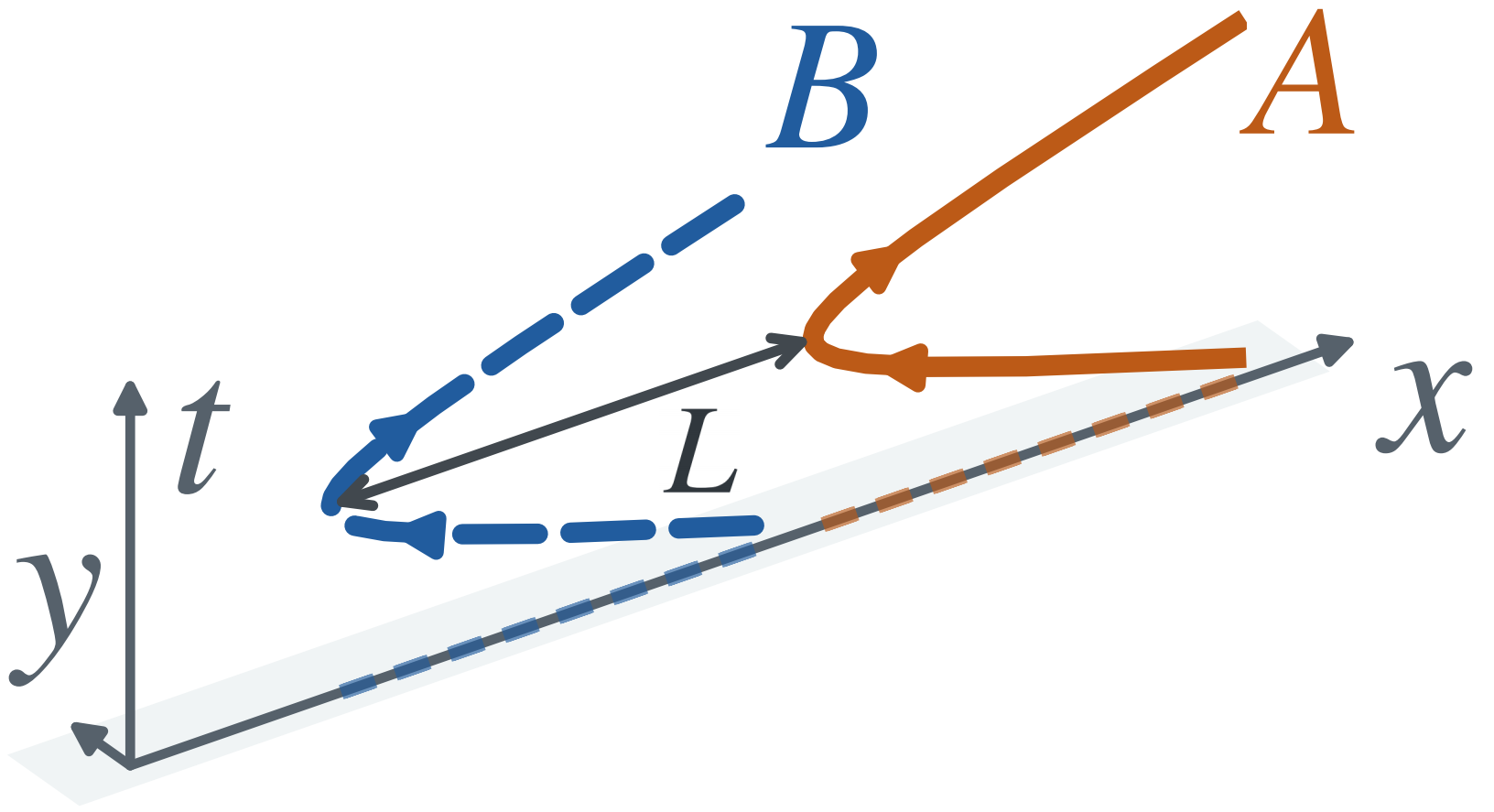}{(a) Linear}\hfill
 \trajectorypanel{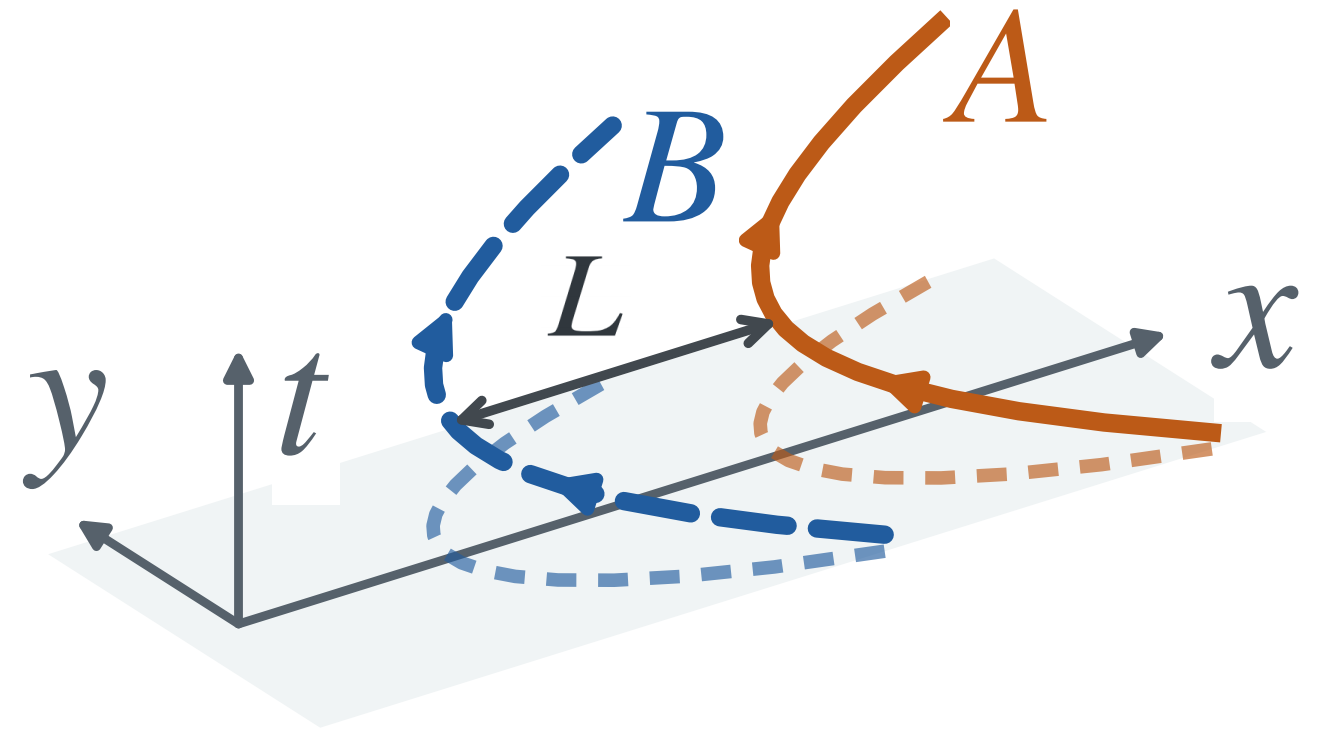}{(b) Catenary}\hfill
 \trajectorypanel{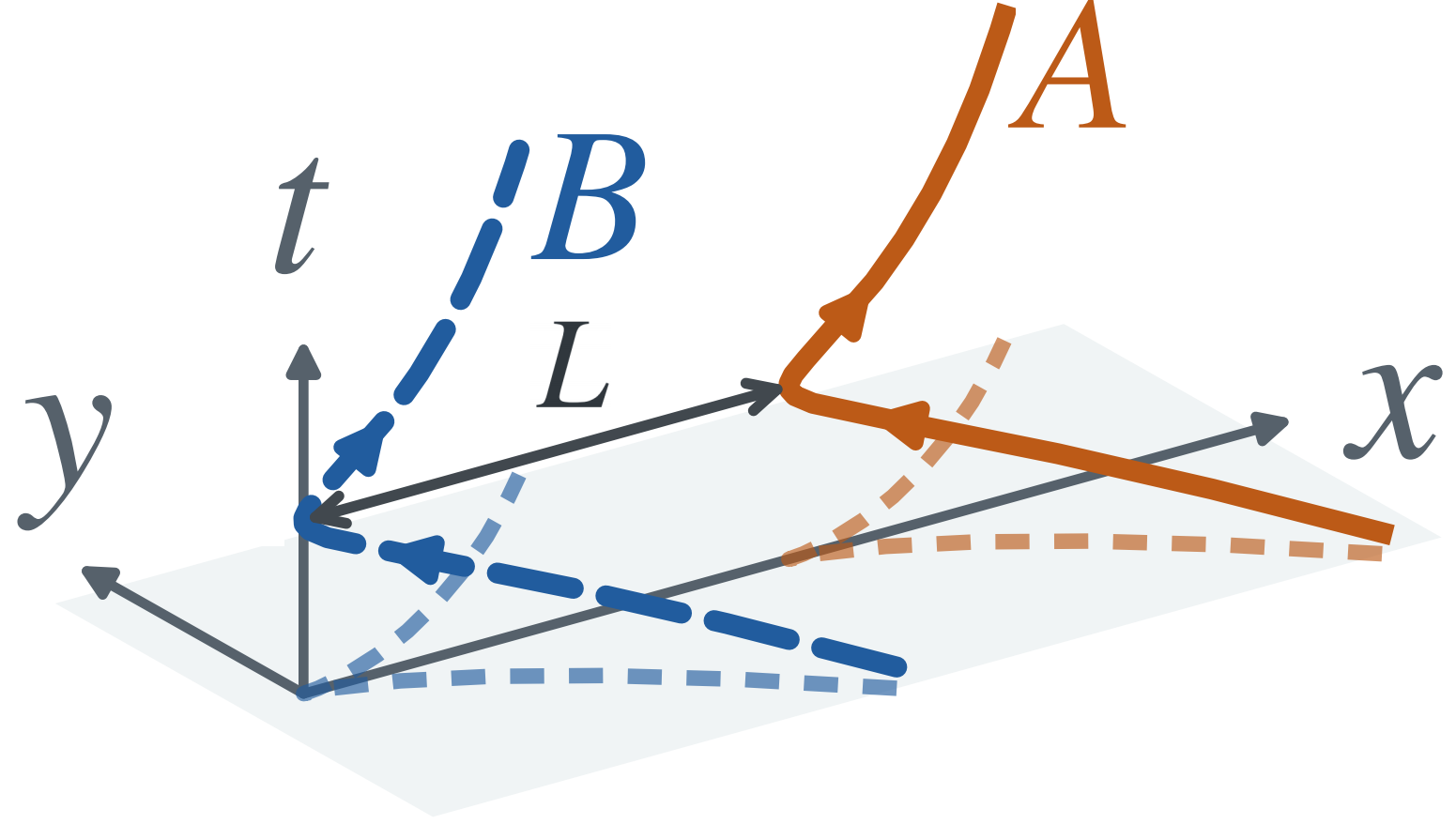}{(c) cusped}\hfill
 \trajectorypanel{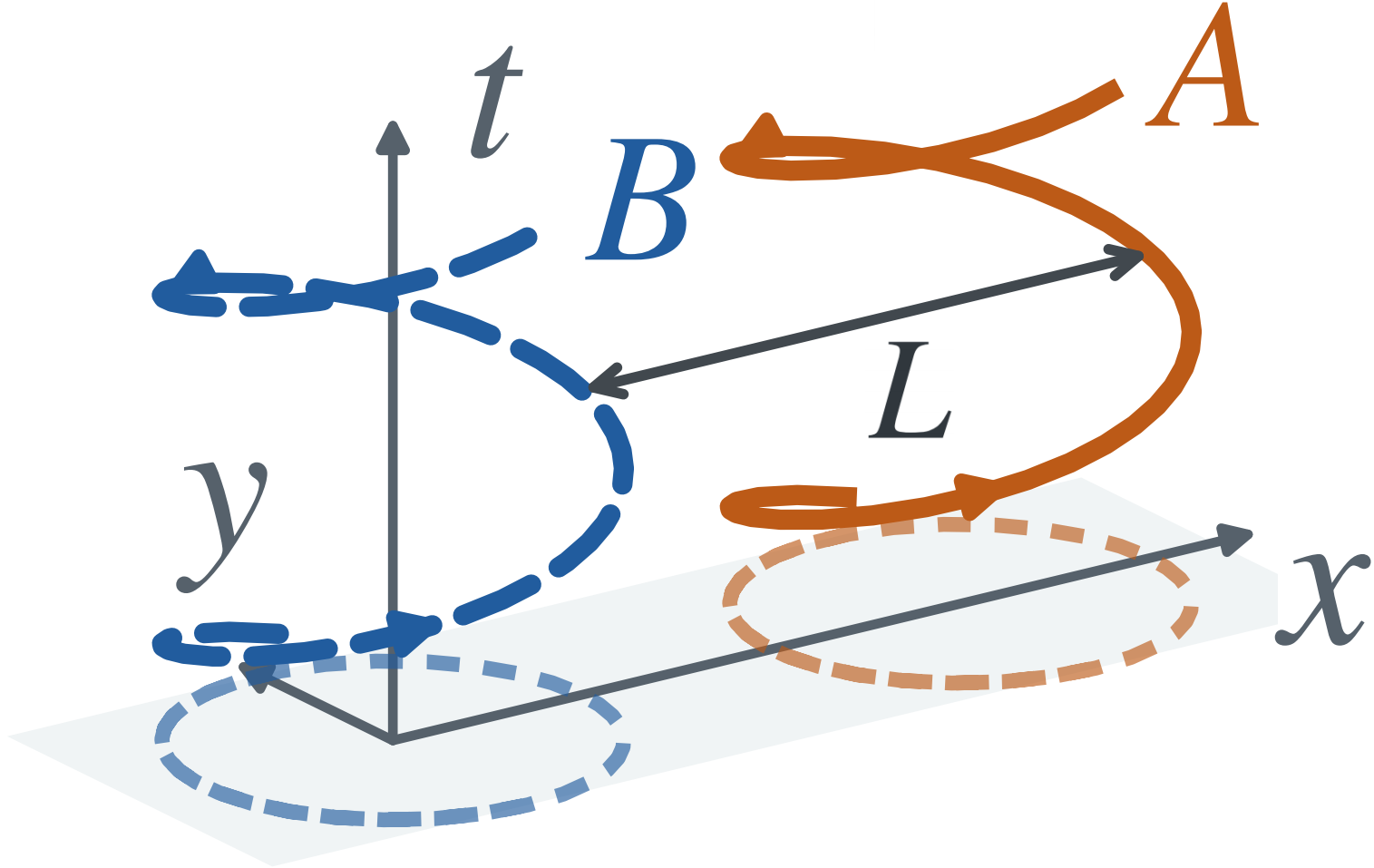}{(d) Circular}
 \caption{Horizontally displaced detector worldlines defined by Eq.~\eqref{eq:horizontal-trajectories}. Detector $A$ is shown in solid orange and detector $B$ in dashed blue. Arrows indicate increasing proper time; the double arrow $L$ joins events at equal Minkowski times. Faint curves show spatial projections. }
 \label{fig:horizontal-trajectories}
\end{figure*}

We now consider two detectors undergoing the same uniform motion, with one trajectory translated by a distance $L>0$ in the $x$ direction. For any of the four trajectories in Sec.~\ref{subsec:single}, we choose
\begin{equation}
 \begin{split}
 x_A^\mu(\tau_A)&=x^\mu(\tau_A)+(0,L,0),\\
 x_B^\mu(\tau_B)&=x^\mu(\tau_B).
 \end{split}
 \label{eq:horizontal-trajectories}
\end{equation}
We refer to these configurations as \emph{horizontally displaced trajectories}. They are the $(2+1)$-dimensional counterparts of the nonstationary configurations in~\cite{Bozanic2023}. Both detectors have the same proper acceleration and torsion, and their proper times are synchronized at $t=0$, when the switching functions attain their maxima. At equal Minkowski times, the detectors have the same velocity and are separated by $L$ along the $x$ axis. Accordingly, $L$ denotes their separation in the chosen inertial frame. Fig.~\ref{fig:horizontal-trajectories} illustrates these pairs of worldlines and the equal-time separation between them.

A spatial translation leaves the Wightman function along an individual trajectory unchanged. The local probabilities are therefore equal, $\mathcal L_{AA}=\mathcal L_{BB}$, and are still given by Eq.~\eqref{eq:local-response}. To evaluate the nonlocal terms, we consider one event on each trajectory. We retain the notation $\Delta\tau$ for the proper-time difference and introduce the sum $\Delta_+\tau$:
\begin{equation}
 \Delta\tau=\tau_A-\tau_B,\qquad
 \Delta_+\tau=\tau_A+\tau_B.
 \label{eq:proper-time-combinations}
\end{equation}
Let $\Delta x=x(\tau_A)-x(\tau_B)$ denote the horizontal coordinate difference on the unshifted worldline. The actual horizontal separation between the two detector events is therefore $x_A-x_B=\Delta x+L$. Their time coordinates remain $t_A$ and $t_B$, and the $y$ coordinates are unchanged by the translation. Expanding $(\Delta x+L)^2$ gives the squared invariant interval between the detector events:
\begin{align}
 &-(t_A-t_B)^2+|\boldsymbol{x}_A-\boldsymbol{x}_B|^2
 \nonumber\\
 &\qquad=\Delta s^2(\Delta\tau)+L^2+2L\Delta x,
 \label{eq:horizontal-interval}
\end{align}
where $\Delta s^2(\Delta\tau)$ is the single-trajectory interval already given in Eq.~\eqref{eq:self-interval-definition}. The displacement thus contributes the constant term $L^2$ and the cross term $2L\Delta x$ to the known single-trajectory result.

To obtain the nonlocal Wightman function between the two trajectories, we evaluate Eq.~\eqref{eq:wightman} at these events and retain its prescription $t_A-t_B\mapsto t_A-t_B-i\epsilon$. The time-dependent part of the denominator expands as
\begin{align*}
 &-(t_A-t_B-i\epsilon)^2
 \\
 &\qquad=-(t_A-t_B)^2+2i\epsilon(t_A-t_B)+\epsilon^2.
\end{align*}
Combining this expansion with Eq.~\eqref{eq:horizontal-interval} yields
\begin{equation}
 \begin{aligned}
 W_{\mathrm{hd}}(\tau_A,\tau_B)
 ={}&\frac{1}{4\pi}\lim_{\epsilon\to0^+}\\
 &\times\Bigl[\Delta s^2(\Delta\tau)+L^2+2L\Delta x\\
 &\qquad{}+2i\epsilon(t_A-t_B)+\epsilon^2\Bigr]^{-1/2}.
 \end{aligned}
 \label{eq:horizontal-W}
\end{equation}
The limit is understood distributionally, with the same square-root branch as in Eq.~\eqref{eq:wightman}. Keeping the $\epsilon$ terms specifies the boundary value of the two-point function when the detector positions approach null separation.

We next evaluate $\Delta x$ for each class of motion. For linear and catenary trajectories, the $x$ coordinates in Eqs.~\eqref{eq:linear-trajectory} and \eqref{eq:catenary-trajectory} have the same hyperbolic-cosine form. Their difference can be evaluated using
\begin{equation}
 \cosh u-\cosh v
 =2\sinh\!\left(\frac{u+v}{2}\right)
 \sinh\!\left(\frac{u-v}{2}\right).
\end{equation}
Taking $u=\sqrt{a^2-b^2}\,\tau_A$ and $v=\sqrt{a^2-b^2}\,\tau_B$ produces one factor depending on $\Delta\tau$ and another on $\Delta_+\tau$. This gives the first line of Eq.~\eqref{eq:horizontal-coordinate-difference} for $0\le\bar b<1$, including linear motion at $b=0$.

For the cusped trajectory, $x(\tau)=a\tau^2/2$, so the coordinate difference follows directly from the difference of two squares:
\begin{equation}
 \Delta x=\frac{a}{2}(\tau_A^2-\tau_B^2)
 =\frac{a}{2}\Delta\tau\,\Delta_+\tau.
\end{equation}
For circular motion, Eq.~\eqref{eq:circular-trajectory} gives $\Delta x=R[\cos(\omega\gamma\tau_A)-\cos(\omega\gamma\tau_B)]$. Here the corresponding identity is
\begin{equation}
 \cos u-\cos v
 =-2\sin\!\left(\frac{u+v}{2}\right)
 \sin\!\left(\frac{u-v}{2}\right).
\end{equation}
With $u=\omega\gamma\tau_A$ and $v=\omega\gamma\tau_B$, it gives the last line of Eq.~\eqref{eq:horizontal-coordinate-difference}. Collecting the three cases, we obtain
\begin{equation}
 \Delta x=\left\{\begin{array}{@{}l@{\;}l@{}}
 \begin{aligned}
 \dfrac{2a}{a^2-b^2}&
 \sinh\!\left(\frac{\sqrt{a^2-b^2}\,\Delta\tau}{2}\right)\\[-0.2em]
 &\times\sinh\!\left(\frac{\sqrt{a^2-b^2}\,\Delta_+\tau}{2}\right),
 \end{aligned}
 &0\le\bar b<1,\\[0.8em]
 \dfrac{a}{2}\Delta\tau\,\Delta_+\tau,&\bar b=1,\\[0.8em]
 \begin{aligned}
 -2R&\sin\!\left(\frac{\omega\gamma\Delta\tau}{2}\right)\\[-0.2em]
 &\times\sin\!\left(\frac{\omega\gamma\Delta_+\tau}{2}\right),
 \end{aligned}
 &\bar b>1.
 \end{array}\right.
 \label{eq:horizontal-coordinate-difference}
\end{equation}
Eq.~\eqref{eq:horizontal-coordinate-difference} shows that the displacement introduces dependence on $\Delta_+\tau$ as well as $\Delta\tau$. A simultaneous translation of both proper-time arguments leaves $\Delta\tau$ unchanged but changes $\Delta_+\tau$, so the nonlocal Wightman function is generally not time-translation invariant. Each detector nevertheless follows a stationary worldline. At equal proper times, $\Delta x=0$ and the interval in Eq.~\eqref{eq:horizontal-interval} reduces to $L^2$, in agreement with the constant horizontal separation.

Substituting Eq.~\eqref{eq:horizontal-W} into Eqs.~\eqref{eq:Lij} and \eqref{eq:M} determines the nonlocal matrix elements $\mathcal L_{AB}$ and $\mathcal M$. In general, the dependence on $\Delta_+\tau$ prevents the Gaussian reduction to a single integral used for the local probability. The time ordering in $\mathcal M$ simplifies because both detectors have the same strictly increasing function $t(\tau)$: $t_A>t_B$ is equivalent to $\tau_A>\tau_B$, so $\Theta(t_A-t_B)=\Theta(\tau_A-\tau_B)$.

For the numerical analysis, we measure times and lengths in units of $\sigma$, and energies and accelerations in units of $\sigma^{-1}$. We normalize the matrix elements by the common factor $\mathcal N:=\lambda^2\sigma$. The dimensionless quantities $\mathcal L_{ij}/\mathcal N$ and $\mathcal M/\mathcal N$ are functions of $a\sigma$, $\bar b$, $\Omega\sigma$, and $L/\sigma$. The concurrence and quantum mutual information are also normalized by $\mathcal N$ in all numerical plots. The local probabilities are independent of $L$, whereas the nonlocal terms depend on the detector separation through Eq.~\eqref{eq:horizontal-interval}. We use these expressions to study the excitation probabilities in Sec.~\ref{sec:transition-temperature} and the concurrence and quantum mutual information in Sec.~V.

For brevity, in the numerical discussion below we refer to $a\sigma$, $\Omega\sigma$, and $L/\sigma$ simply as the acceleration, energy gap, and separation, respectively, without repeatedly specifying that they are dimensionless. Similarly, the plotted quantity $\sigma T_{\mathrm{eff}}$ is referred to as the effective temperature. The notation $\bar b$ continues to denote the torsion-to-acceleration ratio.

\section{Numerical results for a single uniformly accelerating detector}
\label{sec:transition-temperature}

In this section, we first examine the local response, including the transition probability and the effective temperature of a single detector, which is useful to the two detectors case and independent of their separation. The relevant parameters are the acceleration $a\sigma$, the energy gap $\Omega\sigma$, and the torsion-to-acceleration ratio $\bar b$. We report the main features of the dependence of transition probability on these parameters in Sec. \ref{subsec:numerical-probability} and the dependence of effective temperature on these parameters in Sec. \ref{subsec:numerical-temperature}. Then we explain the underlying reasons of these results using the analytical expansion of the transition probability and the effective temperature of a single detector at small- and large-acceleration limits in Sec. \ref{subsec:acceleration-asymptotics} and Monotonicity analysis in the torsion-to-acceleration ratio in Sec. \ref{subsec:torsion-monotonicity}.


\subsection{Transition probability}
\label{subsec:numerical-probability}

\begin{figure}[tbp]
 \centering
 \includegraphics[width=\linewidth]{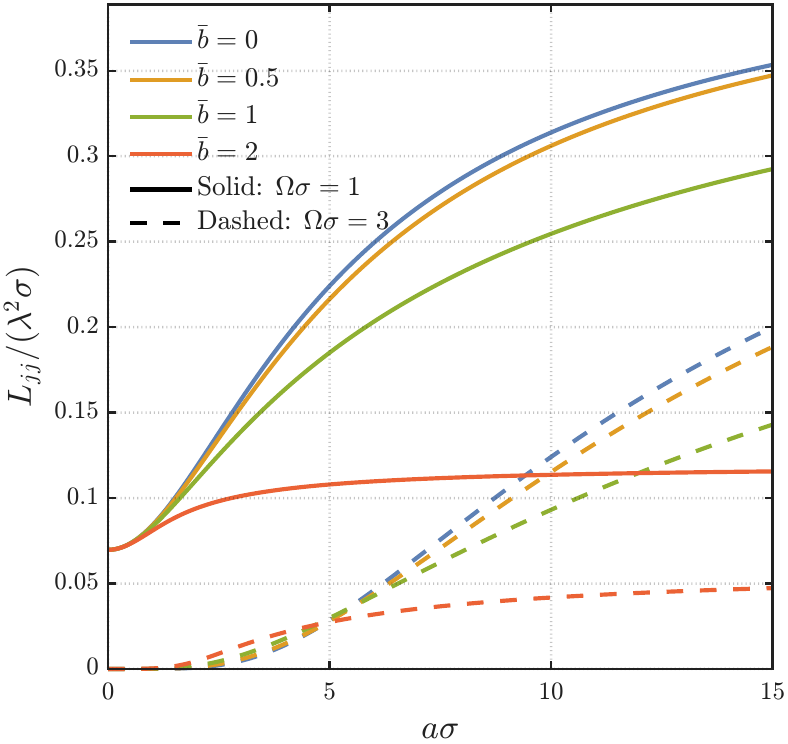}
 \caption{Normalized excitation probability $\mathcal L_{jj}/\mathcal N$ as a function of $a\sigma$. Colors distinguish $\bar b=0$, $0.5$, $1$, and $2$; solid and dashed curves correspond to $\Omega\sigma=1$ and $3$, respectively.}
 \label{fig:probability-acceleration}
\end{figure}

We begin with the transition probabilities in units of $\mathcal N=\lambda^2 \sigma$ in linear, catenary, cusped, and circular motion cases with two different energy gaps $\Omega \sigma=1,3$, represented by $\bar b=0$, $0.5$, $1$, and $2$, respectively, as shown in Fig.~\ref{fig:probability-acceleration}. First we find that the excitation probability increases with $a\sigma$ for all curves in the displayed range. At a fixed gap, the four curves start from the same inertial value~\eqref{eq:inertial}. Specifically, this equation gives $\mathcal L_{jj}/\mathcal N\simeq0.0697014$ and $9.78860\times10^{-6}$ for $\Omega\sigma=1$ and $3$, respectively, illustrating the strong suppression of excitation by the larger gap at small accelerations. For $\Omega\sigma=1$, linear acceleration always gives the largest response, followed by catenary, cusped, and circular motion, and the differences become appreciable as the acceleration increases. In particular, the circular response levels off much earlier than the other three curves.

However, the ordering at $\Omega\sigma=3$ is highly nontrivial, since it is more sensitive to acceleration. At small nonzero $a\sigma$, increasing $\bar b$ enhances the response among these four fixed trajectories, which is opposite to the case where $\Omega\sigma=1$. Moreover, unlike  $\Omega\sigma=1$, when $\Omega\sigma=3$, the ordering reverses as the acceleration grows: the curves cross in the intermediate region around $a\sigma\simeq5$, and linear motion eventually gives the largest probability. A similar reversal was also found for Gaussian-switched detectors in $(3+1)$ dimensions~\cite{Bozanic2023}.

\begin{figure}[tbp]
 \centering
 \includegraphics[width=\linewidth]{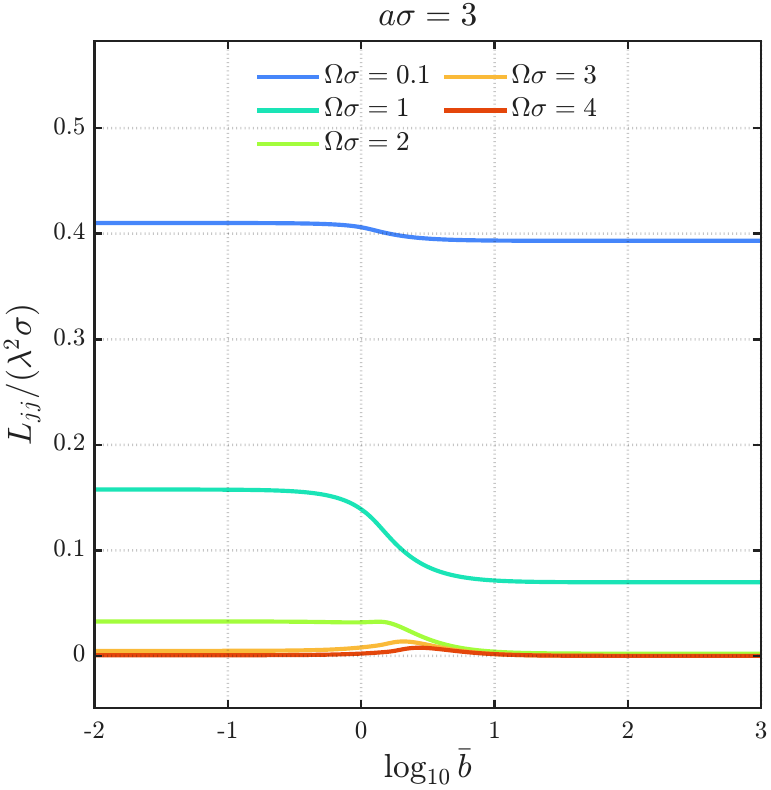}
 \caption{Normalized excitation probability as a function of $\log_{10}\bar b$ at $a\sigma=3$, for $\Omega\sigma=0.1$, $1$, $2$, $3$, and $4$. The origin separates catenary motion from circular motion, with the cusped trajectory at $\bar b=1$.}
 \label{fig:probability-torsion-3}
\end{figure}

\begin{figure}[tbp]
 \centering
 \includegraphics[width=\linewidth]{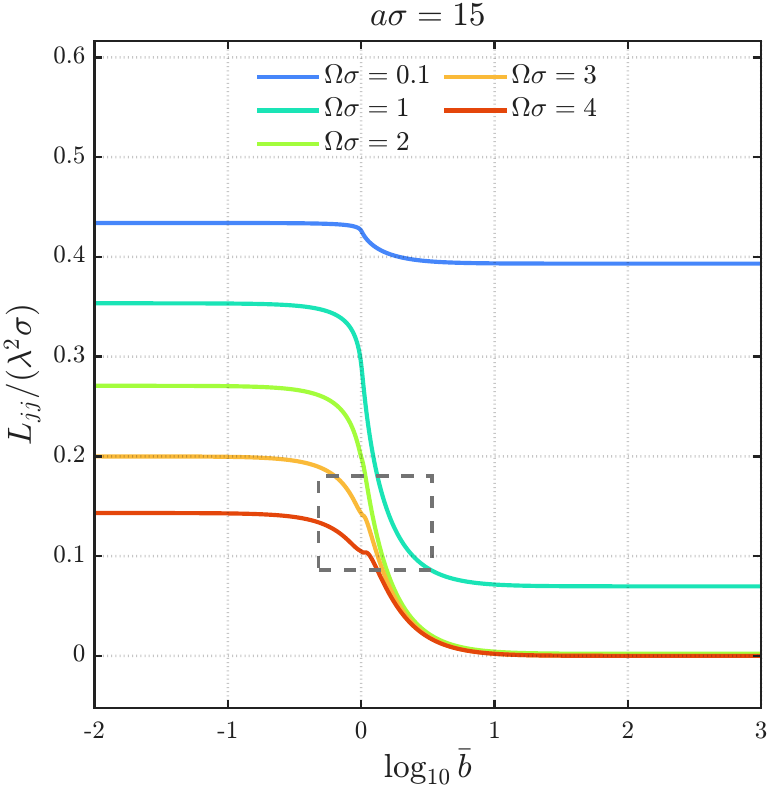}
 \caption{Normalized excitation probability versus $\log_{10}\bar b$ at $a\sigma=15$, for the same gaps as in Fig.~\ref{fig:probability-torsion-3}. The dashed rectangle marks the region enlarged in Fig.~\ref{fig:probability-torsion-zoom}.}
 \label{fig:probability-torsion-15}
\end{figure}

Since each curve corresponds to a specific value of $\bar{b}$, a crossover happened between small- and large-acceleration regime means the monotonicity of transition probability on $\bar{b}$ changes at these two regimes. In order to further track this phenomenon, we plot Fig.\ref{fig:probability-torsion-3} and Fig.\ref{fig:probability-torsion-15} to vary $\bar b$ continuously at $a\sigma=3$ and $a\sigma=15$, which is smaller and larger than $a\sigma\simeq5$, respectively. The horizontal axis is $\log_{10}\bar b$, so the cusped trajectory lies at the origin, with catenary and circular trajectories to its left and right. The nearly horizontal left portions approach the linear-acceleration response as $\bar b\to0$. 

When $a\sigma=3$, as shown in Fig.~\ref{fig:probability-torsion-3}, for $\Omega\sigma=0.1$ and $1$, the transition probability decreases as the trajectory passes through from the linear case to the circular regime. The largest decrease occurs near $\bar b\sim1$, whereas the response becomes almost independent of $\bar b$ sufficiently far from this region on the plotted scale.
For larger gaps, a region appears in which the transition probability increases with $\bar{b}$, and in Fig.~\ref{fig:probability-acceleration}, when $\Omega \sigma=3$, the hierarchy at small acceleration falls into exactly such a region. The $\Omega\sigma=2$ curve appears to be steady as the trajectory passes through the cusped case into the circular regime, while the $\Omega\sigma=3$ and $4$ curves exhibit more pronounced maxima on the circular side. 

\begin{figure}[tbp]
 \centering
 \includegraphics[width=\linewidth]{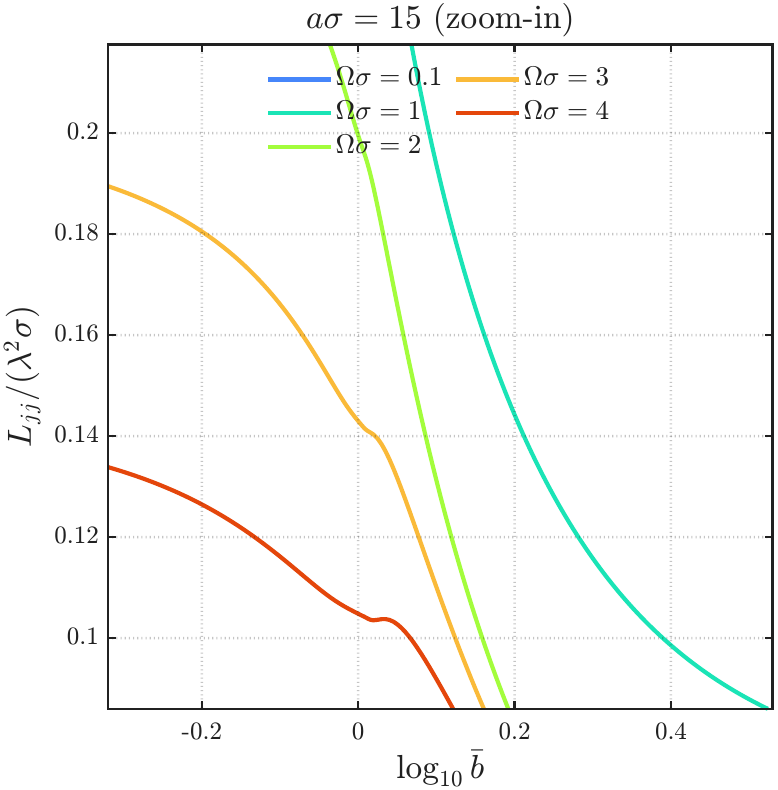}
 \caption{Enlarged view of Fig.~\ref{fig:probability-torsion-15} near the cusped trajectory. At $\Omega\sigma=4$, a local minimum and maximum are visible for $\bar b$ slightly larger than unity. The $\Omega\sigma=3$ curve has a shoulder but remains decreasing.}
 \label{fig:probability-torsion-zoom}
\end{figure}

At $a\sigma=15$, Fig.~\ref{fig:probability-torsion-15} also shows the transition probability decreases as the trajectory passes through from the linear case to the circular regime at small-energy-gaps. Nevertheless, when increasing the energy gap, the $\Omega\sigma=3$ curve develops a shoulder near $\bar b=1$. Although the $\Omega\sigma=4$ curve also appears to decrease at this scale, the enlarged view in Fig.~\ref{fig:probability-torsion-zoom} resolves a shallow local minimum followed by a local maximum just inside the circular regime. This is a small but finite interval of increasing response.

\subsection{Effective temperature}
\label{subsec:numerical-temperature}

The transition probability measures the detector's response to the field, but by itself does not specify a temperature. An operational temperature can be assigned by comparing excitation and de-excitation probability with the same detector gap. Following the detailed-balance prescription used for stationary particle detectors~\cite{JuarezAubryMoustos2019,Biermann2020,Good2020,ZhangYu2020}, we define
\begin{equation}
 T_{\mathrm{eff}}(\Omega)
 =\frac{\Omega}
 {\ln\!\left[\mathcal L_{jj}(-\Omega)/\mathcal L_{jj}(+\Omega)\right]}.
 \label{eq:effective-temperature-definition}
\end{equation}
For a stationary thermal environment, the long-interaction transition rates satisfy detailed balance at the Kubo--Martin--Schwinger (KMS) temperature~\cite{Kubo1957,MartinSchwinger1959}, and the temperature inferred from the ratio of these rates will be independent of the detector's energy gap. However, here Eq.~\eqref{eq:effective-temperature-definition} uses finite-duration probabilities. It is a gap-dependent measure of the excitation--de-excitation asymmetry and does not imply that the detector has equilibrated or that the field along its trajectory is a KMS state. For instance, a Gaussian-switched inertial detector interacting with a massless quantum field in the Minkowski vacuum can still have a nonzero effective temperature because switching produces a nonzero excitation probability, as shown in Eq.~(\ref{eq:inertial-temperature}) in Appendix~\ref{app:small-acceleration}, even though the Minkowski vacuum itself is a zero-temperature quantum state from the perspective of an inertial detector. The reason why we obtain a non-zero effective temperature is the switching effect we impose in the switching function Eq. (\ref{eq:switching}). 

Since the common factor $\mathcal N=\lambda^2 \sigma$ cancels from the ratio in Eq.~(\ref{eq:effective-temperature-definition}) and the effective temperature has the same dimension as the energy gap, the dimensionless effective temperature plotted below is $\sigma T_{\mathrm{eff}}$. Moreover, Eq.~\eqref{eq:response-real} gives a useful identity:
\begin{equation}
 \mathcal L_{jj}(+\Omega)+\mathcal L_{jj}(-\Omega)
 =\frac{\sqrt\pi}{2}\mathcal N.
 \label{eq:probability-sum-rule}
\end{equation}
Consequently, in this massless $(2+1)$-dimensional problem the effective temperature can be obtained directly:
\begin{equation}
 \sigma T_{\mathrm{eff}}
 =\frac{\Omega\sigma}
 {\ln\!\left[(\sqrt\pi/2-\mathcal{P}_{\bar{b}}(\Omega \sigma))/\mathcal{P}_{\bar{b}}(\Omega \sigma)\right]},
 \label{eq:temperature-from-probability}
\end{equation}
where the normalized excitation probability $\mathcal{P}_{\bar{b}}(\Omega \sigma)$ is defined as
\begin{equation}
\mathcal{P}_{\bar{b}}(\Omega \sigma):=\frac{\mathcal{L}_{j j}(+\Omega)}{\mathcal{N}}.
\end{equation}
For these trajectories, the normalized excitation probability satisfies $0<\mathcal P_{\bar b}<\sqrt\pi/4$, and at fixed $\Omega\sigma>0$ the temperature map~\eqref{eq:temperature-from-probability} has a strictly positive derivative with respect to $\mathcal P_{\bar b}$; both statements are proved in Appendix~\ref{app:temperature-monotonicity}. Consequently, at a fixed gap the two observables have the same trajectory ordering, increasing and decreasing intervals, and positions and types of local extrema when varying $a\sigma$ at fixed $\bar b$, or $\bar b$ at fixed $a\sigma$. This correspondence does not constrain comparisons between different energy gaps.

\begin{figure}[tbp]
 \centering
 \includegraphics[width=\linewidth]{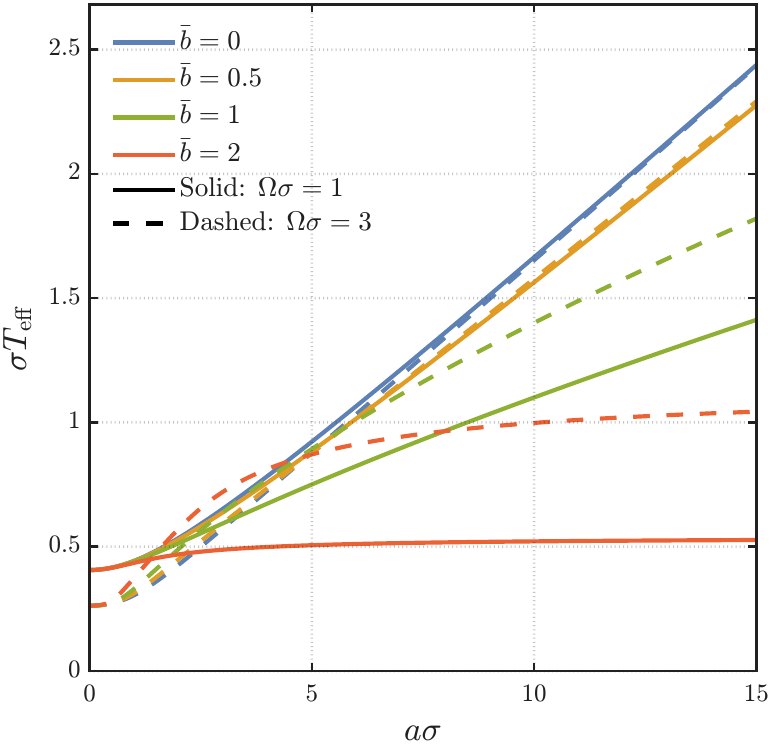}
 \caption{Effective temperature $\sigma T_{\mathrm{eff}}$ versus $a\sigma$, with the same trajectory parameters and line styles as in Fig.~\ref{fig:probability-acceleration}. Linear and catenary motion approach gap-independent linear growth at large acceleration; cusped motion grows more slowly, while circular motion approaches gap-dependent finite limits.}
 \label{fig:temperature-acceleration}
\end{figure}

Fig.~\ref{fig:temperature-acceleration} displays the effective temperature for the same parameters as Fig.~\ref{fig:probability-acceleration}. At each fixed gap, the increase of the excitation probability with $a\sigma$ in Fig.~\ref{fig:probability-acceleration} is therefore accompanied by an increase in the effective temperature over the displayed range. For linear motion, the two gap choices become almost indistinguishable and approach the slope set by the Unruh temperature $T=a/(2\pi)$~\cite{Unruh1976}, namely $\sigma T_{\mathrm{eff}}\sim a\sigma/(2\pi)$ in the large-acceleration regime. The approach to this slope can be understood directly from the
response integral in Eq.~\eqref{eq:local-response}. For linear motion,
Eq.~\eqref{eq:self-interval-linear} implies that the Wightman function
decays on a shorter proper-time scale as the acceleration increases.
At large acceleration, the Gaussian factor in Eq.~\eqref{eq:local-response} is therefore nearly unity over the range of proper-time differences
that contributes appreciably to the integral. Then the response of the detector will approach the long-interaction limit~\cite{Arrechea2021}, where it can
relax to thermal equilibrium. This indicates that,
at large acceleration, the excitation-to-de-excitation
response ratio approaches the stationary thermal
detailed-balance ratio, even at finite switching duration. The large-acceleration
expansion of the excitation probability in
Sec.~\ref{subsec:acceleration-asymptotics} further supports this
interpretation.

Surprisingly, the catenary curves also approach a common linear
growth law with a smaller slope, although the Wightman function
along a catenary trajectory does not satisfy the KMS condition.
The preceding explanation based on a thermal bath therefore does
not apply. Instead, we use the analytic large-acceleration expansion of the effective temperature in Sec.~\ref{subsec:acceleration-asymptotics}
to further support this behavior. On the other hand, the cusped
curves remain more visibly separated over the plotted range and
grow more slowly than the linear and catenary curves, and the
circular curves flatten towards different plateaus for the two gaps.

\begin{figure}[tbp]
 \centering
 \includegraphics[width=\linewidth]{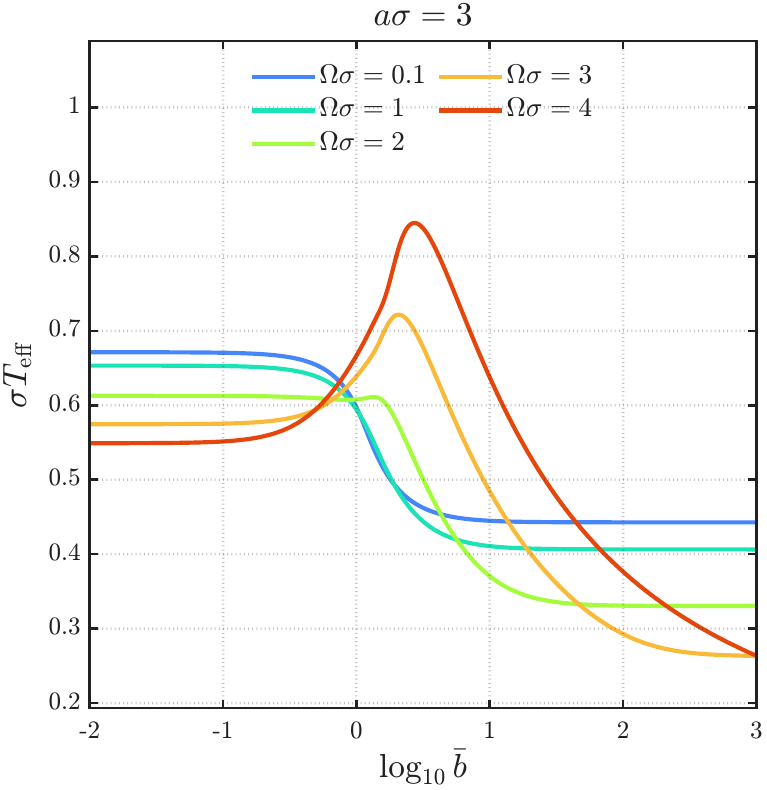}
 \caption{Effective temperature versus $\log_{10}\bar b$ at $a\sigma=3$, for the gaps used in Fig.~\ref{fig:probability-torsion-3}. For each fixed gap, the temperature has its extrema at the same $\bar b$ as the excitation probability. Curves with different gaps can cross.}
 \label{fig:temperature-torsion-3}
\end{figure}

In Fig.~\ref{fig:temperature-torsion-3}, the small-gap curves decrease when increasing the value of $\bar{b}$. At $\Omega\sigma=2$ a nearly steady feature appears before entering deep circular regime, followed by prominent maxima for $\Omega\sigma=3$ and $4$. The positions of local maxima coincide with the probability maxima in Fig.~\ref{fig:probability-torsion-3}, as required by Eq.~\eqref{eq:temperature-from-probability}. Near the catenary end, increasing the gap lowers the temperature for these parameters; near the circular maxima, the ordering is reversed. 

\begin{figure}[tbp]
 \centering
 \includegraphics[width=\linewidth]{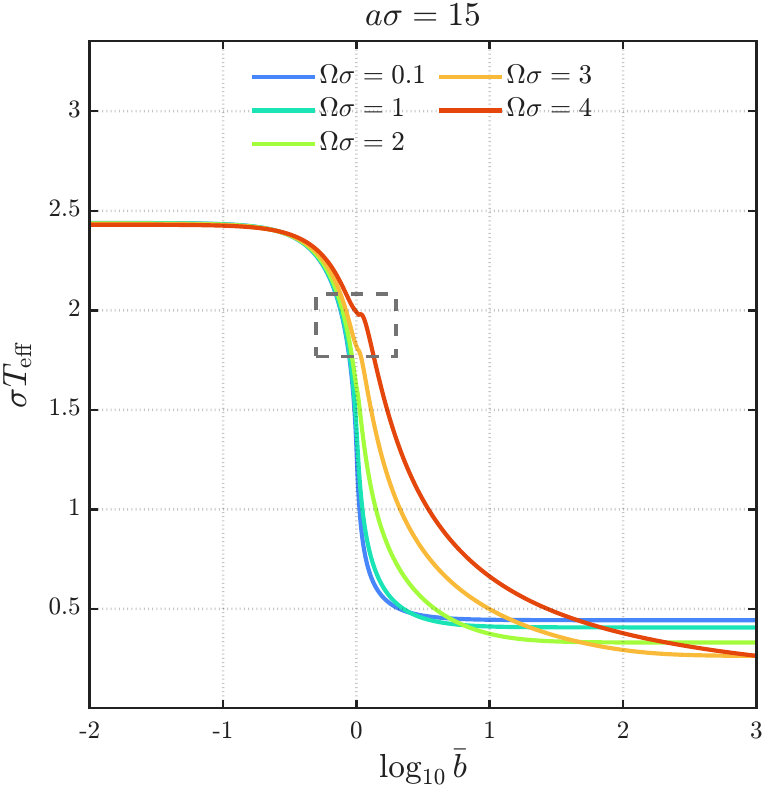}
 \caption{Effective temperature versus $\log_{10}\bar b$ at $a\sigma=15$. The temperatures are nearly gap independent for small $\bar b$ but separate near the cusped trajectory and in the circular regime. The boxed region is enlarged in Fig.~\ref{fig:temperature-torsion-zoom}.}
 \label{fig:temperature-torsion-15}
\end{figure}

\begin{figure}[tbp]
 \centering
 \includegraphics[width=\linewidth]{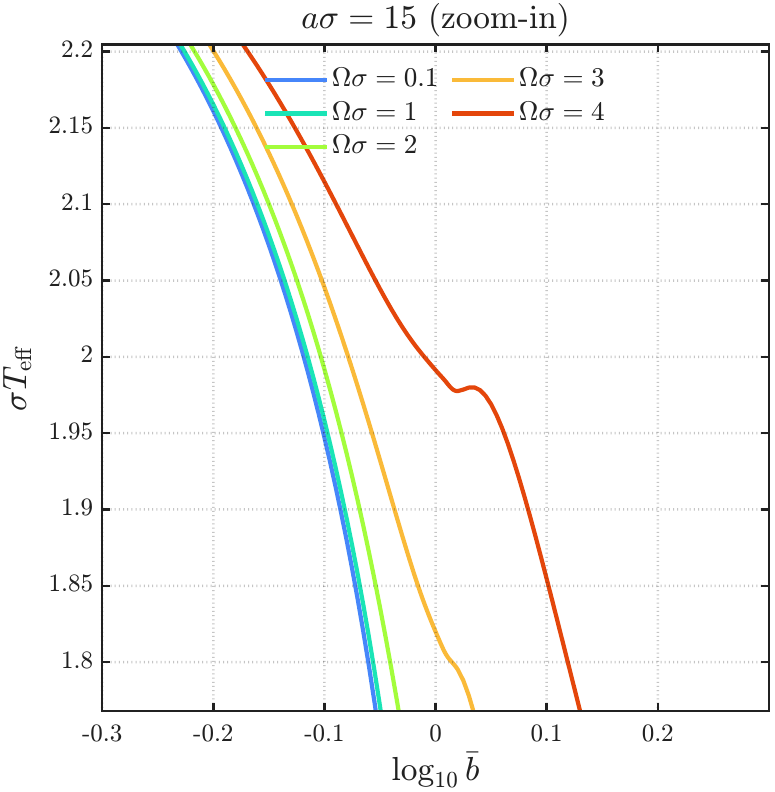}
 \caption{Enlarged view of Fig.~\ref{fig:temperature-torsion-15}. The small interval of increasing temperature for $\Omega\sigma=4$ occurs at exactly the same torsion ratios as the increasing-probability interval in Fig.~\ref{fig:probability-torsion-zoom}.}
 \label{fig:temperature-torsion-zoom}
\end{figure}

The large-acceleration plot in Fig.~\ref{fig:temperature-torsion-15} has a different appearance from the probability plot. Although the excitation transition probabilities for different gaps remain distinct on the catenary side as shown in Fig.~\ref{fig:probability-torsion-15}, their temperatures almost coincide. This exactly matches the energy-independent effective temperature for linear and catenary cases at large-acceleration limit, as shown in Fig.~\ref{fig:temperature-acceleration}. Then, the curves separate near $\bar b=1$ and become strongly gap dependent on the circular side. Fig.~\ref{fig:temperature-torsion-zoom} resolves the small minimum and maximum for $\Omega\sigma=4$, while the $\Omega\sigma=3$ curve has only a shoulder. The temperature feature is therefore the counterpart of the shallow nonmonotonicity in Fig.~\ref{fig:probability-torsion-zoom}.

\subsection{Small- and large-acceleration limits}
\label{subsec:acceleration-asymptotics}

\begin{table*}[tp]
\caption{Acceleration limits of the normalized excitation probability and effective temperature, at fixed $\bar b$ and fixed $\Omega\sigma>0$. The coefficients $\mathcal I$, $\mathcal C$, and $\mathcal R$ are specified in Appendix~\ref{app:acceleration-asymptotics}. The large-acceleration expansions are not uniform near $\bar b=1$.}
\label{tab:acceleration-limits}
\begingroup
\small
\setlength{\tabcolsep}{6pt}
\renewcommand{\arraystretch}{1.2}
\begin{tabular}{@{}p{0.18\textwidth}p{0.09\textwidth}p{0.68\textwidth}@{}}
\toprule
Trajectory & Quantity & Asymptotic expansion \\
\midrule
\multicolumn{3}{@{}l@{}}{\textit{Small acceleration, $a\sigma\ll1$}}\\[0.5em]
All fixed $\bar b$
& $\mathcal P_{\bar b}$
& $\displaystyle\begin{aligned}[t]
 &\frac{\sqrt\pi}{4}\operatorname{erfc}(\Omega\sigma)
 +\frac{(a\sigma)^2(\Omega\sigma)e^{-(\Omega\sigma)^2}}{24}\\[0.25em]
 &\quad-\frac{(7+8\bar b^2)(a\sigma)^4(\Omega\sigma)[3-2(\Omega\sigma)^2]e^{-(\Omega\sigma)^2}}{2880}
 +O\bigl((a\sigma)^6\bigr)
 \end{aligned}$\\
\addlinespace[0.65em]
& $\sigma T_{\mathrm{eff}}$
& $\displaystyle\sigma T_{\mathrm{in}}
 +\frac{(a\sigma)^2(\sigma T_{\mathrm{in}})^2e^{-(\Omega\sigma)^2}}
 {3\sqrt\pi\operatorname{erfc}(\Omega\sigma)\operatorname{erfc}(-\Omega\sigma)}
 +O\bigl((a\sigma)^4\bigr)$\\
\addlinespace[0.4em]
\midrule
\multicolumn{3}{@{}l@{}}{\textit{Large acceleration, $a\sigma\gg1$}}\\[0.5em]
Linear/catenary\newline $0\leq\bar b<1$
& $\mathcal P_{\bar b}$
& $\displaystyle\frac{\sqrt\pi}{4}
 -\frac{(\Omega\sigma)\mathcal I(\bar b)}{2\sqrt\pi(a\sigma)}
 +O\bigl((a\sigma)^{-3}\bigr)$\\
\addlinespace[0.5em]
& $\sigma T_{\mathrm{eff}}$
& $\displaystyle\frac{\pi(a\sigma)}{4\mathcal I(\bar b)}
 +O\bigl((a\sigma)^{-1}\bigr)$\\
\addlinespace[0.9em]
Cusped\newline $\bar b=1$
& $\mathcal P_{\bar{b}}$
& $\displaystyle\frac{\sqrt\pi}{4}
 -\sqrt{\frac3\pi}\,
 \frac{(\Omega\sigma)\ln(a\sigma)+\mathcal C(\Omega\sigma)}{a\sigma}
 +O\!\left(\frac{\ln(a\sigma)}{(a\sigma)^3}\right)$\\
\addlinespace[0.5em]
& $\sigma T_{\mathrm{eff}}$
& $\displaystyle\frac{\pi(a\sigma)}
 {8\sqrt3[\ln(a\sigma)+\mathcal C(\Omega\sigma)/(\Omega\sigma)]}
 \left[1+O\!\left(\frac{[\ln(a\sigma)]^2}{(a\sigma)^2}\right)\right]$\\
\addlinespace[0.9em]
Circular\newline $\bar b>1$
& $\mathcal P_{\bar b}$
& $\displaystyle\mathcal P_\infty
 -\frac{(\Omega\sigma)\mathcal R(\bar b)}{2\sqrt\pi(a\sigma)}
 +O\bigl((a\sigma)^{-2}\bigr)$\\
\addlinespace[0.5em]
& $\sigma T_{\mathrm{eff}}$
& $\displaystyle\sigma T_\infty
 -\frac{4\mathcal R(\bar b)(\sigma T_\infty)^2}
 {\pi(a\sigma)[1-\operatorname{erf}^2(\Omega\sigma)/\gamma^2]}
 +O\bigl((a\sigma)^{-2}\bigr)$\\
\addlinespace[0.4em]
\midrule
\multicolumn{3}{@{}l@{}}{\textit{Reference values} \quad($\gamma=\bar b/\sqrt{\bar b^2-1}$ for circular motion)}\\[0.4em]
\multicolumn{3}{@{}l@{}}{$\displaystyle
 \sigma T_{\mathrm{in}}=
 \frac{\Omega\sigma}{\ln[\operatorname{erfc}(-\Omega\sigma)/\operatorname{erfc}(\Omega\sigma)]},
 \qquad
 \mathcal P_\infty=\frac{\sqrt\pi}{4}
 \left[1-\frac{\operatorname{erf}(\Omega\sigma)}{\gamma}\right]
 $}\\
\addlinespace[0.6em]
\multicolumn{3}{@{}l@{}}{$\displaystyle
 \sigma T_\infty=\frac{\Omega\sigma}
 {\ln[(\gamma+\operatorname{erf}(\Omega\sigma))/(\gamma-\operatorname{erf}(\Omega\sigma))]}
 $}\\[0.4em]
\bottomrule
\end{tabular}
\endgroup
\end{table*}

To explain the features discussed in previous section, an efficient way is to expand the same response integral used in the numerical calculation, at least analytically at certain regions. Below for convenience, we first measure the proper-time difference in units of $a^{-1}$. With $z=au$, define
\begin{equation}
 d_{\bar b}(z):=a\sqrt{-\Delta s^2(z/a)}
 \label{eq:dimensionless-separation}
\end{equation}
as the dimensionless invariant separation determined by Eqs.~\eqref{eq:self-interval} and \eqref{eq:self-interval-cusped}. Then Eq.~\eqref{eq:response-real} becomes
\begin{align}
 \mathcal P_{\bar b}={}&\frac{\sqrt\pi}{4}
 -\frac{1}{2\sqrt\pi}\int_0^\infty dz\,
 e^{-z^2/[4(a\sigma)^2]}
 \frac{\sin[(\Omega\sigma)z/(a\sigma)]}{d_{\bar b}(z)}.
 \label{eq:dimensionless-response}
\end{align}
In all acceleration limits below, $\bar b$ and $\Omega\sigma$ are held fixed. In Appendix~\ref{app:acceleration-asymptotics}, we analytically obtain the expansions of the transition probability and the effective temperature at small- and large-acceleration regimes, and the results are summarized in Table~\ref{tab:acceleration-limits}. In Table~\ref{tab:acceleration-limits}, $T_{\mathrm{in}}$ is the effective temperature when $a \sigma=0$, and $\mathcal P_\infty$ and $T_\infty$ are the asymptotic transition probability and effective temperature in the circular case when $a \sigma \rightarrow \infty$. The positive coefficients $\mathcal I(\bar b)$ and $\mathcal R(\bar b)$ depend only on the trajectory, whereas $\mathcal C(\Omega\sigma)$ is a finite gap-dependent correction. Their integral representations and the derivations of all entries are given in Appendix~\ref{app:acceleration-asymptotics}.

At small acceleration, all trajectories approach the same inertial probability Eq.~(\ref{eq:inertial}) without the normalized factor at a fixed gap. Their leading acceleration correction is positive and independent of $\bar b$, accounting for the common initial rise and close spacing in Figs.~\ref{fig:probability-acceleration} and \ref{fig:temperature-acceleration}. The inertial temperature nevertheless depends on the gap: $\sigma T_{\mathrm{in}}\simeq0.40637$ for $\Omega\sigma=1$ and $0.26285$ for $\Omega\sigma=3$. These finite values reflect the effect of the Gaussian switching, as discussed in Sec.~\ref{subsec:numerical-temperature}.

Torsion first affects the probability at order $(a\sigma)^4$. Its coefficient changes sign at $\Omega\sigma=\sqrt{3/2}$. Consequently, at small acceleration regime, increasing $\bar b$ lowers the probability when $\Omega \sigma=1$ but raises it when $\Omega \sigma=3$. This explains the opposite initial orderings of the two sets of curves in Fig.~\ref{fig:probability-acceleration}. The temperature has the same ordering at a fixed gap because Eq.~\eqref{eq:temperature-from-probability} is increasing in $\mathcal P_{\bar b}$. The expansion is taken at fixed $\bar b$ and does not establish monotonicity over the entire torsion range.

At large acceleration, the probability approaches $\sqrt\pi/4$ for linear, catenary, and cusped motion, whereas circular motion has the smaller, gap-dependent limit $\mathcal P_\infty$. The approach is proportional to $1/(a\sigma)$ for linear and catenary trajectories, but to $\ln(a\sigma)/(a\sigma)$ for the cusped trajectory. Thus the cusped response approaches the common limit more slowly, while the circular response levels off at a distinct value. These different limits account for the separation of the trajectory curves in Fig.~\ref{fig:probability-acceleration}. They also show why the ordering of curves with $\Omega \sigma=3$ at small acceleration must eventually change: the circular probability initially exceeds the other representative responses but has a smaller large-acceleration limit.

The corresponding temperatures fall into three classes. Linear and catenary motion give gap-independent linear growth at leading order. Since $\mathcal I(0)=\pi^2/2$, linear motion recovers the Unruh slope $\sigma T_{\mathrm{eff}}\sim(a\sigma)/(2\pi)$. For $\bar b=0.5$, $\mathcal I(0.5)\simeq5.32034$ gives the smaller slope $0.14762$. More generally, $\mathcal I(\bar b)$ increases through the catenary family, so this slope decreases. The cancellation of the gap in the leading temperature, despite its presence in the probability correction, explains why the catenary temperature curves nearly coincide in Fig.~\ref{fig:temperature-torsion-15} while their excitation probabilities remain distinct in Fig.~\ref{fig:probability-torsion-15}.

Cusped motion instead gives $\sigma T_{\mathrm{eff}}\sim\pi(a\sigma)/[8\sqrt3\ln(a\sigma)]$. Its growth is slower than linear, and the gap-dependent term in the denominator of Table~\ref{tab:acceleration-limits} remains important over the plotted range. In particular, $\mathcal C(1)\simeq-0.42229$ and $\mathcal C(3)\simeq-3.76139$ give different finite-acceleration corrections. The cusped temperature curves can therefore remain visibly separated. This explains both their slower rise and their separation in Fig.~\ref{fig:temperature-acceleration}. For circular motion, both observables approach finite limits from below, since $\mathcal R(\bar b)>0$ as proved in Appendix \ref{app:large-circular}. At $\bar b=2$, $\mathcal R(2)\simeq0.22612$, and the limiting temperatures are $\sigma T_\infty\simeq0.53862$ and $1.13905$ for gaps $1$ and $3$. The factor $\operatorname{erf}(\Omega\sigma)$ retains the Gaussian switching dependence. In the overlapping regime $1\ll\Omega\sigma\ll a\sigma$, it approaches unity and recovers the stationary low-gap circular temperature in $(2+1)$ dimensions~\cite{Biermann2020,ParryLouko2025}.

Finally, the three large-acceleration expansions are not uniform near $\bar b=1$: the cusped result cannot be obtained by simply substituting $\bar b=1$ into either neighboring expression. The analysis in this section can explain the broad trends of the plots, but the narrow minimum and maximum near $\bar b=1$ require a separate analysis of the torsion dependence.

\subsection{Monotonicity in the torsion-to-acceleration ratio}
\label{subsec:torsion-monotonicity}

The acceleration limits explain the broad differences between trajectories at fixed $\bar b$. We now ask when the response, i.e. the transition probability and the effective temperature, viewed as a function of $\bar b$, first develops an increasing interval when varying the energy gap at fixed $a\sigma>0$. We outline the main analysis and results in this section; the detailed proofs are given in Appendix~\ref{app:torsion-monotonicity}.

The two observables can be studied together. As proved in Appendix~\ref{app:temperature-monotonicity}, the normalized transition probability satisfies $0<\mathcal P_{\bar b}<\sqrt\pi/4$ for these trajectories. On this interval, the effective temperature ~\eqref{eq:temperature-from-probability} is strictly increasing with $\mathcal{P}_{\bar{b}}(\Omega \sigma)$ at a fixed positive gap. Differentiation therefore gives
\begin{equation}
 \partial_{\bar b}(\sigma T_{\mathrm{eff}})
 =\frac{d(\sigma T_{\mathrm{eff}})}{d\mathcal P_{\bar b}}
 \partial_{\bar b}\mathcal P_{\bar b} ~~~~~\text{with}
 \qquad
 \frac{d(\sigma T_{\mathrm{eff}})}{d\mathcal P_{\bar b}}>0.
 \label{eq:shared-monotonicity}
\end{equation}
Their increasing and decreasing intervals with $\bar{b}$ thus coincide, as do the positions and types of their local minima and maxima. Therefore, it is sufficient to only analyze $\mathcal P_{\bar b}$.

First, we observe that for every fixed acceleration, there is a nonempty interval of sufficiently small positive gaps in which
\begin{equation}
 \partial_{\bar b}\mathcal P_{\bar b}<0
 \quad\text{for every }\bar b>0.
 \label{eq:small-gap-monotonicity-result}
\end{equation}
At $\bar b=0$ the derivative vanishes, while the response decreases immediately to its right. Appendix~\ref{app:small-gap-monotonicity} proves this statement uniformly over the full torsion-acceleration ratio $\bar{b}$ range. At larger gaps, the oscillating sine factor in Eq.~\eqref{eq:dimensionless-response} allows contributions of opposite signs to compete, and the response need no longer decrease throughout. The pronounced peaks at $a\sigma=3$ and the local minimum and maximum resolved at $a\sigma=15$ in Fig.~\ref{fig:probability-torsion-3} and Fig.~\ref{fig:probability-torsion-15} illustrate this change.

In order to track the threshold of energy gap where non-monotonicity first occurs, we define the critical energy gap by
\begin{align}
 (\Omega\sigma)_c(a\sigma):=
 \inf\bigl\{w>0: &\partial_{\bar b}
 \mathcal P_{\bar b}(a\sigma,w)>0
 \nonumber\\[-0.2em]
 &\text{for some }\bar b\geq0\bigr\}.
 \label{eq:global-critical-gap}
\end{align}
Here $w$ is only a dummy gap argument. Below this value, the response is globally decreasing. At the critical energy gap, its derivative may just reach zero without yet becoming positive. There are two ways for the first increasing interval to appear: immediately to the right of the boundary $\bar b=0$, or around an interior point $\bar b_c>0$.

\begin{figure}[!tbp]
 \centering
 \includegraphics[width=\linewidth]{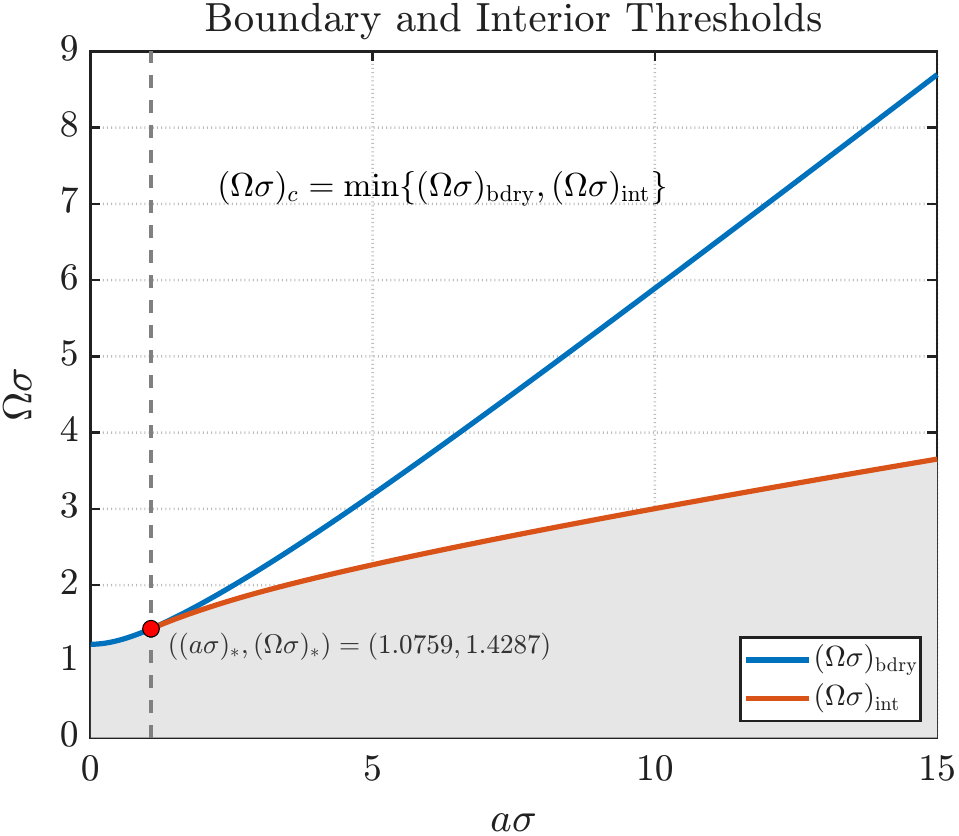}
 \caption{Boundary and interior critical gaps for the first loss of monotonicity in $\bar b$. The lower of the available threshold curves defines $(\Omega\sigma)_c$. Below it, in the shaded region, the response is globally nonincreasing. Nonmonotonic responses are found above it over the computed range. The smaller threshold changes from the boundary value to the interior value at $((a\sigma)_*,(\Omega\sigma)_*)\simeq(1.0759,1.4287)$.}
 \label{fig:monotonicity-thresholds}
\end{figure}

\begin{figure}[!tbp]
 \centering
 \includegraphics[width=\linewidth]{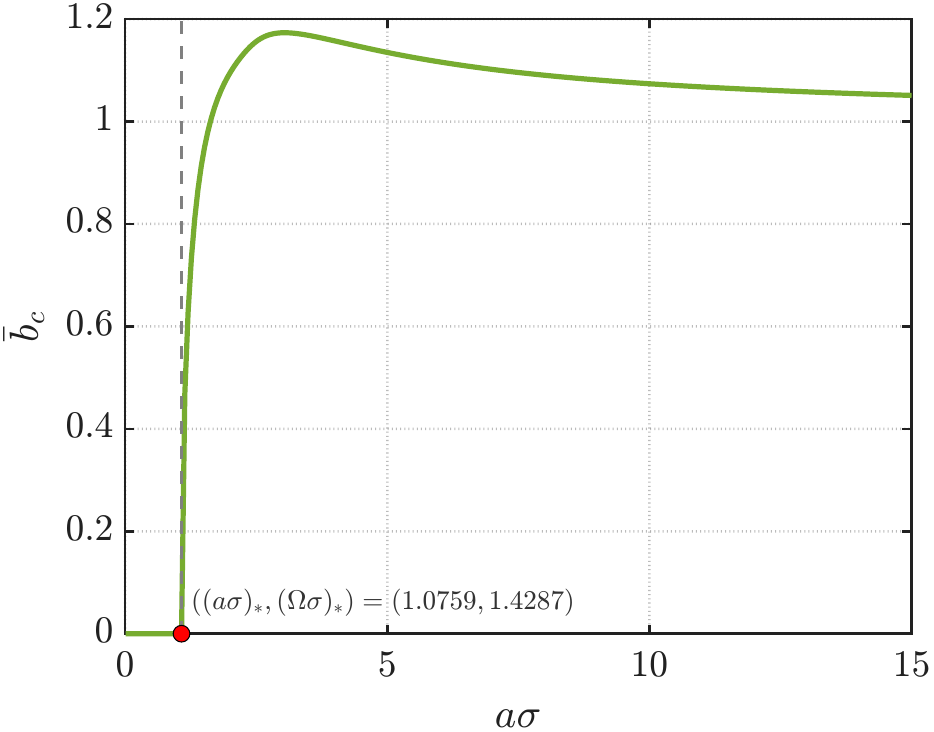}
 \caption{Location $\bar b_c$ where the first increasing interval appears as the gap is raised at fixed acceleration. It is zero while the boundary threshold is lower, becomes positive in the catenary regime, passes through $\bar b_c=1$, and reaches a maximum in the circular regime before decreasing towards unity over the displayed range. This location is evaluated at the critical gap; it is not the position of a minimum or maximum at a larger gap.}
 \label{fig:critical-torsion-ratio}
\end{figure}

In the boundary case, the response begins to increase away from linear motion when its initial curvature changes from negative to positive. Since $\partial_{\bar b}\mathcal P_{\bar b}|_0=0$ at every gap, the boundary threshold is the first such sign-changing root of
\begin{equation}
 \left.\partial_{\bar b}^2\mathcal P_{\bar b}
 \right|_{\bar b=0,\,\Omega\sigma=(\Omega\sigma)_{\mathrm{bdry}}}=0.
 \label{eq:boundary-threshold-condition}
\end{equation}
In the interior case, the first derivative, viewed as a function of $\bar b$, has a local maximum that first reaches zero at $\bar b_c>0$. The interior threshold therefore satisfies
\begin{equation}
 \left.\partial_{\bar b}\mathcal P_{\bar b}\right|_c=0,
 \qquad
 \left.\partial_{\bar b}^2\mathcal P_{\bar b}\right|_c=0,
 \qquad\bar b_c>0,
 \label{eq:interior-threshold-condition}
\end{equation}
at gap $(\Omega\sigma)_{\mathrm{int}}$. Just above this threshold, an interval of increasing behavior will emerge near $\bar b_c$. When both thresholds exist, the critical energy gap is\footnote{%
At fixed $a\sigma$, Eqs.~\eqref{eq:boundary-threshold-condition}
and \eqref{eq:interior-threshold-condition} may yield several
candidate pairs $(\bar b_i,(\Omega\sigma)_i)$. For each pair,
in principle one should check whether increasing the gap around
$\Omega\sigma=(\Omega\sigma)_i$ actually creates an interval of positive slope near
$\bar b_i$. If several candidates pass this check, the smallest
corresponding gap gives $(\Omega\sigma)_c$. In practice, over the range $[0,15]$ of $a \sigma$ shown in Figs.~\ref{fig:monotonicity-thresholds} and \ref{fig:critical-torsion-ratio}, we did not encounter multiple solutions for either threshold, so Eq.~\eqref{eq:threshold-envelope} directly gives $(\Omega \sigma)_c$.}

\begin{equation}
 (\Omega\sigma)_c=
 \min\!\left\{(\Omega\sigma)_{\mathrm{bdry}},
 (\Omega\sigma)_{\mathrm{int}}\right\}.
 \label{eq:threshold-envelope}
\end{equation}

If only one of the two thresholds exists at a given acceleration, that threshold determines $(\Omega\sigma)_c$. Appendix~\ref{app:thresholds} derives the conditions and explains how the two mechanisms meet.

Fig.~\ref{fig:monotonicity-thresholds} compares the numerical results of these two thresholds. At small acceleration the boundary threshold is lower, with $(\Omega\sigma)_{\mathrm{bdry}}\to\sqrt{3/2}$ as $a\sigma\to0^+$, consistently with the fourth-order term in Table~\ref{tab:acceleration-limits}. The two mechanisms meet at
\begin{equation}
 \bigl((a\sigma)_*,(\Omega\sigma)_*\bigr)
 \simeq(1.0759,1.4287).
 \label{eq:threshold-switch-point}
\end{equation}
At larger acceleration, the interior threshold is lower. The response can therefore decrease initially away from $\bar b=0$ yet develop a local minimum and maximum at positive $\bar b$. Both thresholds increase with acceleration over the plotted range, but the interior threshold grows substantially more slowly. Over the computed range, gaps between the two curves give this combination of initial decrease and interior extrema. 

Fig.~\ref{fig:critical-torsion-ratio} shows where the change first occurs. The critical ratio remains at $\bar b_c=0$ below $(a\sigma)_*$, then becomes positive on the catenary side. For example, at $a\sigma=1.5$ it is $\bar b_c\simeq0.9417$. It subsequently crosses the cusped case, reaches about $1.17$ near $a\sigma=3$, and moves back towards unity from the circular side. Thus the first nonmonotonic feature is not confined to circular motion. 

These results explain the representative curves quantitatively. At $a\sigma=3$, the interior and boundary thresholds are approximately $1.9077$ and $2.2065$, with $\bar b_c\simeq1.1734$. The curves with $\Omega \sigma=0.1$ and $\Omega \sigma=1$ in Figs.~\ref{fig:probability-torsion-3} and \ref{fig:temperature-torsion-3} lie below the critical energy gap $(\Omega \sigma)_c(a \sigma=3)$, so they decrease throughout. The curve with $\Omega \sigma=2$ lies between the boundary and interior thresholds, therefore giving a weak local minimum and maximum while the response still decreases near $\bar b=0$. The curves with $\Omega \sigma=3$ and $\Omega \sigma=4$ exceed both thresholds, so the response can increase immediately away from linear motion before reaching the more prominent maxima in the circular regime.

At $a\sigma=15$, the interior threshold is approximately $3.6529$, whereas the boundary threshold is much higher, approximately $8.6935$. The curve with $\Omega \sigma=3$ lies below the critical energy gap, so its shoulder remains monotone. The curve with $\Omega \sigma=4$ lies slightly above the interior threshold and gives the small local minimum and maximum resolved in Figs.~\ref{fig:probability-torsion-zoom} and \ref{fig:temperature-torsion-zoom}. Both the transition probability and the effective temperature reach their minimum at $\bar b\simeq1.0468$ and their maximum at $\bar b\simeq1.0787$. All the detailed derivation and numerical results are given in Appendix~\ref{app:threshold-evaluation}.

\section{Numerical results for two horizontally displaced uniformly accelerating detectors}
\label{Concurrence and quantum mutual information between uniformly accelerating detectors}
We now examine correlations between two uniformly accelerating detectors on the horizontally displaced trajectories introduced in Sec.~\ref{subsec:horizontal}. Entanglement harvesting by accelerated detectors has previously been studied for linear motion~\cite{Salton2015,Liu2022Acceleration} and circular motion~\cite{ZhangYu2020} in $(3+1)$ dimensions. Here we compare the concurrence and quantum mutual information in $(2+1)$ dimensions as functions of $a\sigma$, $\bar b$, and $L/\sigma$, identifying regions where concurrence vanishes while quantum mutual information remains nonzero. We then trace the boundary of entanglement sudden death and examine how the critical separation changes with acceleration, trajectory class, and energy gap. Finally, we distinguish the vacuum-correlation and communication contributions to the nonlocal amplitude and compare the genuine concurrence $C_{AB}^{+}$ with the communication-assisted concurrence $C_{AB}^{-}$. Their positive ranges identify separations at which the vacuum-correlation contribution can overcome local noise even when the communication-assisted concurrence vanishes.

\subsection{Concurrence and quantum mutual information
between uniformly accelerating detectors}
\label{subsec:two-detector-correlations}

\begin{figure*}[p]
 \centering
 \includegraphics[width=\textwidth]{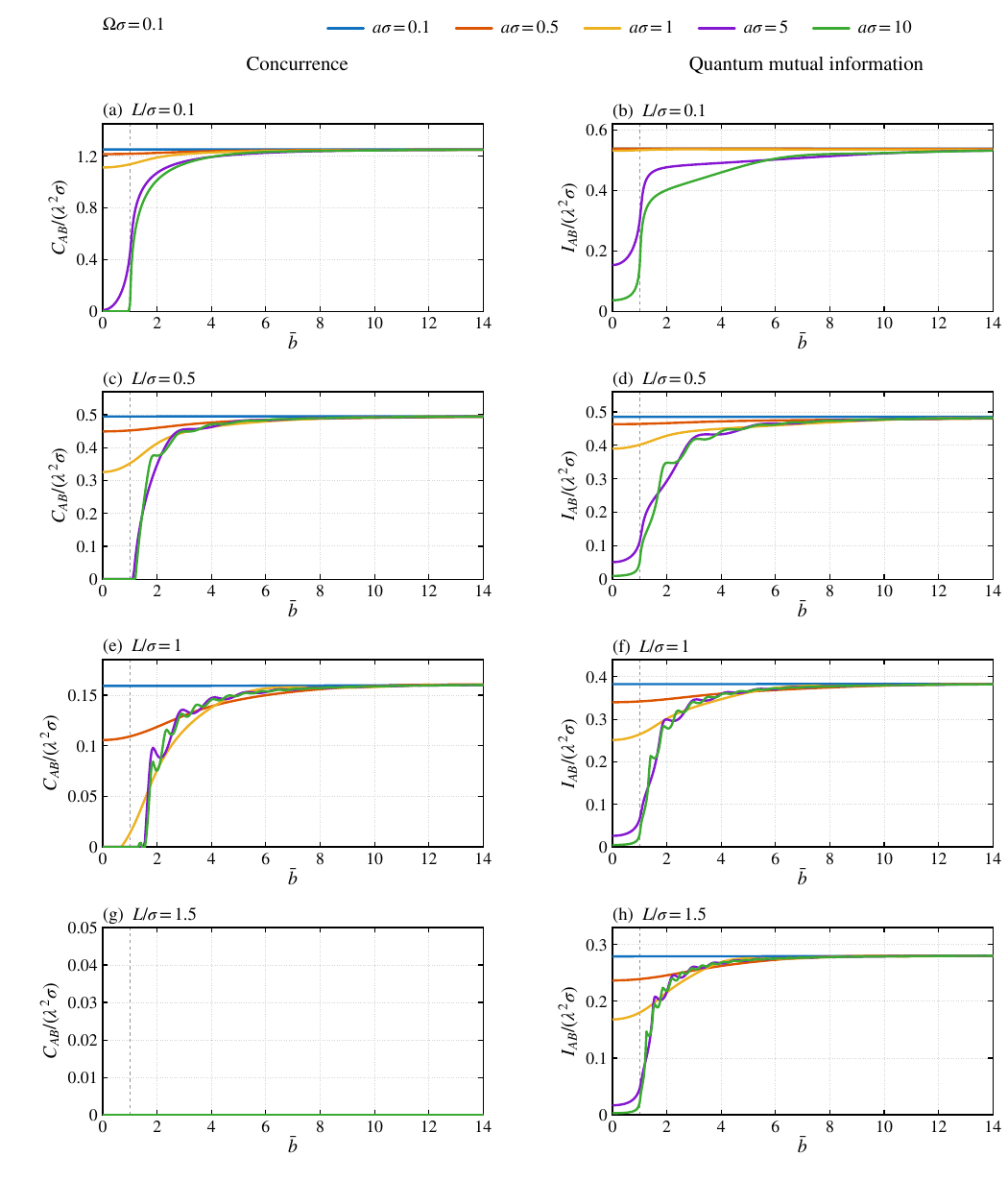}
 \caption{Normalized concurrence $C_{AB}/\mathcal N$ (left column) and
quantum mutual information $I_{AB}/\mathcal N$ (right column) for
horizontally displaced detectors with $\Omega\sigma=0.1$. The paired
panels correspond to $L/\sigma=0.1$ in (a,b), $0.5$ in (c,d), $1$ in
(e,f), and $1.5$ in (g,h). Colors identify the five accelerations in
the shared legend; vertical dashed lines mark $\bar b=1$. Both measures
are evaluated at leading order using Eqs.~\eqref{eq:concurrence} and
\eqref{eq:mutual}. In panel (g), $|\mathcal M|$ does not exceed the local
noise threshold $\sqrt{\mathcal L_{AA}\mathcal L_{BB}}$ anywhere in the
plotted range of $\bar b$, so all five concurrence curves coincide at
zero.}
 \label{fig:two-detector-correlations}
\end{figure*}

We fix $\Omega\sigma=0.1$ and compare the two correlation measures at $a\sigma=0.1$, $0.5$, $1$, $5$, and $10$. Figure~\ref{fig:two-detector-correlations} places the concurrence in the left column and the quantum mutual information in the right column, with each row showing the same separation. From top to bottom, $L/\sigma=0.1$, $0.5$, $1$, and $1.5$. The horizontal axis is $\bar b$ itself, with linear motion at zero, catenary motion for $0<\bar b<1$, the cusped trajectory at $\bar b=1$, and circular motion for $\bar b>1$.

At the smallest separation, $L/\sigma=0.1$ [Figs.~\ref{fig:two-detector-correlations}(a,b)], both measures depend only weakly on $\bar b$ at the smaller accelerations. The $a\sigma=0.1$ curves are nearly constant, with $C_{AB}/\mathcal N\simeq1.25$ and $I_{AB}/\mathcal N\simeq0.54$. At $a\sigma=0.5$ and $1$, the concurrence starts lower and increases gradually, while the quantum mutual information changes comparatively little. The dependence on the trajectory becomes much stronger at $a\sigma=5$ and $10$. Near linear motion, the former has only a small positive concurrence and the latter has zero concurrence, but their quantum mutual information remains finite, at approximately $0.15$ and $0.037$, respectively. Both measures rise rapidly near $\bar b=1$ and then approach the smaller-acceleration curves more gradually in the circular regime. The quantum mutual information curves for $a\sigma=5$ and $10$ also cross, so their ordering near $\bar b=0$ does not hold at every larger torsion ratio.

Two features of the first pair of panels also occur in the remaining panels at different detector separations. First, the curves at small accelerations, such as those for $a\sigma=0.1$, remain nearly flat as $\bar b$ varies in all panels of Fig.~\ref{fig:two-detector-correlations}. This shared feature can be understood as followings: The horizontal translation leaves the detectors' velocities unchanged, so they share an instantaneous rest frame at the switching center $\tau=0$. Their common initial velocity at the switching center is along the $y$ direction, and the boost to this frame preserves the horizontal separation $L$. At fixed $\bar b$, their trajectories in this frame have the small-acceleration expansion
\begin{equation}
 x_j^\mu(\tau)=x_j^\mu(0)+\dot x_j^\mu(0)\tau
 +\mathcal O(a\tau^2),\qquad j=A,B.
 \label{eq:weak-acceleration-correlation-trajectories}
\end{equation}
By construction, $\dot x_j^\mu(0)=(1,0,0)$ in this instantaneous rest frame. Acceleration enters through the quadratic and higher-order terms in proper time, so the first two terms describe two inertial detectors at rest. Since the Gaussian switching suppresses contributions from $|\tau|\gg\sigma$, the acceleration-induced displacement during the main interaction interval is of order $a\sigma^2$. Relative to the interaction length scale $\sigma$, this correction is of order $a\sigma$. When $a\sigma\ll1$, the detectors therefore remain close to this common inertial motion throughout the main part of the interaction. Lorentz invariance of the Minkowski vacuum makes the leading inertial correlations independent of the shared initial velocity, and hence of $\bar b$, while differences between the trajectory classes enter through the acceleration corrections. This provides a qualitative explanation for the weak dependence of both measures on $\bar b$ at small acceleration. The expansion is understood at fixed $\bar b$.

Second, at each fixed separation, the curves for different accelerations tend towards common values as $\bar b$ becomes large. This behavior has a geometrical explanation that applies to both correlation measures. For circular motion at fixed $a$, the limit $\bar b\to\infty$ gives
\begin{equation}
 \begin{gathered}
 v=R\omega=\frac1{\bar b}\longrightarrow0,\qquad
 R=\frac1{a(\bar b^2-1)}\longrightarrow0,\\[0.3em]
 \gamma=\frac{\bar b}{\sqrt{\bar b^2-1}}\longrightarrow1.
 \end{gathered}
 \label{eq:circular-correlation-limit}
\end{equation}
In Eq.~\eqref{eq:circular-trajectory}, the time coordinate $t=\gamma\tau$ therefore tends to $\tau$, while the spatial coordinates vanish because their magnitudes are bounded by the shrinking radius. Consequently, $x_B^\mu(\tau)\to(\tau,0,0)$ and $x_A^\mu(\tau)\to(\tau,L,0)$, describing two inertial detectors at rest with separation $L$. Their velocities also tend to those of these static trajectories. The proper acceleration nevertheless remains fixed: the relation $a=\gamma^2v^2/R$ is compatible with this limit because $v^2$ and $R$ both scale as $\bar b^{-2}$. The proper-time angular frequency $\omega\gamma$ grows while the radius shrinks, so convergence of the coordinates and velocities is compatible with nonvanishing second proper-time derivatives. Since the Wightman function in Eq.~\eqref{eq:wightman} depends only on the spacetime separation between the interaction events, the response of the detectors cannot distinguish the difference of inertial motion and circular motion when $\bar{b} \rightarrow \infty$. Neither argument relies on the particular choice $L/\sigma=0.1$: changing the horizontal translation alters the separation but leaves the individual velocities and circular radii unchanged. The weak dependence on $\bar b$ at small acceleration and the approach towards common inertial values at large $\bar b$ therefore also occur in the subsequent rows. 

At $L/\sigma=0.5$ [Figs.~\ref{fig:two-detector-correlations}(c,d)], both correlation measures are smaller, but the reduction is particularly strong for concurrence. The curves for $a\sigma=5$ and $10$ now have zero concurrence throughout the catenary regime and first become positive just above $\bar b=1$. Their quantum mutual information remains nonzero and increases through this same region. The curves at the three smaller accelerations retain positive concurrence throughout the displayed range. At large $\bar b$, the five curves approach approximately $C_{AB}/\mathcal N\simeq0.50$ and $I_{AB}/\mathcal N\simeq0.48$, respectively. Small oscillations are visible during the rise of both measures for the larger accelerations.

The effect of increasing the separation has a particularly direct interpretation here: $\mathcal L_{jj}$ is independent of $L$, so the additional suppression at fixed acceleration and torsion is due to the nonlocal terms. The weaker field correlations at larger separation are illustrated by $W_{\mathrm{hd}}(\tau,\tau)=1/(4\pi L)$, which follows from Eq.~\eqref{eq:horizontal-W} at equal proper times. The full nonlocal matrix elements involve unequal times as well, but this equal-time expression makes the distance dependence of the underlying field correlations explicit. As $|\mathcal M|$ decreases towards the unchanged local-noise level, concurrence reaches zero; quantum mutual information can remain finite while $|\mathcal L_{AB}|$ is reduced. This explains why increasing $L/\sigma$ enlarges the zero-concurrence regions in the plots without producing corresponding intervals of zero quantum mutual information.

For $L/\sigma=1$ [Figs.~\ref{fig:two-detector-correlations}(e,f)], even the $a\sigma=1$ curve starts with zero concurrence, although it becomes positive within the catenary regime. Its quantum mutual information remains positive throughout. The principal rise of the $a\sigma=5$ and $10$ curves occurs farther into the circular regime, around $\bar b\gtrsim1.5$, and small oscillations are more apparent in both panels. The curves cross during their approach to the large-$\bar b$ levels, approximately $0.16$ for concurrence and $0.38$ for quantum mutual information. Consequently, the suppression with increasing acceleration seen near linear motion cannot be used to order the correlations throughout the circular regime.

The oscillations can be interpreted by examining how circular motion enters the nonlocal Wightman function Eq.~(\ref{eq:horizontal-W}). Using the circular expression in Eq.~\eqref{eq:horizontal-coordinate-difference}, the oscillating term in Eq.~(\ref{eq:horizontal-W}) is
\begin{equation}
 \begin{aligned}
 2L\Delta x={}&-4LR
 \sin\!\left(\frac{\sqrt{b^2-a^2}\,\Delta\tau}{2}\right)\\[-0.2em]
 &\times\sin\!\left(\frac{\sqrt{b^2-a^2}\,\Delta_+\tau}{2}\right).
 \end{aligned}
 \label{eq:circular-correlation-displacement}
\end{equation}
Here $\Delta\tau=\tau_A-\tau_B$ and $\Delta_+\tau=\tau_A+\tau_B$ are the proper-time difference and sum defined earlier. The product of the Gaussian switching functions mainly weights $|\tau_A|$ and $|\tau_B|$ of order $\sigma$, so both proper-time combinations entering the sine factors are also of this order. The relevant dimensionless frequency is consequently
\begin{equation}
 \sigma\sqrt{b^2-a^2}
 =(a\sigma)\sqrt{\bar b^2-1}.
 \label{eq:circular-correlation-phase}
\end{equation}
As $\bar b$ varies, the changing sine factors modify the invariant separation in Eq.~\eqref{eq:horizontal-W}, and hence the Wightman function integrated in both $\mathcal M$ and $\mathcal L_{AB}$. These variations can produce small increases and decreases in the resulting correlations. At a fixed $\bar{b}$, increasing $a\sigma$ produces a larger phase variation during the interaction; consequently the phases change more rapidly with $\bar b$. This is consistent with the more closely spaced oscillations of the $a\sigma=10$ curves. The time integrations average these variations, however, so the phase dependence provides a qualitative explanation of the observed oscillations rather than determining their positions or amplitudes by itself. Similar oscillations in the nonstationary $(3+1)$-dimensional setting have been related to the trigonometric terms in the nonlocal Wightman function~\cite{Bozanic2023}. Here the correlation amplitudes are determined by the inverse-square-root Wightman function~\eqref{eq:horizontal-W} appropriate to $(2+1)$ dimensions.

At the largest separation, $L/\sigma=1.5$ [Figs.~\ref{fig:two-detector-correlations}(g,h)], all five concurrence curves vanish over the plotted range. No entanglement is detected at this perturbative order, even though increasing $\bar b$ restored positive concurrence when choosing shorter separations in the previous panels. The quantum mutual information nevertheless remains finite and retains a strong dependence on the trajectory at the larger accelerations. Near $\bar b=0$, it ranges from approximately $0.28$ at $a\sigma=0.1$ to $0.003$ at $a\sigma=10$. The latter curve rises markedly as the trajectory enters the circular regime, and all five curves nearly coincide around $0.28$ at the right of the panel. Thus the nonlocal correlations can still change substantially with the trajectory while remaining insufficient to overcome the concurrence threshold.

\begin{figure}[tbp]
 \centering
 \includegraphics[width=\linewidth]{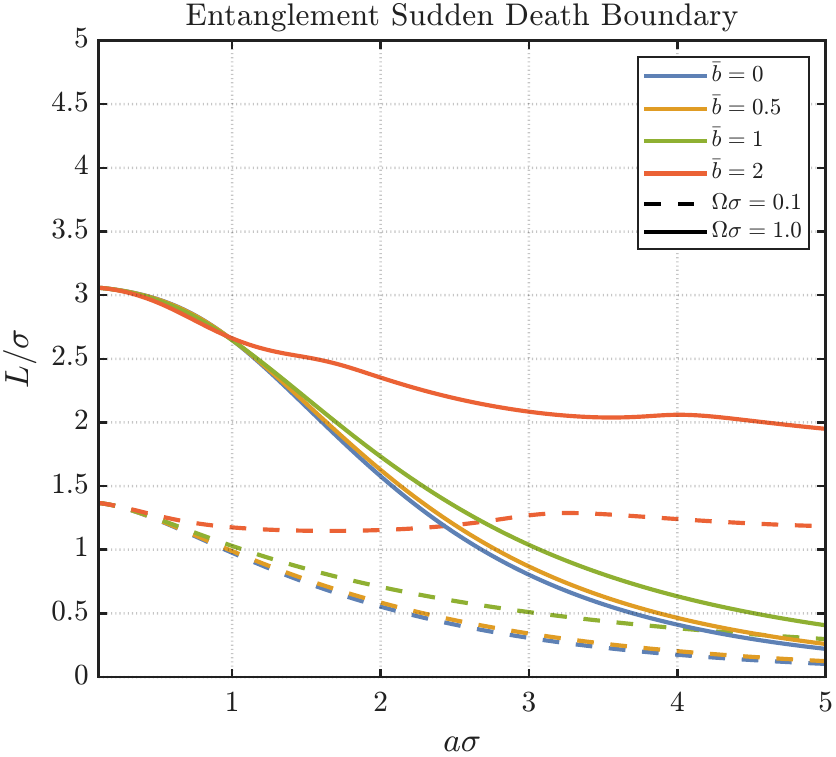}
 \caption{Boundaries between positive and vanishing leading-order concurrence for horizontally displaced detectors. Blue, orange, green, and red curves correspond to $\bar b=0$, $0.5$, $1$, and $2$, respectively. Dashed curves have $\Omega\sigma=0.1$, while solid curves have $\Omega\sigma=1$. Each curve satisfies $|\mathcal M|=\mathcal L_{jj}$; concurrence is positive below and zero above the corresponding boundary.}
 \label{fig:entanglement-boundary}
\end{figure}

The threshold for nonzero concurrence at larger separations and accelerations can be characterized more directly by tracing the boundary between $C_{AB}>0$ and $C_{AB}=0$. Fig.~\ref{fig:entanglement-boundary} shows the numerically determined boundary in the $(a\sigma,L/\sigma)$ plane for the four representative trajectories $\bar b=0$, $0.5$, $1$, and $2$, with energy gaps $\Omega\sigma=0.1$ and $1$. According to Eq.~\eqref{eq:concurrence}, each curve is determined by $|\mathcal M|=\mathcal L_{jj}$, with positive concurrence below the boundary and zero concurrence above it at the perturbative order considered here.

At small acceleration, the four trajectories give almost the same critical separation for each gap, consistent with their nearly inertial motion during the finite interaction discussed above. The boundaries start near $L/\sigma\simeq1.4$ for $\Omega\sigma=0.1$ and $3.1$ for $\Omega\sigma=1$. Thus, in this regime, the larger gap permits positive concurrence over a wider range of separations. The important role of the energy gap in determining the harvesting range has also been demonstrated for linearly accelerated detectors in $(3+1)$ dimensions~\cite{Liu2022Acceleration}; the boundaries here correspond to the $(2+1)$-dimensional horizontally displaced configuration.

For linear, catenary, and cusped motion, the critical separation decreases strongly as $a\sigma$ increases, so the region of positive concurrence becomes progressively narrower. Once the curves separate, the linear boundary is the lowest, the catenary boundary lies slightly above it, and the cusped boundary allows the largest separation among these three cases. This ordering is consistent with the decrease in local excitation with increasing $\bar b$ shown by the solid $\Omega\sigma=1$ curves in Fig.~\ref{fig:probability-acceleration}.\footnote{The same ordering of the local excitation probabilities holds for $\Omega\sigma=0.1$. As shown in Fig.~\ref{fig:monotonicity-thresholds}, both $\Omega\sigma=0.1$ and $1$ lie below the critical gap $(\Omega\sigma)_c(a\sigma)$ throughout the acceleration range considered here. The excitation probability therefore decreases monotonically with $\bar b$ at both gaps.}
A smaller $\mathcal L_{jj}$ lowers the threshold that $|\mathcal M|$ must exceed, thereby favoring positive concurrence. Moreover, the correspondence between the hierarchy of local excitation probabilities and the ordering of the critical separations suggests that the reduction in local excitation plays a more important role in extending the entangled region, although the nonlocal term $|\mathcal M|$ also influences the boundary positions. In addition, since $\mathcal L_{jj}$ is independent of $L$, the shrinking entangled region means that the detectors must be brought closer together for the nonlocal term to exceed the local-noise level at higher accelerations.

Circular motion behaves differently as $a\sigma$ increases. Its boundaries remain well above those of the other three trajectories at moderate and large accelerations, reaching approximately $L/\sigma\simeq1.2$ and $1.95$ near the right edge of the plot for $\Omega\sigma=0.1$ and $1$, respectively. Their dependence on acceleration also becomes comparatively weak in this region. The large-acceleration results in Sec.~\ref{subsec:acceleration-asymptotics} provide useful insight into this behavior. At fixed $\bar b$ and $\Omega\sigma$, the local excitation probability approaches the common limit $\lambda^2\sigma\sqrt{\pi}/4$ for linear, catenary, and cusped motion. For circular motion, it instead approaches the smaller, gap-dependent limit
\begin{equation*}
 \mathcal L_{jj}\longrightarrow
 \frac{\lambda^2\sigma\sqrt{\pi}}{4}
 \left[
  1-\frac{\sqrt{\bar b^2-1}}{\bar b}
  \operatorname{erf}(\Omega\sigma)
 \right].
\end{equation*}
At $\bar b=2$, the ratio of this circular limit to the common limit of the other three trajectories is
$1-(\sqrt{3}/2)\operatorname{erf}(\Omega\sigma)$, which is approximately $0.90$ for $\Omega\sigma=0.1$ and $0.27$ for $\Omega\sigma=1$. By lowering the threshold for positive concurrence, this smaller local response helps explain why circular motion supports entanglement over a larger range of separations.

The circular boundaries are also nonmonotonic. The dashed curve first decreases, rises again around $a\sigma\simeq3$, and then exhibits further undulations; the solid curve decreases overall but has smaller oscillations superimposed on this trend. These features are consistent with the trigonometric displacement term in Eq.~\eqref{eq:circular-correlation-displacement}. At fixed $\bar b=2$, the dimensionless frequency in Eq.~\eqref{eq:circular-correlation-phase} is $\sqrt3\,a\sigma$, so changing the acceleration changes the phases entering the nonlocal response. The resulting changes in enhancement and cancellation within the integral can modulate $|\mathcal M|$ and shift the concurrence boundary.

\subsection{Harvesting genuine entanglement from the quantum vacuum}
\label{subsec:Harvesting-genuine-entanglement-from-the-quantum-vacuum}

\begin{figure*}[tp]
 \centering
 \begin{minipage}[t]{0.485\textwidth}
  \centering
  \includegraphics[width=\linewidth]{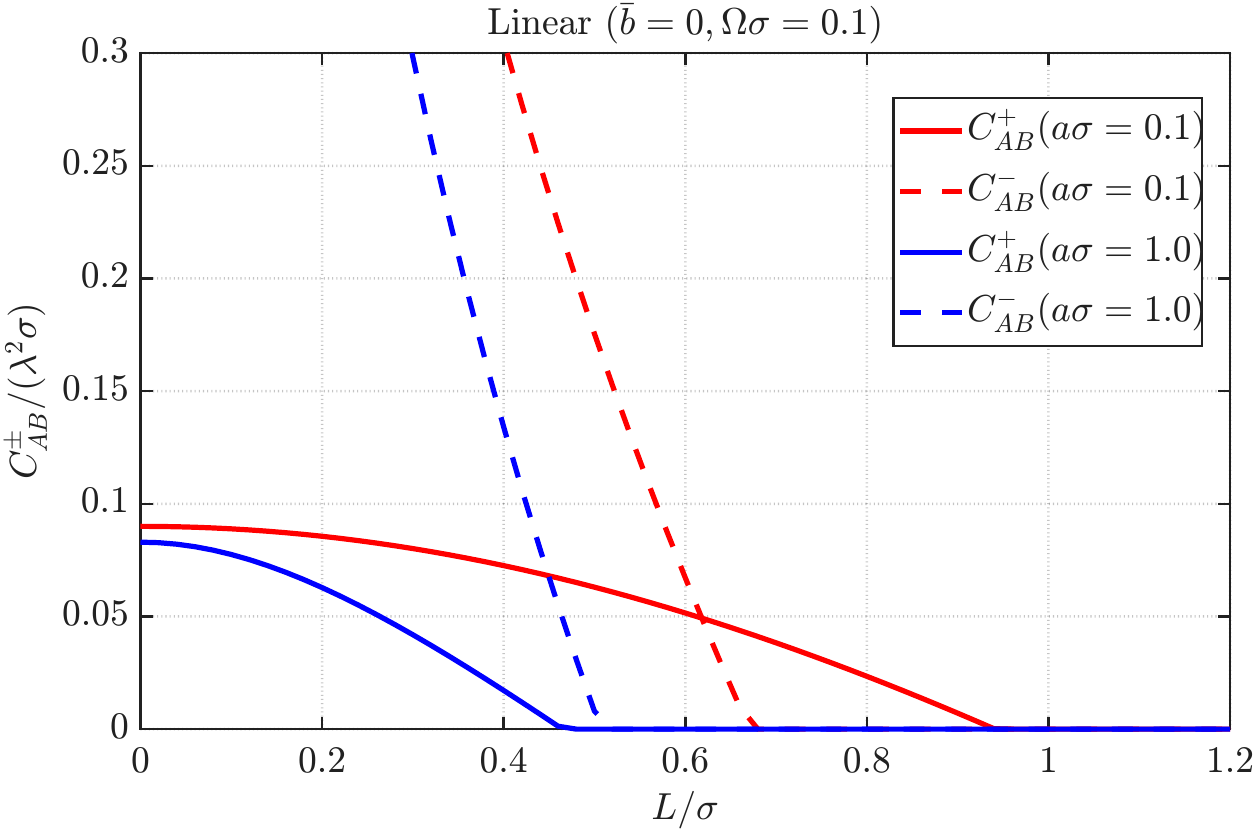}
  \par\vspace{-0.3em}{\footnotesize (a) Linear}
 \end{minipage}\hfill
 \begin{minipage}[t]{0.485\textwidth}
  \centering
  \includegraphics[width=\linewidth]{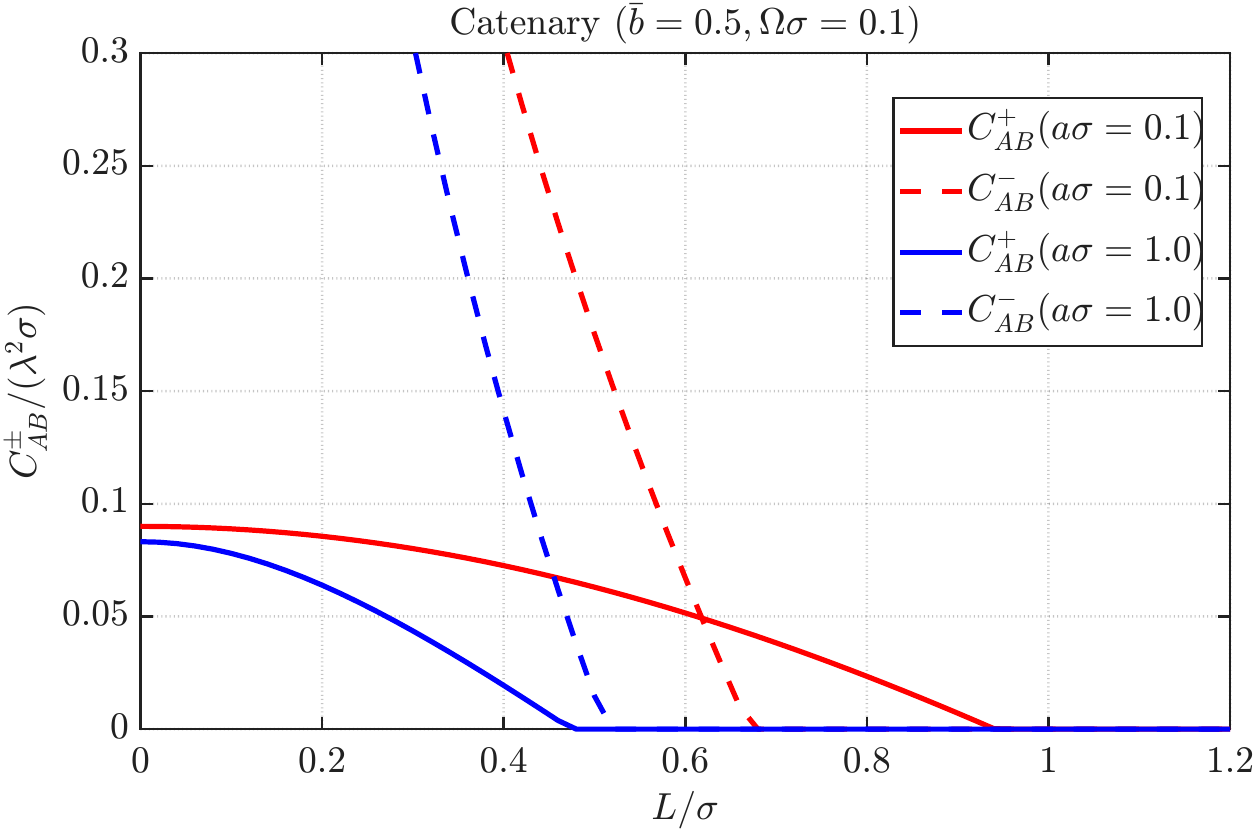}
  \par\vspace{-0.3em}{\footnotesize (b) Catenary}
 \end{minipage}
 \par\vspace{0.6em}
 \begin{minipage}[t]{0.485\textwidth}
  \centering
  \includegraphics[width=\linewidth]{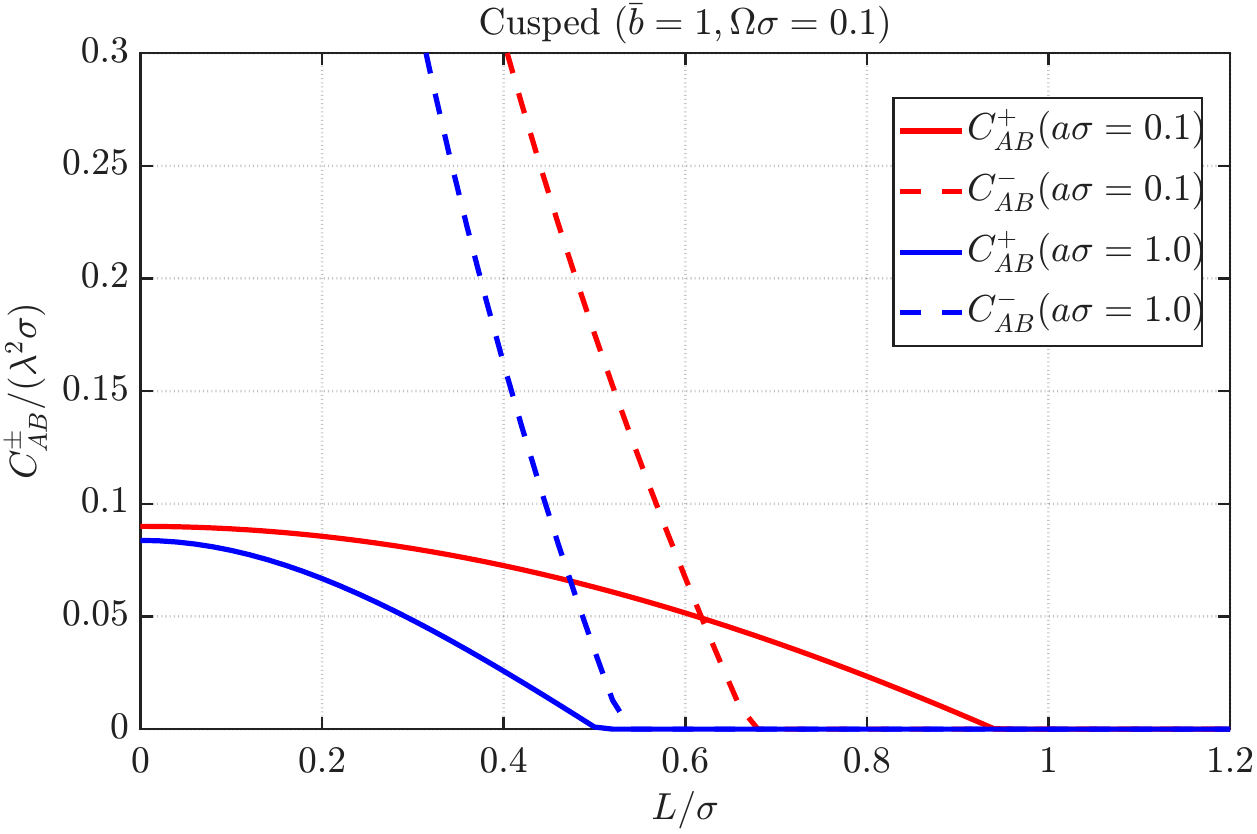}
  \par\vspace{-0.3em}{\footnotesize (c) cusped}
 \end{minipage}\hfill
 \begin{minipage}[t]{0.485\textwidth}
  \centering
  \includegraphics[width=\linewidth]{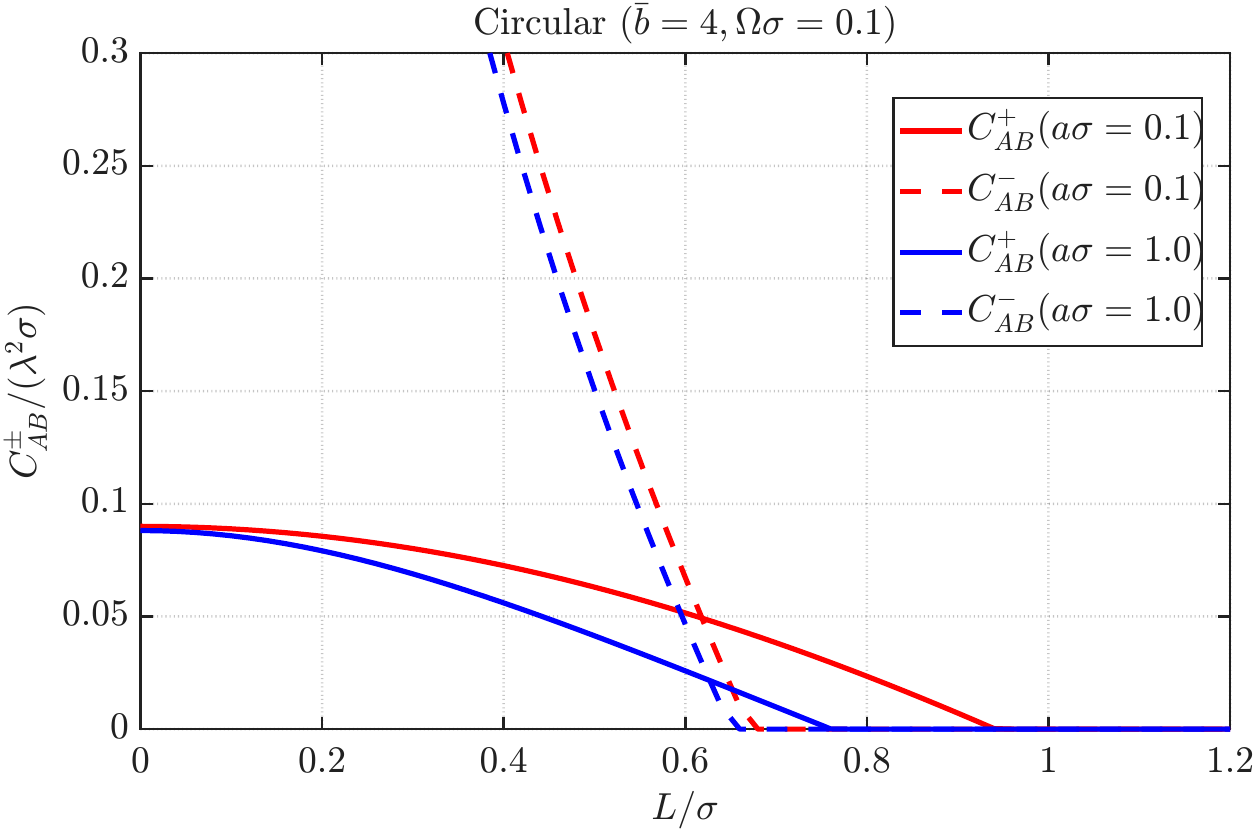}
  \par\vspace{-0.3em}{\footnotesize (d) Circular}
 \end{minipage}
 \caption{Harvested and communication-assisted concurrence, normalized by $\mathcal N=\lambda^2\sigma$, at $\Omega\sigma=0.1$ as functions of $L/\sigma$. Panels (a)--(d) correspond to $\bar b=0$, $0.5$, $1$, and $4$. Solid and dashed curves show $C_{AB}^+$ and $C_{AB}^-$, respectively; red curves have $a\sigma=0.1$ and blue curves have $a\sigma=1$. The short-distance portions of the dashed curves extend above the displayed vertical range.}
 \label{fig:genuine-small-gap}
\end{figure*}

Positive concurrence does not by itself identify the origin of the detectors' entanglement. When their interactions allow causal contact, the detectors can become entangled both by extracting vacuum correlations and by communicating through the field. Distinguishing these contributions is therefore necessary before attributing the acquired entanglement to genuine harvesting~\cite{ETjoa2018}. 

\begin{samepage}
For a real scalar field, the two parts of the Wightman function satisfy
\begin{subequations}
\label{eq:genuine-field-decomposition}
\begin{align}
 2\operatorname{Re}W(x,x')
 &=\langle0|\{\hat\phi(x),\hat\phi(x')\}|0\rangle,
 \label{eq:genuine-anticommutator}\\
 2i\operatorname{Im}W(x,x')
 &=\langle0|[\hat\phi(x),\hat\phi(x')]|0\rangle.
 \label{eq:genuine-commutator}
\end{align}
\end{subequations}
\end{samepage}
Based on the analysis in ~\cite{ETjoa2018}, the symmetric part contains the state-dependent field correlations. By contrast, the commutator of the free field is a multiple of the identity and is independent of the field state; it governs the causal influence that one detector can exert on the other. Microcausality makes it vanish for spacelike-separated events. In $(2+1)$ dimensions, however, the massless-field commutator also has support inside the light cone, whereas in $(3+1)$-dimensional Minkowski spacetime its support is confined to the light cone~\cite{Jonsson2015,ETjoa2018}. Using the decomposition of the Wightman function into its real and imaginary parts in Eqs.~\eqref{eq:genuine-anticommutator} and \eqref{eq:genuine-commutator}, we can write Eq.~\eqref{eq:M} as
\begin{equation}
 \mathcal M=\mathcal M^++i\mathcal M^-,
 \label{eq:genuine-M-decomposition}
\end{equation}
where
\begin{subequations}
\label{eq:genuine-M-parts}
\begin{align}
 \mathcal M^+={}&-\lambda^2\int_{\mathbb R}d\tau_A
 \int_{\mathbb R}d\tau_B\,
 \chi(\tau_A)\chi(\tau_B)
 \nonumber\\[-0.3em]
 &\quad\times e^{i\Omega(\tau_A+\tau_B)}
 \operatorname{Re}W_{\mathrm{hd}}(\tau_A,\tau_B),
 \label{eq:genuine-M-plus}\\
 \mathcal M^-={}&-\lambda^2\int_{\mathbb R}d\tau_A
 \int_{\mathbb R}d\tau_B\,
 \chi(\tau_A)\chi(\tau_B)
 \nonumber\\[-0.3em]
 &\quad\times e^{i\Omega(\tau_A+\tau_B)}
 \operatorname{sgn}(\Delta\tau)
 \operatorname{Im}W_{\mathrm{hd}}(\tau_A,\tau_B).
 \label{eq:genuine-M-minus}
\end{align}
\end{subequations}
\begin{figure*}[tp]
 \centering
 \begin{minipage}[t]{0.485\textwidth}
  \centering
  \includegraphics[width=\linewidth]{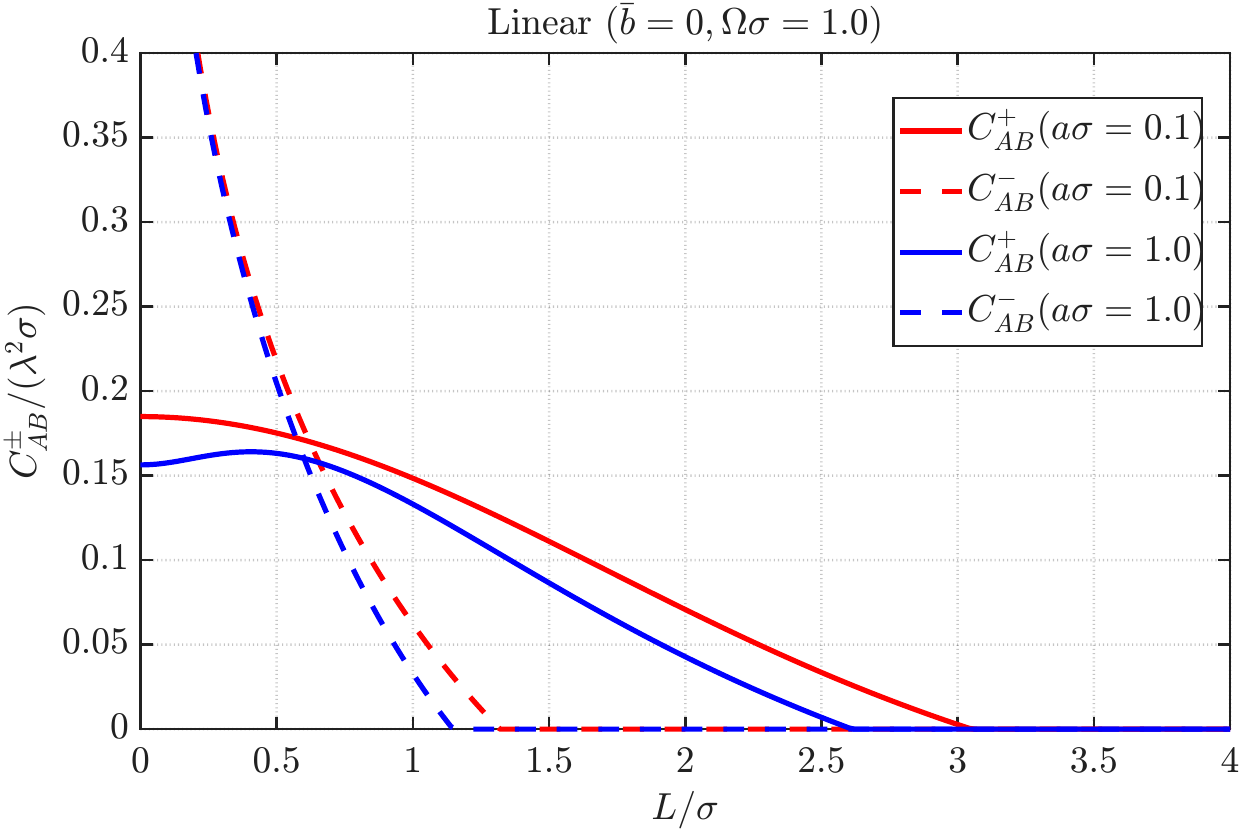}
  \par\vspace{-0.3em}{\footnotesize (a) Linear}
 \end{minipage}\hfill
 \begin{minipage}[t]{0.485\textwidth}
  \centering
  \includegraphics[width=\linewidth]{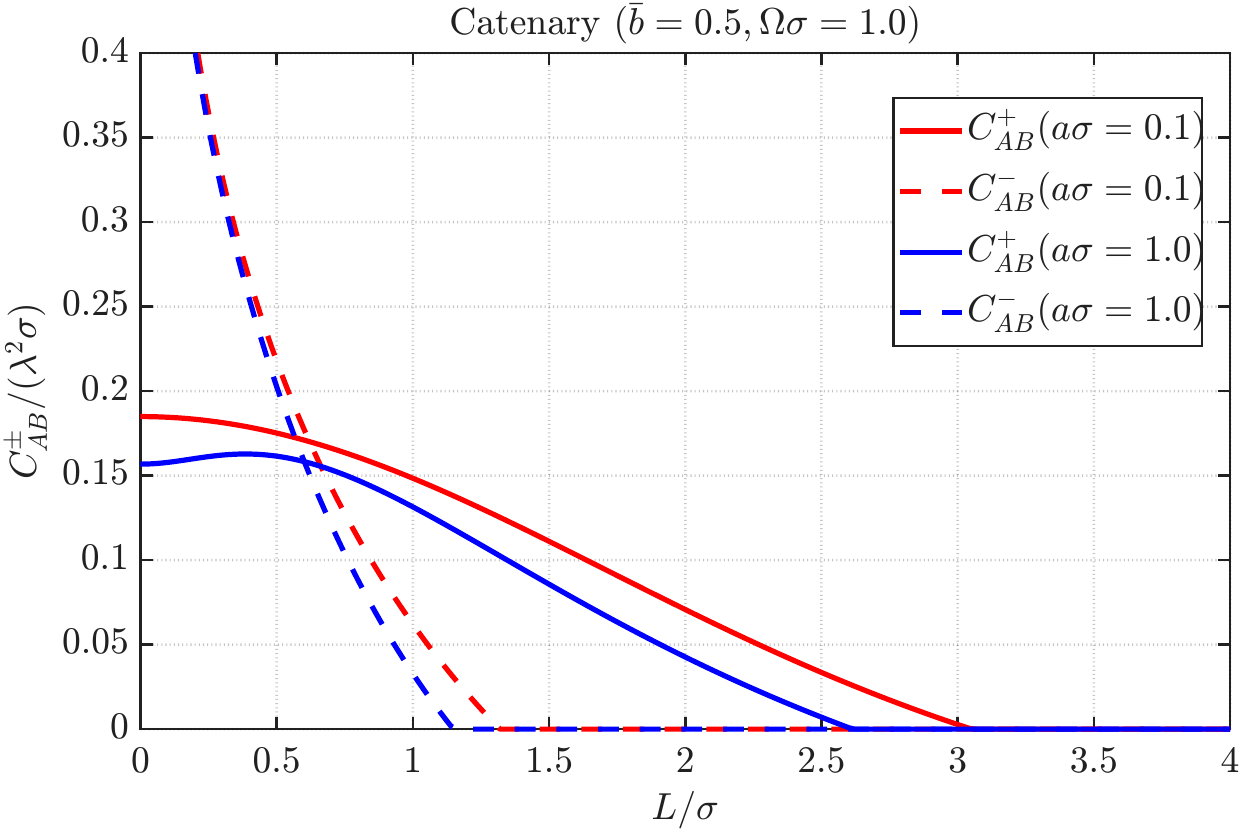}
  \par\vspace{-0.3em}{\footnotesize (b) Catenary}
 \end{minipage}
 \par\vspace{0.6em}
 \begin{minipage}[t]{0.485\textwidth}
  \centering
  \includegraphics[width=\linewidth]{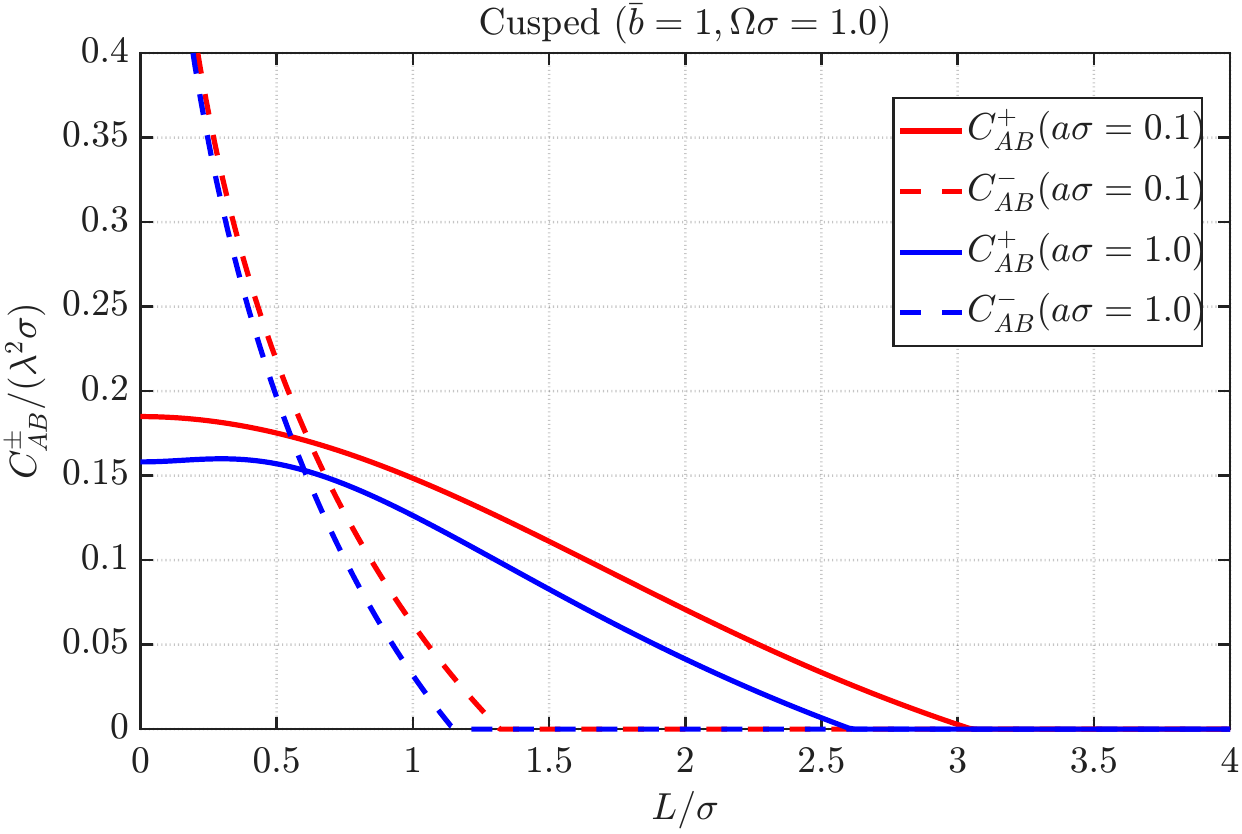}
  \par\vspace{-0.3em}{\footnotesize (c) cusped}
 \end{minipage}\hfill
 \begin{minipage}[t]{0.485\textwidth}
  \centering
  \includegraphics[width=\linewidth]{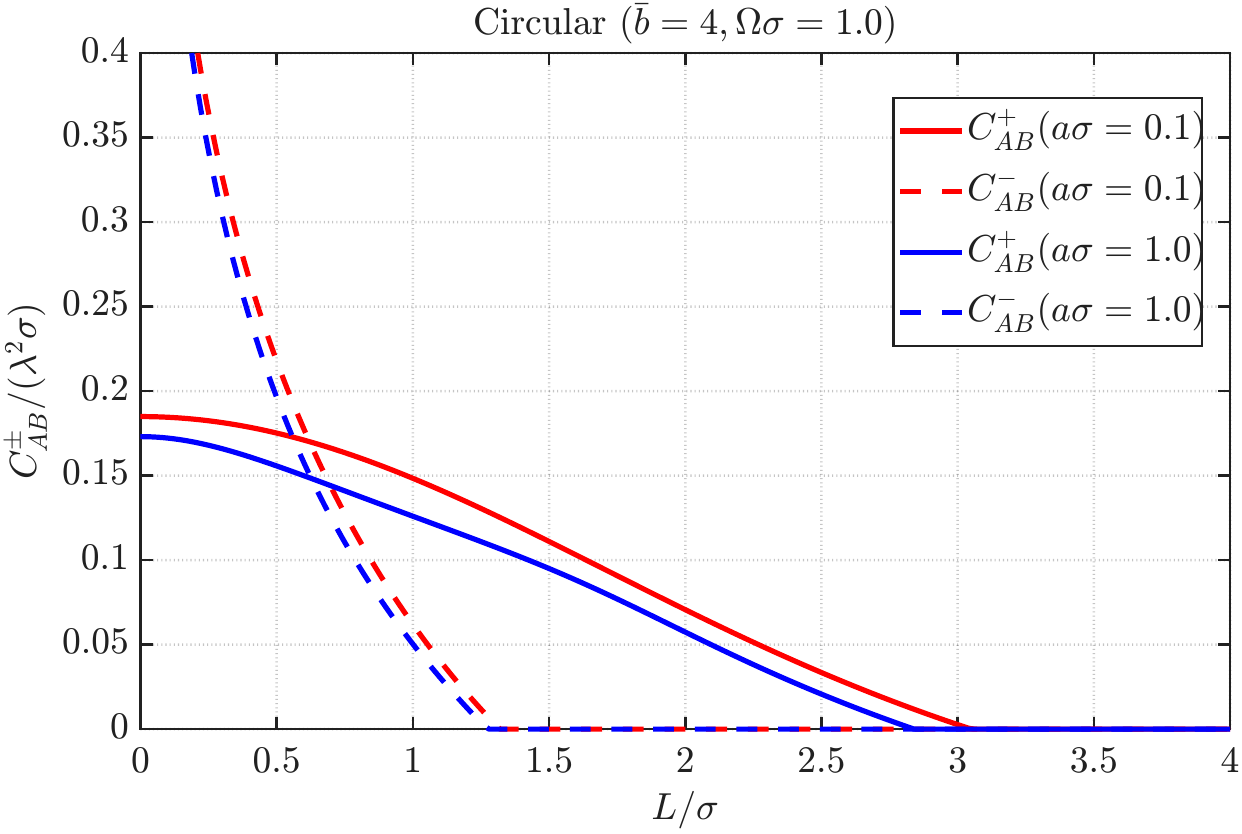}
  \par\vspace{-0.3em}{\footnotesize (d) Circular}
 \end{minipage}
 \caption{The same concurrence diagnostics and trajectory parameters as in Fig.~\ref{fig:genuine-small-gap}, now with $\Omega\sigma=1$. The horizontal range is enlarged to display the longer positive range of $C_{AB}^+$. Colors and line styles have the same meaning as in Fig.~\ref{fig:genuine-small-gap}.}
 \label{fig:genuine-unit-gap}
\end{figure*}

The first integral retains the anticommutator contribution, while the second retains the commutator contribution with the sign required by time ordering.

For the trajectories considered here, simultaneous reversal of both proper times leaves $\operatorname{Re}W_{\mathrm{hd}}$ and $\operatorname{sgn}(\Delta\tau)\operatorname{Im}W_{\mathrm{hd}}$ unchanged. Since the switching functions are even, the sine part of $e^{i\Omega(\tau_A+\tau_B)}$ integrates to zero, making both $\mathcal M^+$ and $\mathcal M^-$ real in the convention of Eq.~\eqref{eq:genuine-M-decomposition}. Consequently,
\begin{equation}
 |\mathcal M|^2=|\mathcal M^+|^2+|\mathcal M^-|^2.
 \label{eq:genuine-quadrature}
\end{equation}
The role of such symmetries in fixing the relative phase of the two contributions is discussed more generally in ~\cite{Zambianco2024}.

We then define the genuine concurrence $C_{AB}^{+}$ and communication-assisted concurrence $C_{AB}^{-}$ as~\cite{ETjoa2018,Bozanic2023}
\begin{equation}
 C_{AB}^{\pm}
 :=2\max\!\left\{0,|\mathcal M^{\pm}|-\mathcal L_{jj}\right\},
 \label{eq:genuine-concurrences}
\end{equation}
at the same perturbative order as Eq.~\eqref{eq:concurrence}. The full local transition probability is retained in both expressions. Thus $C_{AB}^+>0$ means that vacuum correlations alone are strong enough to overcome local noise and allow genuine entanglement to be harvested from the vacuum, while $C_{AB}^->0$ means that the communication contribution alone does so. These quantities are not additive components of $C_{AB}$: the amplitudes combine according to Eq.~\eqref{eq:genuine-quadrature} before the concurrence threshold is applied. In particular, both quantities can vanish while the total concurrence still remains positive.

Figs.~\ref{fig:genuine-small-gap} and \ref{fig:genuine-unit-gap} show $C_{AB}^{\pm}$ for $\Omega\sigma=0.1$ and $1$, respectively. Each figure compares linear ($\bar b=0$), catenary ($\bar b=0.5$), cusped ($\bar b=1$), and circular ($\bar b=4$) motion at $a\sigma=0.1$ and $1$. The vertical axes show $C_{AB}^{\pm}/\mathcal N$, with $\mathcal N=\lambda^2\sigma$. Solid curves show $C_{AB}^+$ and dashed curves show $C_{AB}^-$. Red and blue distinguish the smaller and larger accelerations.

Overall, the corresponding red curves are almost indistinguishable across the four trajectories for $\Omega\sigma=0.1$ and $a\sigma=0.1$ in Fig.~\ref{fig:genuine-small-gap}. The genuine concurrence, represented by all red solid curves, generally decreases from approximately $0.09$ near zero separation and vanishes around $L/\sigma\simeq0.95$. By contrast, the communication-assisted concurrence is larger at short distances but decreases much more rapidly, reaching zero near $L/\sigma\simeq0.68$. There is consequently an interval, approximately $0.68\lesssim L/\sigma\lesssim0.94$, in which $C_{AB}^+>0$ while $C_{AB}^-=0$. In this interval, the vacuum correlations alone suffice to overcome the
local excitation noise, whereas the communication contribution alone does
not: $C_{AB}^-=0$ means $|\mathcal M^-|\leq\mathcal L_{jj}$, although a nonzero
$\mathcal M^-$ can still affect the total concurrence through
Eq.~\eqref{eq:genuine-quadrature}.

Increasing the acceleration to $a\sigma=1$ suppresses both the genuine and communication-assisted concurrence, but the contraction of their positive ranges depends strongly on the trajectory. In the linear and catenary cases, $C_{AB}^+$ vanishes near $L/\sigma\simeq0.48$; the cusped case extends slightly farther, to about $0.51$. For all three trajectories, $C_{AB}^-$ remains positive over a short
interval after $C_{AB}^+$ has vanished. Thus none of these three
configurations exhibits a separation interval with both $C_{AB}^+>0$
and $C_{AB}^-=0$ at these parameters. However, circular motion has the opposite ordering: its genuine concurrence persists to approximately $L/\sigma\simeq0.77$, while the communication-assisted concurrence vanishes around $0.66$. It therefore retains an interval with $C_{AB}^+>0$ and $C_{AB}^-=0$ even at the larger acceleration. The circular dashed curves are also much closer to each other than their counterparts in the other three panels. Thus the change from $a\sigma=0.1$ to $1$ is appreciably less restrictive for the circular configuration at this energy gap.

Increasing the gap to $\Omega\sigma=1$ changes this comparison substantially. At $a\sigma=0.1$, the four genuine-concurrence curves again nearly coincide, but their small-separation values increase to approximately $0.185$ and their positive ranges extend to $L/\sigma\simeq3.05$. The communication-assisted curves vanish much earlier, near $L/\sigma\simeq1.31$. At $a\sigma=1$, the genuine concurrence remains positive up to about $L/\sigma\simeq2.6$ for linear, catenary, and cusped motion, and to $2.83$ for circular motion. The corresponding communication-assisted concurrence vanishes near $1.15$ for the first three trajectories and $1.27$ for the circular case. Hence all four trajectories now exhibit broad intervals with $C_{AB}^+>0$ and $C_{AB}^-=0$ at both accelerations. The extent of these intervals is considerably greater than at the smaller energy gap.

The genuine concurrence at $a\sigma=1$ also acquires a weak nonmonotonic dependence on separation in the first three panels. It initially rises to a shallow maximum near $L/\sigma\simeq0.4$ for linear and catenary motion, and near $0.3$ for cusped motion, before decreasing to zero. This maximum is least pronounced for the cusped trajectory. The circular curve instead decreases throughout the plotted range. In all four panels, the larger acceleration gives a smaller genuine concurrence than $a\sigma=0.1$. The dashed curves decrease rapidly with distance and show much less sensitivity to the trajectory than the solid curves over most of their positive range at this gap.

These results show that vacuum correlations can supply a nonlocal contribution sufficient to overcome local noise for every trajectory class considered here. Whether this remains possible when the communication-assisted concurrence vanishes depends on both the energy gap and the acceleration: at the smaller gap it is retained by circular motion at $a\sigma=1$, while at the larger gap it occurs over a broad separation range for all four trajectories. Thus causal contact does not prevent the vacuum-correlation contribution from exceeding local noise. The separate thresholds of $C_{AB}^+$ and $C_{AB}^-$ resolve distinctions that are not visible in the total-concurrence boundary alone.

\section{Conclusions}
\label{app:conclusions}

Within leading-order perturbation theory, we have investigated how uniform acceleration and torsion affect Gaussian-switched UDW detectors coupled to a real massless scalar field in the $(2+1)$-dimensional Minkowski vacuum. The analysis covers linear, catenary, cusped, and circular motion, examining single-detector excitation probabilities and effective temperatures alongside concurrence and quantum mutual information for two horizontally displaced detectors.

For a single detector, the small-acceleration expansions at fixed $\bar b$ and $\Omega\sigma$ give a common inertial limit and a positive leading correction independent of the trajectory class. Torsion first enters the excitation probability at order $(a\sigma)^4$, and the torsion-dependent correction changes sign at $\Omega\sigma=\sqrt{3/2}$. This explains why increasing $\bar b$ can either enhance or suppress the weak-acceleration response, depending on the gap, and why the ordering of the representative trajectory curves can reverse as acceleration increases. A similar reversal was observed numerically for Gaussian-switched detectors in $(3+1)$ dimensions~\cite{Bozanic2023}; the present expansion identifies the term governing the initial ordering in $(2+1)$ dimensions.

At large acceleration, with $\bar b$ and $\Omega\sigma$ fixed, the normalized excitation probability approaches $\sqrt\pi/4$ for linear, catenary, and cusped motion, whereas circular motion approaches a smaller, gap-dependent limit. The effective temperatures nevertheless have three distinct behaviors. Linear and catenary motion give gap-independent linear growth at leading order, with linear motion recovering the Unruh slope; cusped motion instead gives growth proportional to $a/\ln(a\sigma)$, while circular motion has a finite limit. Thus a common limit of the excitation probability does not imply the same temperature behavior. In particular, catenary motion has a gap-independent leading temperature even though the field correlations along its worldline do not satisfy the KMS condition, while the circular limit retains the dependence on the finite Gaussian interaction.

In this model, the effective temperature is a strictly increasing function of the excitation probability at a fixed energy gap. Their monotonicity in $\bar b$ and the positions and types of their extrema therefore coincide. We proved that both observables decrease throughout the torsion-to-acceleration ratio $\bar b$ range for sufficiently small gaps at every fixed nonzero acceleration. As the gap increases, the first increasing interval can appear either next to linear motion at $\bar b=0$ or at a positive value of $\bar b$. The corresponding threshold conditions characterize the first departure from monotonicity and account for both pronounced maxima and shallow local extrema near the cusped trajectory. They also show that the first loss of monotonicity need not occur in the circular regime. 

For two horizontally displaced detectors, both correlation measures depend weakly on $\bar b$ at small acceleration. Larger accelerations produce stronger trajectory dependence and crossings between curves, with oscillations in the circular regime. Similar oscillations occur in the nonstationary $(3+1)$-dimensional configurations of~\cite{Bozanic2023}. Quantum mutual information remains nonzero in regions where the nonlocal entangling contribution cannot overcome local excitation and the concurrence vanishes. The boundary of positive concurrence reveals a strong contraction of the entangled separation range for linear, catenary, and cusped motion as acceleration increases. In the representative circular configuration, this range remains appreciably wider at moderate and large accelerations, and the boundary varies more weakly with acceleration while exhibiting small oscillations. The smaller circular local excitation favors this wider positive-concurrence region, although the nonlocal response also determines the boundary. At small acceleration, the larger gap also increases the critical separation, consistent with the gap dependence of the harvesting range reported for linearly accelerated detectors in $(3+1)$ dimensions~\cite{Liu2022Acceleration}.

Separating the anticommutator and commutator contributions further shows why the total concurrence alone is insufficient to identify the origin of the acquired entanglement~\cite{ETjoa2018}. For all four trajectory classes, we found separation intervals with $C_{AB}^{+}>0$ and $C_{AB}^{-}=0$: the vacuum-correlation contribution exceeds the full local-noise level, while the communication contribution alone does not. At the smaller gap, increasing acceleration removes these intervals for linear, catenary, and cusped motion in the representative settings, whereas circular motion retains one. At the larger gap, broad intervals persist for all four classes at both accelerations examined. Bozanic \textit{et al.} demonstrated genuine harvesting during causal contact for stationary linear and circular configurations in $(3+1)$ dimensions~\cite{Bozanic2023}. Here it occurs for all four classes in the horizontally displaced setting, with the gap strongly affecting the separation range over which vacuum correlations can overcome local noise.

Our results are relevant to experimental efforts to probe quantum fields using localized detectors in planar analogue systems. Proposals based on quasi-two-dimensional Bose--Einstein condensates and thin superfluid-helium films use moving laser beams to detect acceleration-dependent field fluctuations~\cite{Gooding2020,Bunney2024}, while a recent nondestructive optical readout scheme offers access to unequal-time density correlations along stationary paths~\cite{Gooding2026}. The field dimension and coplanar trajectories considered here therefore provide a natural theoretical setting for connecting such measurements with detector-based descriptions of quantum correlations. For implementations with the detector coupling studied here, the entanglement boundaries and intervals of genuine harvesting identify favorable choices of energy gap, acceleration, and separation. An experimental test of genuine harvesting in these regimes would probe the transfer of preexisting vacuum entanglement to accelerating systems even when causal communication is possible, complementing measurements of the local excitation associated with acceleration effects.

\begin{acknowledgments}
S.W.H. and J.F. are supported by the National Natural Science Foundation of China (Nos. 12475061, 12075178) and Natural Science Basic Research Plan in Shaanxi Province of China (No. 2026JC-QYCX-002). This work was supported in part by the Natural Sciences and Engineering Research Council of Canada.
\end{acknowledgments}

\appendix

\section{Derivations of the acceleration limits}
\label{app:acceleration-asymptotics}

This appendix derives the expansions summarized in
Table~\ref{tab:acceleration-limits}, starting from the dimensionless
separation and response in Eqs.~\eqref{eq:dimensionless-separation}
and \eqref{eq:dimensionless-response}. Unless stated otherwise,
$\bar b$ and $\Omega\sigma>0$ are held fixed; constants in the
remainder estimates may depend on these parameters. At small
acceleration, the Gaussian switching selects the near-coincidence
behavior of the separation. At large acceleration, the behavior
at large separations instead distinguishes three cases, according
to whether $z/d_{\bar b}(z)$ is integrable, has a logarithmically
divergent integral, or tends to a nonzero constant.

\subsection{Small-acceleration limit}
\label{app:small-acceleration}

For $a\sigma\ll1$, the Gaussian in
Eq.~\eqref{eq:dimensionless-response} restricts the main
contribution to $z=O(a\sigma)$. Expanding the squared separation determined by Eqs.~\eqref{eq:self-interval}, \eqref{eq:self-interval-cusped} and \eqref{eq:dimensionless-separation} at the origin gives
\begin{equation}
 d_{\bar b}^2(z)=z^2+\frac{z^4}{12}
 +\frac{1-\bar b^2}{360}z^6+O(z^8).
 \label{eq:separation-small-z}
\end{equation}
The corresponding inverse separation is
\begin{equation}
 \frac1{d_{\bar b}(z)}=\frac1z-\frac z{24}
 +\frac{7+8\bar b^2}{5760}z^3+O(z^5).
 \label{eq:inverse-separation-small-z}
\end{equation}
Each power of $z$ then produces a Gaussian moment in Eq. \eqref{eq:dimensionless-response}. Setting $z=(a \sigma) v$ , we obtain the following formulas can be obtained:
\begin{align}
 \int_0^\infty e^{-v^2/4}
 \frac{\sin[(\Omega\sigma)v]}v\,dv
 &=\frac\pi2\operatorname{erf}(\Omega\sigma),
 \nonumber\\
 \int_0^\infty v e^{-v^2/4}\sin[(\Omega\sigma)v]dv
 &=2\sqrt\pi(\Omega\sigma)e^{-(\Omega\sigma)^2},
 \nonumber\\
 \int_0^\infty v^3 e^{-v^2/4}\sin[(\Omega\sigma)v]dv
 &=4\sqrt\pi(\Omega\sigma)
 \nonumber\\[-0.2em]
 &\quad\times[3-2(\Omega\sigma)^2]e^{-(\Omega\sigma)^2}.
 \label{eq:small-acceleration-gaussian-moments}
\end{align}
The first follows by integrating the Gaussian cosine transform with respect to the gap; differentiating that transform once and three times gives the other two. Inserting Eq.~\eqref{eq:inverse-separation-small-z} into Eq. \eqref{eq:dimensionless-response} therefore yields
\begin{align}
 \mathcal P_{\bar b}={}&\frac{\sqrt\pi}{4}\operatorname{erfc}(\Omega\sigma)
 +\frac{(a\sigma)^2(\Omega\sigma)e^{-(\Omega\sigma)^2}}{24}
 \nonumber\\
 &-\frac{(7+8\bar b^2)(a\sigma)^4(\Omega\sigma)}{2880}
 \bigl[3-2(\Omega\sigma)^2\bigr]e^{-(\Omega\sigma)^2}
 \nonumber\\
 &+O\bigl((a\sigma)^6\bigr).
 \label{eq:small-acceleration-probability}
\end{align}

The zeroth-order term is the common inertial response. The
leading acceleration correction is positive and independent of
torsion, explaining the common initial rise and close spacing
of the trajectory curves in
Fig.~\ref{fig:probability-acceleration}. Torsion first enters at
fourth order. More explicitly,
\begin{align}
 \partial_{\bar b}\mathcal P_{\bar b}
 ={}&-\frac{\bar b(a\sigma)^4(\Omega\sigma)}{180}
 \bigl[3-2(\Omega\sigma)^2\bigr]e^{-(\Omega\sigma)^2}
 \nonumber\\
 &+O\bigl((a\sigma)^6\bigr).
 \label{eq:small-acceleration-torsion-derivative}
\end{align}
For fixed finite $\bar b>0$, the leading torsion dependence changes sign at $\Omega\sigma=\sqrt{3/2}$. Thus the response
decreases with $\bar b$ for $\Omega\sigma=1$ and increases for
$\Omega\sigma=3$ at sufficiently small acceleration, accounting
for the opposite orderings in
Fig.~\ref{fig:probability-acceleration}. The expansion is not uniform
when $\bar b$ grows on the scale $1/(a\sigma)$, so it does not
establish monotonicity over the entire torsion range.

For the effective temperature, substituting the zeroth-order probability
into Eq.~\eqref{eq:temperature-from-probability} gives
\begin{equation}
 \sigma T_{\mathrm{in}}
 =\frac{\Omega\sigma}
 {\ln\!\left[\operatorname{erfc}(-\Omega\sigma)/
 \operatorname{erfc}(\Omega\sigma)\right]}.
 \label{eq:inertial-temperature}
\end{equation}
The values $\sigma T_{\mathrm{in}}\simeq0.40637$ and $0.26285$
for $\Omega\sigma=1$ and $3$, respectively, explain the distinct
intercepts in Fig.~\ref{fig:temperature-acceleration}.
Inserting Eq.~\eqref{eq:small-acceleration-probability} into Eq.~\eqref{eq:temperature-from-probability} gives
\begin{align}
 \sigma T_{\mathrm{eff}}={}&\sigma T_{\mathrm{in}}
 +\frac{(a\sigma)^2(\Omega\sigma)^2e^{-(\Omega\sigma)^2}}
 {3\sqrt\pi\operatorname{erfc}(\Omega\sigma)
 \operatorname{erfc}(-\Omega\sigma)}
 \nonumber\\
 &\times\frac{1}
 {\left\{\ln\!\left[\operatorname{erfc}(-\Omega\sigma)/
 \operatorname{erfc}(\Omega\sigma)\right]\right\}^2}
 +O\bigl((a\sigma)^4\bigr).
 \label{eq:small-acceleration-temperature}
\end{align}
This correction is again positive and independent of $\bar b$. The first torsion-dependent temperature term has the same sign as the fourth-order probability term, since they share the same monotonicity.

\subsection{Large-acceleration limit}
\label{app:large-acceleration}

As $a\sigma$ grows, the Gaussian cutoff in $z$ moves to larger
separations. Expanding the sine in
Eq.~\eqref{eq:dimensionless-response} first produces the
factor $(\Omega\sigma)z/(a\sigma)$, so the leading coefficient
involves the integral of $z/d_{\bar b}(z)$. Its convergence is
determined by
\begin{equation}
 d_{\bar b}(z)\sim
 \begin{cases}
 \displaystyle\frac{e^{\sqrt{1-\bar b^2}\,z/2}}{1-\bar b^2},
 &0\leq\bar b<1,\\[0.5em]
 z^2/\sqrt{12},&\bar b=1,\\[0.2em]
 \gamma z,&\bar b>1,
 \end{cases}
 \label{eq:large-z-separation}
\end{equation}
where $\gamma=\bar b/\sqrt{\bar b^2-1}$ in the circular case.
Thus $z/d_{\bar b}(z)$ is integrable for linear and catenary
motion, behaves as $\sqrt{12}/z$ for cusped motion, and tends
to $1/\gamma$ for circular motion. These differences explain
why the three cases require separate expansions. They also
correspond to the different large-separation decay rates of the
stationary Wightman functions discussed in~\cite{ParryLouko2025}.

\subsubsection{Linear and catenary motion}
\label{app:large-linear-catenary}

For each fixed $0\leq\bar b<1$, define
\begin{equation}
 \begin{split}
 \mathcal I(\bar b)&:=\int_0^\infty\frac{z\,dz}{d_{\bar b}(z)}
 \\
 &=2\int_0^\infty\frac{r\,dr}
 {\sqrt{\sinh^2r-\bar b^2r^2}}.
 \end{split}
 \label{eq:catenary-moment}
\end{equation}
The second line follows from
$r=\sqrt{1-\bar b^2}\,z/2$. In the first line, the integrand
tends to unity at the origin and decays exponentially at
infinity, so $\mathcal I(\bar b)$ is finite. Moreover,
\begin{equation}
 \left|\frac{(a\sigma)e^{-z^2/[4(a\sigma)^2]}}
 {d_{\bar b}(z)}
 \sin\!\left(\frac{(\Omega\sigma)z}{a\sigma}\right)\right|
 \leq\frac{(\Omega\sigma)z}{d_{\bar b}(z)}.
 \label{eq:catenary-dominated-bound}
\end{equation}
The bound is integrable and independent of $a\sigma$.
Dominated convergence therefore gives
\begin{equation}
 \lim_{a\sigma\to\infty}(a\sigma)
 \left(\frac{\sqrt\pi}{4}-\mathcal P_{\bar b}\right)
 =\frac{(\Omega\sigma)\mathcal I(\bar b)}{2\sqrt\pi}.
 \label{eq:catenary-leading-limit}
\end{equation}

To obtain the correction and its remainder, expand the full
Gaussian--sine factor:
\begin{equation}
 \begin{split}
 &e^{-z^2/[4(a\sigma)^2]}
 \sin\!\left(\frac{(\Omega\sigma)z}{a\sigma}\right)
 \\
 &\quad=\frac{(\Omega\sigma)z}{a\sigma}
 -\frac{z^3}{(a\sigma)^3}
 \left[\frac{\Omega\sigma}{4}
 +\frac{(\Omega\sigma)^3}{6}\right]
 +O\!\left(\frac{z^5}{(a\sigma)^5}\right).
 \end{split}
 \label{eq:large-acceleration-weight-expansion}
\end{equation}
Consequently,
\begin{equation}
 \begin{split}
 \mathcal P_{\bar b}={}&\frac{\sqrt\pi}{4}
 -\frac{(\Omega\sigma)\mathcal I(\bar b)}{2\sqrt\pi(a\sigma)}
 \\
 &+\frac1{2\sqrt\pi(a\sigma)^3}
 \left[\frac{\Omega\sigma}{4}
 +\frac{(\Omega\sigma)^3}{6}\right]
 \int_0^\infty\frac{z^3\,dz}{d_{\bar b}(z)}
 \\
 &+O\bigl((a\sigma)^{-5}\bigr).
 \end{split}
 \label{eq:large-acceleration-catenary-probability}
\end{equation}

The effective temperature depends on the ratio between the
excitation and de-excitation responses. Through its definition~\eqref{eq:temperature-from-probability}, and using
$\ln[(1+x)/(1-x)]=2x+O(x^3)$ with $x=1-4\mathcal P_{\bar b}/\sqrt\pi$ gives
\begin{equation}
 \begin{split}
 \ln\frac{\sqrt\pi/2-\mathcal P_{\bar b}}{\mathcal P_{\bar b}}
 ={}&\frac8{\sqrt\pi}
 \left(\frac{\sqrt\pi}{4}-\mathcal P_{\bar b}\right)
 \\
 &+O\!\left[\left(\frac{\sqrt\pi}{4}
 -\mathcal P_{\bar b}\right)^3\right].
 \end{split}
 \label{eq:temperature-log-expansion}
\end{equation}
Substituting Eq.~\eqref{eq:large-acceleration-catenary-probability}
yields
\begin{equation}
 \ln\frac{\sqrt\pi/2-\mathcal P_{\bar b}}{\mathcal P_{\bar b}}
 =\frac{4(\Omega\sigma)\mathcal I(\bar b)}{\pi(a\sigma)}
 +O\bigl((a\sigma)^{-3}\bigr).
 \label{eq:catenary-temperature-logarithm}
\end{equation}
Dividing $\Omega\sigma$ by this expression gives
\begin{equation}
 \sigma T_{\mathrm{eff}}
 =\frac{\pi(a\sigma)}{4\mathcal I(\bar b)}
 +O\bigl((a\sigma)^{-1}\bigr).
 \label{eq:large-acceleration-catenary-temperature}
\end{equation}
The gap cancels at leading order. For linear motion, expanding
$1/\sinh r=2\sum_{n=0}^\infty e^{-(2n+1)r}$ gives
\begin{equation}
 \mathcal I(0)
 =4\sum_{n=0}^\infty\frac1{(2n+1)^2}
 =\frac{\pi^2}{2},
 \label{eq:linear-moment-value}
\end{equation}
and hence $\sigma T_{\mathrm{eff}}\sim(a\sigma)/(2\pi)$,
recovering the Unruh slope. For $\bar b=0.5$, numerical
integration gives $\mathcal I(0.5)\simeq5.32034$.
The coefficient in
Eq.~\eqref{eq:large-acceleration-catenary-temperature} agrees
with the stationary small-gap drifted-Rindler temperature in~\cite{ParryLouko2025}. Finally, the
integrand in the second line of Eq.~\eqref{eq:catenary-moment}
increases with $\bar b$, so $\mathcal I(\bar b)$ increases and
the leading temperature slope decreases through the catenary
family, corresponding to the decrease of the effective temperature in Fig.~\ref{fig:temperature-torsion-15} at this region.

\subsubsection{Cusped motion}
\label{app:large-cusped}

At $\bar b=1$, $d_1(z)=z\sqrt{1+z^2/12}$, so
$z/d_1(z)\sim\sqrt{12}/z$ as $z\to\infty$.
The first moment used above therefore diverges logarithmically.
Consequently, the leading sine expansion used for linear and
catenary motion cannot be integrated directly, so alternative method is needed. In the following, we first isolate
the contribution responsible for this divergence.

Set $z=(a\sigma)v$ and introduce the small expansion parameter
$\xi=\sqrt{12}/(a\sigma)$. Since
$d_1((a\sigma)v)=(a\sigma)^2v\sqrt{v^2+\xi^2}/\sqrt{12}$,
Eq.~\eqref{eq:dimensionless-response} becomes exactly
\begin{equation}
 \frac{\sqrt\pi}{4}-\mathcal P_1
 =\sqrt{\frac3\pi}\frac1{a\sigma}
 \int_0^\infty
 \frac{e^{-v^2/4}\sin[(\Omega\sigma)v]}
 {v\sqrt{v^2+\xi^2}}\,dv,
 \label{eq:cusped-rescaled-response}
\end{equation}
where $\mathcal P_1$ is the value of $\mathcal P_{\bar b}$
at $\bar b=1$. The factor
$e^{-v^2/4}\sin[(\Omega\sigma)v]/v$ tends to $\Omega\sigma$
at the origin. To isolate the logarithmic divergence as
$\xi\to0^+$, we first split the integration range into
$(0,1)$ and $(1,\infty)$. On $(0,1)$, we add and subtract
$\Omega\sigma$ in this factor. The constant contribution
integrates to $(\Omega\sigma)\operatorname{arsinh}(1/\xi)$,
while the subtracted remainder on $(0,1)$ and the integral
over $(1,\infty)$ give the other two terms:
\begin{equation}
 \begin{split}
 &\int_0^\infty
 \frac{e^{-v^2/4}\sin[(\Omega\sigma)v]}
 {v\sqrt{v^2+\xi^2}}\,dv
 \\
 &\quad=(\Omega\sigma)\operatorname{arsinh}(1/\xi)
 \\
 &\qquad+\int_0^1
 \frac{e^{-v^2/4}\sin[(\Omega\sigma)v]/v-\Omega\sigma}
 {\sqrt{v^2+\xi^2}}\,dv
 \\
 &\qquad+\int_1^\infty
 \frac{e^{-v^2/4}\sin[(\Omega\sigma)v]}
 {v\sqrt{v^2+\xi^2}}\,dv.
 \end{split}
 \label{eq:cusped-subtraction-identity}
\end{equation}
The first term contains the divergent part. Using the
logarithmic representation of the inverse hyperbolic sine gives
\begin{equation}
 \begin{split}
 \operatorname{arsinh}(1/\xi)
 &=\ln\!\left(\frac{1+\sqrt{1+\xi^2}}\xi\right)
 \\
 &=\ln(a\sigma)-\ln\sqrt3
 +O\bigl((a\sigma)^{-2}\bigr).
 \end{split}
 \label{eq:cusped-logarithm}
\end{equation}

We next estimate the error incurred by replacing
$1/\sqrt{v^2+\xi^2}$ with $1/v$ in each of the two
remaining integrals. For the integral over $(0,1)$, denote
the subtracted numerator by
\begin{equation}
 f(v):=\frac{e^{-v^2/4}\sin[(\Omega\sigma)v]}{v}
 -\Omega\sigma.
 \label{eq:cusped-subtracted-numerator}
\end{equation}
Its expansion near the origin is
\begin{equation}
 f(v)=-\left[\frac{\Omega\sigma}{4}
 +\frac{(\Omega\sigma)^3}{6}\right]v^2+O(v^4).
 \label{eq:cusped-subtracted-numerator-expansion}
\end{equation}
Thus $f(v)/v=O(v)$, so the integral obtained after the
replacement is finite at the origin. To show that it
approximates the original integral, we must also bound
their difference. We first rewrite $f(v)$ as
\begin{equation}
 \begin{split}
 f(v)={}&(\Omega\sigma)\bigl(e^{-v^2/4}-1\bigr)
 \\
 &+\frac{e^{-v^2/4}}{v}
 \bigl\{\sin[(\Omega\sigma)v]-(\Omega\sigma)v\bigr\}.
 \end{split}
 \label{eq:cusped-numerator-decomposition}
\end{equation}
Using $0\leq1-e^{-v^2/4}\leq v^2/4$,
$e^{-v^2/4}\leq1$, and $|\sin x-x|\leq|x|^3/6$
gives
\begin{equation}
 |f(v)|\leq Cv^2,\qquad
 C=\frac{\Omega\sigma}{4}+\frac{(\Omega\sigma)^3}{6},
 \label{eq:cusped-numerator-bound}
\end{equation}
where $C$ is independent of $\xi$ at a fixed positive gap.
Rationalizing the difference between the kernels gives
\begin{equation}
 \begin{split}
 0&\leq\frac1v-\frac1{\sqrt{v^2+\xi^2}}
 \\
 &=\frac{\xi^2}
 {v\sqrt{v^2+\xi^2}\bigl(\sqrt{v^2+\xi^2}+v\bigr)}
 \\
 &\leq\min\!\left\{\frac1v,\frac{\xi^2}{2v^3}\right\}.
 \end{split}
 \label{eq:cusped-replacement-bound}
\end{equation}
The first upper bound follows from the nonnegativity of
$1/\sqrt{v^2+\xi^2}$. The second follows from
$\sqrt{v^2+\xi^2}\geq v$, which makes the denominator
in the middle line at least $2v^3$.
The absolute difference between the original subtracted
integral and its proposed limit therefore satisfies
\begin{equation}
 \begin{split}
 &\left|\int_0^1\frac{f(v)}{\sqrt{v^2+\xi^2}}\,dv
 -\int_0^1\frac{f(v)}v\,dv\right|
 \\
 &\quad\leq\int_0^1|f(v)|
 \left(\frac1v-\frac1{\sqrt{v^2+\xi^2}}\right)dv
 \\
 &\quad\leq C\int_0^1v^2
 \left(\frac1v-\frac1{\sqrt{v^2+\xi^2}}\right)dv.
 \end{split}
 \label{eq:cusped-small-v-error-integral}
\end{equation}
If directly using the upper bound $\xi^2/(2v^3)$ throughout $(0,1)$, it
would give a divergent bound proportional to
$\int_0^1dv/v$. Since $0<\xi<1$ at large-acceleration limit, we instead split the error integral at $v=\xi$: we use the upper bound $1/v$
on $(0,\xi)$ and $\xi^2/(2v^3)$ on $(\xi,1)$. This gives
\begin{equation}
 \begin{split}
 &C\int_0^1v^2
 \left(\frac1v-\frac1{\sqrt{v^2+\xi^2}}\right)dv
 \\
 &\quad\leq C\int_0^\xi v\,dv
 +\frac{C\xi^2}{2}\int_\xi^1\frac{dv}{v}
 \\
 &\quad=\frac{C\xi^2}{2}
 +\frac{C\xi^2}{2}\ln(1/\xi).
 \end{split}
 \label{eq:cusped-small-v-error}
\end{equation}
The two intervals therefore contribute errors of order
$\xi^2$ and $\xi^2\ln(1/\xi)$, respectively. In particular,
the logarithm in the error estimate comes from integrating
$1/v$ between $\xi$ and $1$.

For the integral over $(1,\infty)$, the replacement changes
the integrand by
\begin{equation}
 \frac{e^{-v^2/4}\sin[(\Omega\sigma)v]}{v}
 \left(\frac1{\sqrt{v^2+\xi^2}}-\frac1v\right).
 \label{eq:cusped-tail-integrand-difference}
\end{equation}
Using $|\sin[(\Omega\sigma)v]|\leq1$ and the second
kernel bound in Eq.~\eqref{eq:cusped-replacement-bound},
the absolute value of this difference is at most
$\xi^2e^{-v^2/4}/(2v^4)$. Hence the absolute error
in the tail integral is bounded by
\begin{equation}
 \begin{split}
 &\left|\int_1^\infty
 \frac{e^{-v^2/4}\sin[(\Omega\sigma)v]}
 {v\sqrt{v^2+\xi^2}}\,dv
 \right.
 \\
 &\qquad\left.-\int_1^\infty
 \frac{e^{-v^2/4}\sin[(\Omega\sigma)v]}{v^2}\,dv\right|
 \\
 &\quad\leq\frac{\xi^2}{2}\int_1^\infty
 \frac{e^{-v^2/4}}{v^4}\,dv=O(\xi^2).
 \end{split}
 \label{eq:cusped-large-v-error}
\end{equation}
The last integral is finite and independent of $\xi$,
which establishes the stated order. Combining both
intervals, the total replacement error is
$O(\xi^2\ln(1/\xi))$. The inverse hyperbolic sine in
Eq.~\eqref{eq:cusped-logarithm} contributes a further
$O(\xi^2)$ error, so the same order controls the full
expansion. Since $\xi=\sqrt{12}/(a\sigma)$, the integral on the
left-hand side of Eq.~\eqref{eq:cusped-subtraction-identity}
has the expansion
\begin{equation}
 (\Omega\sigma)\ln(a\sigma)+\mathcal C(\Omega\sigma)
 +O\!\left(\frac{\ln(a\sigma)}{(a\sigma)^2}\right),
 \label{eq:cusped-integral-asymptotics}
\end{equation}
where
\begin{equation}
 \begin{split}
 \mathcal C(\Omega\sigma)={}&-(\Omega\sigma)\ln\sqrt3
 \\
 &+\int_0^1
 \frac{e^{-v^2/4}\sin[(\Omega\sigma)v]/v-\Omega\sigma}{v}\,dv
 \\
 &+\int_1^\infty
 \frac{e^{-v^2/4}\sin[(\Omega\sigma)v]}{v^2}\,dv.
 \end{split}
 \label{eq:cusped-finite-part}
\end{equation}
The first integrand is $O(v)$ at the origin by
Eq.~\eqref{eq:cusped-numerator-bound}, while the second
is integrable at infinity. Thus $\mathcal C$ is finite
for every fixed gap.

Combining Eqs.~\eqref{eq:cusped-rescaled-response} and
\eqref{eq:cusped-integral-asymptotics} gives
\begin{equation}
 \begin{split}
 \mathcal P_1={}&\frac{\sqrt\pi}{4}
 -\sqrt{\frac3\pi}
 \frac{(\Omega\sigma)\ln(a\sigma)+\mathcal C(\Omega\sigma)}{a\sigma}
 \\
 &+O\!\left(\frac{\ln(a\sigma)}{(a\sigma)^3}\right).
 \end{split}
 \label{eq:large-acceleration-cusped-probability}
\end{equation}
The probability still tends to $\sqrt\pi/4$, but its leading
correction contains $\ln(a\sigma)$. The integral remainder
in Eq.~\eqref{eq:cusped-integral-asymptotics} acquires one
additional inverse power of $a\sigma$ from the prefactor
in Eq.~\eqref{eq:cusped-rescaled-response}.

For the effective temperature, the logarithm expansion in
Eq.~\eqref{eq:temperature-log-expansion} gives
\begin{equation}
 \begin{split}
 &\ln\frac{\sqrt\pi/2-\mathcal P_1}{\mathcal P_1}
 \\
 &\quad=\frac{8\sqrt3}{\pi(a\sigma)}
 \bigl[(\Omega\sigma)\ln(a\sigma)+\mathcal C(\Omega\sigma)\bigr]
 \\
 &\qquad+O\!\left(\frac{[\ln(a\sigma)]^3}{(a\sigma)^3}\right).
 \end{split}
 \label{eq:cusped-temperature-logarithm}
\end{equation}
Here the probability remainder contributes
$O(\ln(a\sigma)/(a\sigma)^3)$, while the cubic term in
the logarithm contributes
$O([\ln(a\sigma)]^3/(a\sigma)^3)$.
At fixed $\Omega\sigma>0$, the leading logarithm is of
order $\ln(a\sigma)/(a\sigma)$, so its relative remainder
is $O([\ln(a\sigma)]^2/(a\sigma)^2)$.
Substitution into Eq.~\eqref{eq:temperature-from-probability}
therefore gives
\begin{equation}
 \begin{split}
 \sigma T_{\mathrm{eff}}={}&
 \frac{\pi(a\sigma)}
 {8\sqrt3[\ln(a\sigma)+\mathcal C(\Omega\sigma)/(\Omega\sigma)]}
 \\
 &\times\left[1+O\!\left(
 \frac{[\ln(a\sigma)]^2}{(a\sigma)^2}\right)\right].
 \end{split}
 \label{eq:large-acceleration-cusped-temperature}
\end{equation}
Numerical evaluation of Eq.~\eqref{eq:cusped-finite-part}
gives $\mathcal C(\Omega \sigma=1)\simeq-0.42229$ and
$\mathcal C(\Omega \sigma=3)\simeq-3.76139$.
Their different ratios to the energy gap account for the visible
separation of the cusped temperature curves in
Fig.~\ref{fig:temperature-acceleration}.

\subsubsection{Circular motion}
\label{app:large-circular}

For fixed $\bar b>1$, the circular separation is
\begin{equation}
 d_{\bar b}^{\,2}(z)=\gamma^2z^2
 -\frac{4\sin^2\!\left(\sqrt{\bar b^2-1}\,z/2\right)}
 {(\bar b^2-1)^2}.
 \label{eq:circular-separation-for-asymptotics}
\end{equation}
The sine term remains bounded as $z\to\infty$, so
$d_{\bar b}(z)\sim\gamma z$ at this region. Thus the moment
$\int_0^\infty z/d_{\bar b}(z)\,dz$ diverges, preventing
termwise integration of the fixed-$z$ expansion used for
linear and catenary motion.
To retain the contribution from $z$ of order $a\sigma$,
we use the same trick as in the cusped case to set
$z=(a\sigma)v$ in Eq.~\eqref{eq:dimensionless-response}:
\begin{equation}
 \frac{\sqrt\pi}{4}-\mathcal P_{\bar b}
 =\frac1{2\sqrt\pi}\int_0^\infty
 e^{-v^2/4}\sin[(\Omega\sigma)v]\,
 \frac{a\sigma}{d_{\bar b}((a\sigma)v)}\,dv.
 \label{eq:circular-rescaled-response}
\end{equation}
For fixed $v>0$, the last factor tends to $1/(\gamma v)$
at large-$z$ region.
To take this limit under the integral, note that
$\sin^2x\leq x^2$ and
$\gamma^2-(\bar b^2-1)^{-1}=1$ in
Eq.~\eqref{eq:circular-separation-for-asymptotics} imply
$d_{\bar b}(z)\geq z$. The absolute integrand is therefore
bounded by $e^{-v^2/4}|\sin[(\Omega\sigma)v]|/v$.
Then, the following integral can be evaluated by writing
$\sin[(\Omega\sigma)v]/v
=\int_0^{\Omega\sigma}\cos(yv)\,dy$
and using the Gaussian cosine transform:
\begin{equation}
 \begin{split}
 \int_0^\infty e^{-v^2/4}
 \frac{\sin[(\Omega\sigma)v]}v\,dv
 &=\sqrt\pi\int_0^{\Omega\sigma}e^{-y^2}\,dy
 \\
 &=\frac\pi2\operatorname{erf}(\Omega\sigma).
 \end{split}
 \label{eq:gaussian-sine-over-v}
\end{equation}
Consequently,
\begin{equation}
 \mathcal P_\infty
 :=\lim_{a\sigma\to\infty}\mathcal P_{\bar b}
 =\frac{\sqrt\pi}{4}
 \left[1-\frac{\operatorname{erf}(\Omega\sigma)}\gamma\right].
 \label{eq:circular-probability-plateau}
\end{equation}
The surviving $1/(\gamma z)$ part of the inverse separation
therefore lowers the circular limit below $\sqrt\pi/4$.

To determine the approach to this limit, we subtract that
part before expanding the remaining integral. Define
\begin{equation}
 \begin{split}
 g_{\bar b}(z)
 &:=\frac{z}{d_{\bar b}(z)}-\frac1\gamma
 \\
 &=\frac{\gamma^2z^2-d_{\bar b}^{\,2}(z)}
 {\gamma d_{\bar b}(z)[\gamma z+d_{\bar b}(z)]}.
 \end{split}
 \label{eq:circular-remainder-kernel}
\end{equation}
The numerator is bounded and nonnegative by
Eq.~\eqref{eq:circular-separation-for-asymptotics}.
Since the denominator is strictly positive for $z>0$,
we have $g_{\bar b}(z)\geq0$ throughout the integration
range. Moreover, the denominator grows as $2\gamma^3z^2$,
so $g_{\bar b}(z)=O(z^{-2})$ at infinity.
At the origin, $d_{\bar b}(z)/z\to1$, and hence
$g_{\bar b}(z)\to1-1/\gamma>0$.
The kernel is therefore integrable, and we define
\begin{equation}
 \mathcal R(\bar b):=\int_0^\infty g_{\bar b}(z)\,dz.
 \label{eq:circular-remainder-moment}
\end{equation}
Since $g_{\bar b}(z)\geq0$, it immediately gives $\mathcal R(\bar b)>0$.
The remaining response is therefore
\begin{equation}
 \begin{split}
 \mathcal P_{\bar b}-\mathcal P_\infty
 ={}&-\frac1{2\sqrt\pi}\int_0^\infty
 g_{\bar b}(z)e^{-z^2/[4(a\sigma)^2]}
 \\
 &\hspace{2em}\times
 \frac{\sin[(\Omega\sigma)z/(a\sigma)]}{z}\,dz.
 \end{split}
 \label{eq:circular-subtracted-response}
\end{equation}
At fixed $z$, we aim to replace the Gaussian--sine
factor with $(\Omega\sigma)/(a\sigma)$, and the
corresponding error in this replacement is defined as
\begin{equation}
 \begin{split}
 r_{a\sigma}(z):={}&
 e^{-z^2/[4(a\sigma)^2]}
 \frac{\sin[(\Omega\sigma)z/(a\sigma)]}{z}
 -\frac{\Omega\sigma}{a\sigma}
 \\
 ={}&\frac{e^{-z^2/[4(a\sigma)^2]}}z
 \left\{\sin\!\left[\frac{(\Omega\sigma)z}{a\sigma}\right]
 -\frac{(\Omega\sigma)z}{a\sigma}\right\}
 \\
 &+\frac{\Omega\sigma}{a\sigma}
 \left(e^{-z^2/[4(a\sigma)^2]}-1\right).
 \end{split}
 \label{eq:circular-replacement-error}
\end{equation}
The constant part integrates to
$(\Omega\sigma)\mathcal R/(a\sigma)$, so the response becomes
\begin{equation}
 \begin{split}
 \mathcal P_{\bar b}-\mathcal P_\infty
 ={}&-\frac{(\Omega\sigma)\mathcal R(\bar b)}
 {2\sqrt\pi(a\sigma)}
 \\
 &-\frac1{2\sqrt\pi}\int_0^\infty
 g_{\bar b}(z)r_{a\sigma}(z)\,dz.
 \end{split}
 \label{eq:circular-response-remainder}
\end{equation}
It remains to bound the last integral.
Using $|\sin x-x|\leq |x|^3/6$ and
$0\leq1-e^{-x}\leq x$ for $x\geq0$ gives
\begin{equation}
 |r_{a\sigma}(z)|
 \leq
 \left[\frac{\Omega\sigma}{4}
 +\frac{(\Omega\sigma)^3}{6}\right]
 \frac{z^2}{(a\sigma)^3}.
 \label{eq:circular-small-z-error-bound}
\end{equation}
This bound grows with $z$, so we use it on $(0,a\sigma)$
and estimate the tail separately. To make both estimates
explicit, note that $z^2g_{\bar b}(z)$ is continuous, tends
to zero at the origin, and is $O(1)$ at infinity. Hence there
is a constant $K_{\bar b}>0$, independent of $a\sigma$, such
that $0\leq z^2g_{\bar b}(z)\leq K_{\bar b}$ for all $z>0$.
On the first interval this gives
\begin{equation}
 \begin{split}
 &\left|\int_0^{a\sigma}
 g_{\bar b}(z)r_{a\sigma}(z)\,dz\right|
 \\
 &\quad\leq
 \frac{(\Omega\sigma)/4+(\Omega\sigma)^3/6}{(a\sigma)^3}
 \int_0^{a\sigma}z^2g_{\bar b}(z)\,dz
 \\
 &\quad\leq
 \frac{K_{\bar b}[(\Omega\sigma)/4+(\Omega\sigma)^3/6]}
 {(a\sigma)^3}\int_0^{a\sigma}dz
 \\
 &\quad=\frac{K_{\bar b}[(\Omega\sigma)/4+(\Omega\sigma)^3/6]}
 {(a\sigma)^2}.
 \end{split}
 \label{eq:circular-small-z-integrated-error}
\end{equation}
The factor $a\sigma$ thus comes from integrating a constant
bound over an interval of length $a\sigma$.
For $z>a\sigma$, the bounds
$e^{-z^2/[4(a\sigma)^2]}\leq1$ and $|\sin x|\leq1$
instead give
$|r_{a\sigma}(z)|\leq1/z+(\Omega\sigma)/(a\sigma)
\leq(1+\Omega\sigma)/(a\sigma)$.
Since $g_{\bar b}(z)\leq K_{\bar b}/z^2$, the tail satisfies
\begin{equation}
 \begin{split}
 &\left|\int_{a\sigma}^\infty
 g_{\bar b}(z)r_{a\sigma}(z)\,dz\right|
 \\
 &\quad\leq\frac{1+\Omega\sigma}{a\sigma}
 \int_{a\sigma}^\infty g_{\bar b}(z)\,dz
 \\
 &\quad\leq\frac{K_{\bar b}(1+\Omega\sigma)}{a\sigma}
 \int_{a\sigma}^\infty\frac{dz}{z^2}
 =\frac{K_{\bar b}(1+\Omega\sigma)}{(a\sigma)^2}.
 \end{split}
 \label{eq:circular-tail-integrated-error}
\end{equation}
Here the integral of $z^{-2}$ equals $1/(a\sigma)$, giving
the second inverse power of $a\sigma$ in the tail error.
Combining the two error estimates in
Eq.~\eqref{eq:circular-response-remainder} yields
\begin{equation}
 \mathcal P_{\bar b}=\mathcal P_\infty
 -\frac{(\Omega\sigma)\mathcal R(\bar b)}{2\sqrt\pi(a\sigma)}
 +O\bigl((a\sigma)^{-2}\bigr).
 \label{eq:large-acceleration-circular-probability}
\end{equation}

The effective temperature in the circular case follows
by substituting
Eq.~\eqref{eq:circular-probability-plateau} into
Eq.~\eqref{eq:temperature-from-probability}:
\begin{equation}
 \sigma T_\infty
 =\frac{\Omega\sigma}
 {\ln\!\left[\dfrac{\gamma+\operatorname{erf}(\Omega\sigma)}
 {\gamma-\operatorname{erf}(\Omega\sigma)}\right]}.
 \label{eq:large-acceleration-circular-temperature}
\end{equation}
For the correction, write
$\delta\mathcal P=\mathcal P_{\bar b}-\mathcal P_\infty$.
Expanding the numerator and denominator logarithms separately
gives
\begin{equation}
 \begin{split}
 \ln\frac{\sqrt\pi/2-\mathcal P_{\bar b}}{\mathcal P_{\bar b}}
 ={}&\frac{\Omega\sigma}{\sigma T_\infty}
 \\
 &-\left[\frac1{\sqrt\pi/2-\mathcal P_\infty}
 +\frac1{\mathcal P_\infty}\right]\delta\mathcal P
 \\
 &+O\bigl((\delta\mathcal P)^2\bigr).
 \end{split}
 \label{eq:circular-temperature-log-expansion}
\end{equation}
Using Eq.~\eqref{eq:large-acceleration-circular-probability}
and the identity
$\mathcal P_\infty(\sqrt\pi/2-\mathcal P_\infty)
=(\pi/16)[1-\operatorname{erf}^2(\Omega\sigma)/\gamma^2]$,
we obtain the inverse-temperature expansion
\begin{equation}
 \begin{split}
 \frac1{\sigma T_{\mathrm{eff}}}={}&\frac1{\sigma T_\infty}
 \\
 &+\frac{4\mathcal R(\bar b)}
 {\pi(a\sigma)[1-\operatorname{erf}^2(\Omega\sigma)/\gamma^2]}
 \\
 &+O\bigl((a\sigma)^{-2}\bigr).
 \end{split}
 \label{eq:circular-inverse-temperature-expansion}
\end{equation}
Taking the reciprocal to first order then gives
\begin{equation}
 \begin{split}
 \sigma T_{\mathrm{eff}}={}&\sigma T_\infty
 -\frac{4\mathcal R(\bar b)(\sigma T_\infty)^2}
 {\pi(a\sigma)[1-\operatorname{erf}^2(\Omega\sigma)/\gamma^2]}
 \\
 &+O\bigl((a\sigma)^{-2}\bigr).
 \end{split}
 \label{eq:circular-temperature-first-correction}
\end{equation}
Both Taylor remainders are of order $(a\sigma)^{-2}$,
since $\delta\mathcal P=O((a\sigma)^{-1})$ and the limiting
logarithm is positive at fixed $\bar b>1$ and $\Omega\sigma>0$.
The negative first corrections show that both limits are
approached from below at sufficiently large acceleration.
For $\bar b=2$, numerical integration gives
$\mathcal R(2)\simeq0.22612$. The factor $\operatorname{erf}(\Omega\sigma)$ retains the
Gaussian switching dependence. In the overlap
$1\ll\Omega\sigma\ll a\sigma$, it tends to unity, and
Eq.~\eqref{eq:large-acceleration-circular-temperature}
recovers the stationary small-gap result
$T\sim\Omega/\ln[(\gamma+1)/(\gamma-1)]$
\cite{Biermann2020,ParryLouko2025}.

\section{Derivations for the torsion-dependence analysis}
\label{app:torsion-monotonicity}

This appendix gives the derivations used in
Sec.~\ref{subsec:torsion-monotonicity}. Using the response integral
and invariant separation from
Sec.~\ref{subsec:acceleration-asymptotics}, we first establish the
positive-temperature domain and the relation between the extrema of
the two observables. We then derive the torsion derivative and analyze
the small-gap regime, before determining the boundary and interior
thresholds for the first loss of decreasing behavior.

\subsection{Positive temperature and shared extrema}
\label{app:temperature-monotonicity}

Equation~\eqref{eq:temperature-from-probability} expresses $\sigma T_{\mathrm{eff}}$ as a function of $\mathcal P_{\bar b}$. Before differentiating this relation, we establish its physical domain,
\begin{equation}
 0<\mathcal P_{\bar b}<\frac{\sqrt\pi}{4}.
 \label{eq:positive-temperature-domain}
\end{equation}
The upper bound follows from the integral in Eq.~\eqref{eq:dimensionless-response}. Its sign is not fixed by positivity of the inverse separation alone, because the sine oscillates. The additional property needed here is that the inverse separation, including its Gaussian weight, decreases with $z$.

This property can be checked directly from the invariant interval. For $0\leq\bar b<1$, differentiation with respect to $z$ gives
\begin{align}
 \frac{d}{dz}d_{\bar b}^2(z)
 &=\frac{2}{(1-\bar b^2)^{3/2}}
       (\sinh y-\bar b^2y)>0,
 \nonumber\\[-0.3em]
 &\hspace{4em}y=\sqrt{1-\bar b^2}\,z>0.
 \label{eq:separation-increases-catenary}
\end{align}
Indeed, $\sinh y>y\geq\bar b^2y$. For $\bar b>1$, the corresponding expression is
\begin{align}
 \frac{d}{dz}d_{\bar b}^2(z)
 &=\frac{2}{(\bar b^2-1)^{3/2}}
       (\bar b^2y-\sin y)>0,
 \nonumber\\[-0.3em]
 &\hspace{4em}y=\sqrt{\bar b^2-1}\,z>0,
 \label{eq:separation-increases-circular}
\end{align}
since $\bar b^2y-\sin y\geq(\bar b^2-1)y>0$. At $\bar b=1$, $d_1^2=z^2+z^4/12$ is also strictly increasing. Therefore
\begin{equation}
 f(z):=\frac{e^{-z^2/[4(a\sigma)^2]}}{d_{\bar b}(z)}
 \label{eq:decreasing-weighted-inverse-separation}
\end{equation}
is positive and strictly decreasing for $z>0$.

To determine the sign of its sine transform, temporarily write $k=(\Omega\sigma)/(a\sigma)>0$ and pair each positive half period with the following negative one. Changing variables in each pair gives
\begin{align}
 \int_0^\infty f(z)\sin(kz)\,dz
 ={}&\frac1k\sum_{n=0}^{\infty}\int_0^\pi dy\,\sin y
 \nonumber\\
 &\times\biggl[
 f\!\left(\frac{2n\pi+y}{k}\right)
 \nonumber\\
 &\hspace{2em}-f\!\left(\frac{(2n+1)\pi+y}{k}\right)
 \biggr]>0.
 \label{eq:positive-paired-sine-transform}
\end{align}
Every bracket is positive because its first argument is smaller than its second. The apparent singularity at the first endpoint causes no difficulty: $d_{\bar b}(z)\sim z$, so $f(z)\sin(kz)$ has a finite limit there. Equation~\eqref{eq:dimensionless-response} subtracts this positive integral from $\sqrt\pi/4$, proving the upper bound in Eq.~\eqref{eq:positive-temperature-domain}.

The strict lower bound follows independently from positivity of the detector response. More explicitly, define the stationary spectrum by $\widehat W_{\bar b}(E)=\int_{\mathbb R}du\,e^{-iEu}W_{\bar b}(u)$. The positive-type property of the Wightman function makes this spectrum nonnegative, and Gaussian switching gives
\begin{equation}
 \mathcal P_{\bar b}
 =\sigma\int_{\mathbb R}dE\,
 e^{-\sigma^2(E-\Omega)^2}\widehat W_{\bar b}(E)>0.
 \label{eq:spectral-gaussian-convolution}
\end{equation}
The Gaussian is strictly positive and the spectrum is nonzero. Together, the two bounds imply that the logarithm in Eq.~\eqref{eq:temperature-from-probability} is positive, so the effective temperature is well defined and positive for every positive finite gap considered here.

Differentiating that equation at fixed gap now gives
\begin{align}
 \frac{d(\sigma T_{\mathrm{eff}})}{d\mathcal P_{\bar b}}
 ={}&\frac{(\Omega\sigma)(\sqrt\pi/2)}
 {\mathcal P_{\bar b}(\sqrt\pi/2-\mathcal P_{\bar b})}
 \nonumber\\
 &\times\frac{1}
 {\left\{\ln[(\sqrt\pi/2-\mathcal P_{\bar b})/
 \mathcal P_{\bar b}]\right\}^2}>0.
 \label{eq:temperature-probability-derivative}
\end{align}
The first derivative obeys the chain rule~\eqref{eq:shared-monotonicity} in the main text. This proves that its sign is preserved.

For clarity, differentiating once more gives the general relation
\begin{align}
 \partial_{\bar b}^2(\sigma T_{\mathrm{eff}})
 ={}&\frac{d(\sigma T_{\mathrm{eff}})}{d\mathcal P_{\bar b}}
       \partial_{\bar b}^2\mathcal P_{\bar b}
 \nonumber\\
 &+\frac{d^2(\sigma T_{\mathrm{eff}})}{d\mathcal P_{\bar b}^2}
       \bigl(\partial_{\bar b}\mathcal P_{\bar b}\bigr)^2.
 \label{eq:temperature-second-chain-rule}
\end{align}
At any ratio where $\partial_{\bar b}\mathcal P_{\bar b}=0$, the second term vanishes. At those ratios, the second derivatives therefore differ only by the positive factor in Eq.~\eqref{eq:temperature-probability-derivative}; an ordinary minimum remains a minimum and an ordinary maximum remains a maximum. This proportionality of second derivatives is not asserted away from zeros of the first derivative. It is consequently sufficient to analyze $\mathcal P_{\bar b}$ below.

\subsection{The torsion derivative of the invariant separation}
\label{app:geometric-kernel}

The first step is to determine how the invariant separation changes with $\bar b$ while $z$ is held fixed. For $0\leq\bar b<1$, expansion of the hyperbolic function in the exact interval gives
\begin{equation}
 d_{\bar b}^2(z)=z^2+
 2\sum_{n=2}^{\infty}
 \frac{(1-\bar b^2)^{n-2}z^{2n}}{(2n)!}.
 \label{eq:hyperbolic-separation-series}
\end{equation}
The $n=2$ term is independent of $\bar b$. Differentiating the remaining terms therefore gives
\begin{equation}
 \partial_{\bar b}d_{\bar b}^2(z)
 =-4\bar b\sum_{n=3}^{\infty}
 \frac{(n-2)(1-\bar b^2)^{n-3}z^{2n}}{(2n)!}<0
 \label{eq:catenary-separation-torsion-derivative}
\end{equation}
for $0<\bar b<1$ and $z>0$. Each term in the sum is positive, which fixes the sign without any cancellation. At $\bar b=1$, only the $n=3$ term survives in this derivative, giving the continuous limit
\begin{equation}
 \left.\partial_{\bar b}d_{\bar b}^2(z)\right|_{\bar b=1}
 =-\frac{z^6}{180}<0.
 \label{eq:cusped-separation-torsion-derivative}
\end{equation}

For circular motion, the separation is
\begin{equation}
 d_{\bar b}^2(z)=\frac{\bar b^2z^2}{\bar b^2-1}
 -\frac{4\sin^2[z\sqrt{\bar b^2-1}/2]}{(\bar b^2-1)^2}.
 \label{eq:circular-separation-for-torsion}
\end{equation}
Set $x=z\sqrt{\bar b^2-1}/2$. In differentiating at fixed $z$, one must include $\partial_{\bar b}x=\bar b x/(\bar b^2-1)$. Collecting the resulting terms gives
\begin{align}
 \partial_{\bar b}d_{\bar b}^2(z)
 &=-\frac{8\bar b}{(\bar b^2-1)^3}G(x),
 \nonumber\\
 G(x)&:=x^2+x\sin x\cos x-2\sin^2x.
 \label{eq:circular-separation-torsion-derivative}
\end{align}
It remains to show $G(x)>0$. For $0<x<\pi$,
\begin{equation}
 G''(x)=4\sin x\,(\sin x-x\cos x)>0.
 \label{eq:circular-G-convexity}
\end{equation}
Here $\sin x>0$, while the second factor starts from zero and has derivative $x\sin x>0$. Since $G(0)=G'(0)=0$, two integrations imply $G(x)>0$ throughout this interval. For $x\geq\pi$, the bounds $\sin x\cos x\geq-1/2$ and $\sin^2x\leq1$ give
\begin{equation}
 G(x)\geq x^2-\frac{x}{2}-2>0.
 \label{eq:circular-G-large-x-bound}
\end{equation}
This completes the sign proof for circular motion. Thus $\partial_{\bar b}d_{\bar b}^2(z)<0$ for every $\bar b>0$ and $z>0$.

To transfer this geometric result to the response, define
\begin{equation}
 \mathcal K_{\bar b}(z):=
 -\frac{\partial_{\bar b}d_{\bar b}^2(z)}{d_{\bar b}^3(z)}>0.
 \label{eq:torsion-kernel}
\end{equation}
The factor and sign in the response derivative follow from
\begin{equation}
 \partial_{\bar b}\frac1{d_{\bar b}(z)}
 =-\frac{\partial_{\bar b}d_{\bar b}^2(z)}{2d_{\bar b}^3(z)}
 =\frac12\mathcal K_{\bar b}(z).
 \label{eq:inverse-separation-torsion-derivative}
\end{equation}
Substituting this identity into the derivative of Eq.~\eqref{eq:dimensionless-response} yields
\begin{align}
 \partial_{\bar b}\mathcal P_{\bar b}
 ={}&-\frac1{4\sqrt\pi}\int_0^\infty dz\,
 e^{-z^2/[4(a\sigma)^2]}\mathcal K_{\bar b}(z)
 \nonumber\\
 &\hspace{4em}\times\sin\!\left(\frac{(\Omega\sigma)z}{a\sigma}\right).
 \label{eq:torsion-derivative-transform}
\end{align}
Although $\mathcal K_{\bar b}>0$, this integral need not be positive at an arbitrary gap because the sine takes both signs. At a small gap, however, the sine is approximately linear over the separations that contribute appreciably through the Gaussian, and its leading contribution is positive.

\subsection{Uniform decrease at small gaps}
\label{app:small-gap-monotonicity}

We first examine the small-gap regime at fixed $a\sigma>0$.
The monotonicity statement at small-gap limit is that there exists a positive
$\delta(a\sigma)$ that depends on $a \sigma$ such that
\begin{equation}
 0<\Omega\sigma<\delta(a\sigma)
 \quad\Longrightarrow\quad
 \partial_{\bar b}\mathcal P_{\bar b}<0
 \qquad\text{for all }\bar b>0.
 \label{eq:appendix-small-gap-statement}
\end{equation}
To prove such a statement, we first rewrite
$\sin[(\Omega\sigma)z/(a\sigma)]
=(\Omega\sigma)z/(a\sigma)+r(z)$ where $r(z)=\sin[(\Omega\sigma)z/(a\sigma)]-(\Omega\sigma)z/(a\sigma)$.
Equation~\eqref{eq:torsion-derivative-transform} then becomes
\begin{equation}
 \begin{split}
 \partial_{\bar b}\mathcal P_{\bar b}
 ={}&-\frac{\Omega\sigma}{4\sqrt\pi(a\sigma)}
 \int_0^\infty dz\,z\,
 e^{-z^2/[4(a\sigma)^2]}\mathcal K_{\bar b}(z)\\
 &-\frac1{4\sqrt\pi}\int_0^\infty dz\,
 e^{-z^2/[4(a\sigma)^2]}\mathcal K_{\bar b}(z)r(z).
 \end{split}
 \label{eq:small-gap-integral-decomposition}
\end{equation}
Denote the positive integral in the first term as
\begin{equation}
 M_1(\bar b):=\int_0^\infty dz\,z\,
 e^{-z^2/[4(a\sigma)^2]}\mathcal K_{\bar b}(z)>0.
 \label{eq:kernel-moments}
\end{equation}
By its definition it is easy to check its positivity. Then we define the second term by
\begin{equation}
 R_{\bar b}:=-\frac1{4\sqrt\pi}\int_0^\infty dz\,
 e^{-z^2/[4(a\sigma)^2]}\mathcal K_{\bar b}(z)r(z).
 \label{eq:small-gap-integral-remainder}
\end{equation}
The derivative therefore takes the form
\begin{equation}
 \partial_{\bar b}\mathcal P_{\bar b}
 =-\frac{\Omega\sigma}{4\sqrt\pi(a\sigma)}M_1(\bar b)
 +R_{\bar b}.
 \label{eq:small-gap-response-decomposition}
\end{equation}
Since the first term is always negative, we now want to determine the sign and size of the second term $R_{\bar b}$.
For $x\geq0$, the identity
$x-\sin x=\int_0^x(1-\cos t)dt$, together with
$0\leq1-\cos t\leq t^2/2$, gives
$0\leq x-\sin x\leq x^3/6$.
Consequently,
\begin{equation}
 0\leq-r(z)\leq
 \frac{(\Omega\sigma)^3z^3}{6(a\sigma)^3}.
 \label{eq:small-gap-sine-remainder}
\end{equation}
Since $\mathcal K_{\bar b}(z)>0$ for $\bar b>0$,
this makes $R_{\bar b}>0$: the remainder opposes the
negative linear contribution. Moreover, its magnitude can be bounded by
\begin{equation}
 \begin{split}
 |R_{\bar b}|
 &\leq\frac1{4\sqrt\pi}\int_0^\infty dz\,
 e^{-z^2/[4(a\sigma)^2]}\mathcal K_{\bar b}(z)|r(z)|\\
 &\leq\frac{(\Omega\sigma)^3}{24\sqrt\pi(a\sigma)^3}
 \int_0^\infty dz\,z^3
 e^{-z^2/[4(a\sigma)^2]}\mathcal K_{\bar b}(z)\\
 &=\frac{(\Omega\sigma)^3}{24\sqrt\pi(a\sigma)^3}
 M_3(\bar b),
 \end{split}
 \label{eq:small-gap-error-bound}
\end{equation}
where the last integral defines a second positive quantity,
\begin{equation}
 M_3(\bar b):=\int_0^\infty dz\,z^3
 e^{-z^2/[4(a\sigma)^2]}\mathcal K_{\bar b}(z)>0.
 \label{eq:third-kernel-moment}
\end{equation}
Both $M_1$ and $M_3$ are independent of the gap.
Combining the decomposition and the remainder estimate gives
\begin{equation}
 \begin{split}
 \partial_{\bar b}\mathcal P_{\bar b}
 \leq{}&-\frac{\Omega\sigma}{4\sqrt\pi(a\sigma)}M_1(\bar b)\\
 &+\frac{(\Omega\sigma)^3}{24\sqrt\pi(a\sigma)^3}M_3(\bar b).
 \end{split}
 \label{eq:small-gap-derivative-upper-bound}
\end{equation}
The derivative is therefore strictly negative whenever the
positive correction is smaller than the magnitude of the
negative term, namely,
\begin{equation}
 \frac{(\Omega\sigma)^3}{24\sqrt\pi(a\sigma)^3}M_3(\bar b)
 <\frac{\Omega\sigma}{4\sqrt\pi(a\sigma)}M_1(\bar b).
 \label{eq:small-gap-dominance-condition}
\end{equation}
Dividing by the positive quantity
$(\Omega\sigma)M_3(\bar b)/[24\sqrt\pi(a\sigma)^3]$
gives the sufficient condition
\begin{equation}
 (\Omega\sigma)^2
 <6(a\sigma)^2\frac{M_1(\bar b)}{M_3(\bar b)}.
 \label{eq:small-gap-monotonicity}
\end{equation}
The sufficient condition
\eqref{eq:small-gap-monotonicity} still depends on $\bar b$ through
$M_1(\bar b)/M_3(\bar b)$. To obtain a single small-gap interval that
works for every $\bar b>0$, we must show that this ratio is bounded
away from zero throughout the torsion range. Positivity at each
fixed $\bar b$ is not sufficient, since the ratio could still tend
to zero as $\bar b\to0^+$ or $\bar b\to\infty$. We therefore examine
these two limits before treating the intervening range. 

Before treating these two limits, we want to prove a bound that is repeatedly used in the following estimates:
\begin{equation}
 d_{\bar b}(z)\geq z, ~~~~~\text{for}\qquad z>0,\quad \bar b\geq0.
 \label{eq:small-gap-separation-lower-bound}
\end{equation}
For $0\leq\bar b<1$, this follows from
Eq.~\eqref{eq:hyperbolic-separation-series}: every term added
to $z^2$ is nonnegative. At $\bar b=1$, the explicit expression
$d_1^2(z)=z^2+z^4/12$ gives the same conclusion.
For $\bar b>1$, set $x=z\sqrt{\bar b^2-1}/2$ and rewrite
Eq.~\eqref{eq:circular-separation-for-torsion} as
\begin{equation}
 \begin{split}
 d_{\bar b}^2(z)
 &=z^2+\frac{z^2}{\bar b^2-1}
   -\frac{4\sin^2x}{(\bar b^2-1)^2}\\
 &=z^2+\frac{4(x^2-\sin^2x)}{(\bar b^2-1)^2}
 \geq z^2,
 \end{split}
 \label{eq:small-gap-circular-separation-bound}
\end{equation}
where the inequality follows from $|\sin x|\leq x$.
Taking the positive square root establishes
Eq.~\eqref{eq:small-gap-separation-lower-bound} for all four
trajectory classes.

Let us first consider $\bar b\to0^+$. Combining
Eqs.~\eqref{eq:catenary-separation-torsion-derivative} and
\eqref{eq:torsion-kernel} gives
\begin{equation}
 \frac{\mathcal K_{\bar b}(z)}{\bar b}
 =
 \frac{4}{d_{\bar b}^3(z)}
 \sum_{k=3}^{\infty}
 \frac{(k-2)(1-\bar b^2)^{k-3}z^{2k}}{(2k)!}.
 \label{eq:small-gap-kernel-near-linear}
\end{equation}
For each fixed $z>0$, $d_{\bar b}(z)\to d_0(z)=2\sinh(z/2)$.
The series converges uniformly for $0\leq\bar b\leq1$, since its
terms are bounded by the summable series obtained by setting
$\bar b=0$. Consequently,
\begin{equation}
 \begin{split}
 k_0(z)
 &:=
 \lim_{\bar b\to0^+}\frac{\mathcal K_{\bar b}(z)}{\bar b}\\
 &=
 \frac{4}{[2\sinh(z/2)]^3}
 \sum_{k=3}^{\infty}\frac{(k-2)z^{2k}}{(2k)!}>0.
 \end{split}
 \label{eq:small-gap-linear-kernel-limit}
\end{equation}
Every term in the last series is positive for $z>0$.

To pass this limit through the moment integrals, we need an
integrable bound independent of $\bar b$. For $0<\bar b<1$,
the inequalities $(1-\bar b^2)^{k-3}\leq1$ and
$d_{\bar b}(z)\geq z$ imply
\begin{equation}
 0<
 k_0(z)
 \leq H(z),
 \qquad
 H(z):=\frac{4}{z^3}
 \sum_{k=3}^{\infty}\frac{(k-2)z^{2k}}{(2k)!}.
 \label{eq:small-gap-linear-kernel-bound}
\end{equation}

For $0<z\leq1$, each power $z^{2k-3}$ is at most $z^3$, so
\begin{equation}
 H(z)\leq C_0z^3,
 \qquad
 C_0:=4\sum_{k=3}^{\infty}\frac{k-2}{(2k)!}<\infty.
 \label{eq:small-gap-linear-bound-origin}
\end{equation}
For $z\geq1$, replacing $k-2$ by $k$ and extending the sum gives
\begin{equation}
 \begin{split}
 H(z)
 &\leq\frac{4}{z^3}
 \sum_{k=1}^{\infty}\frac{kz^{2k}}{(2k)!}\\
 &=\frac{2\sinh z}{z^2}
 \leq\frac{e^z}{z^2}.
 \end{split}
 \label{eq:small-gap-linear-bound-tail}
\end{equation}
It follows that
$z^n e^{-z^2/[4(a\sigma)^2]}H(z)$ is integrable for $n=1,3$:
near zero it is bounded by a constant times $z^{n+3}$, while
at infinity the Gaussian dominates the exponential factor $e^z$.
Dominated convergence therefore yields
\begin{equation}
 \begin{split}
 \lim_{\bar b\to0^+}\frac{M_n(\bar b)}{\bar b}
 &=\int_0^\infty dz\,z^n
 e^{-z^2/[4(a\sigma)^2]}k_0(z)\\
 &:=m_n,\qquad 0<m_n<\infty,\quad n=1,3.
 \end{split}
 \label{eq:small-gap-linear-moment-limits}
\end{equation}
Thus both moments vanish linearly with $\bar b$, and their common
factor cancels in the ratio:
\begin{equation}
 \lim_{\bar b\to0^+}\frac{M_1(\bar b)}{M_3(\bar b)}
 =\lim_{\bar b\to0^+}
 \frac{M_1(\bar b)/\bar b}{M_3(\bar b)/\bar b}
 =\frac{m_1}{m_3}>0.
 \label{eq:small-gap-linear-ratio-limit}
\end{equation}
At $\bar b=0$, both moments themselves vanish. We therefore
understand their ratio at this endpoint as the continuous
extension defined by Eq.~\eqref{eq:small-gap-linear-ratio-limit}.

Next consider $\bar b\to\infty$. With
$x=z\sqrt{\bar b^2-1}/2$, the circular expression
\eqref{eq:circular-separation-torsion-derivative} gives
\begin{equation}
 \begin{split}
 \bar b^3\mathcal K_{\bar b}(z)
 &=\frac{8\bar b^4G(x)}
 {(\bar b^2-1)^3d_{\bar b}^3(z)}\\
 &=2\frac{\bar b^4}{(\bar b^2-1)^2}
 \frac{z^2}{d_{\bar b}^3(z)}
 \frac{G(x)}{x^2}.
 \end{split}
 \label{eq:small-gap-circular-scaled-kernel}
\end{equation}
For each fixed $z>0$, $x\to\infty$ and
\begin{equation}
 \frac{G(x)}{x^2}
 =1+\frac{\sin x\cos x}{x}
 -2\frac{\sin^2x}{x^2}
 \longrightarrow1.
 \label{eq:small-gap-circular-G-limit}
\end{equation}
Moreover, Eq.~\eqref{eq:circular-separation-for-torsion} gives
$d_{\bar b}(z)\to z$: the coefficient of $z^2$ tends to unity,
and the bounded sine term is suppressed by
$(\bar b^2-1)^{-2}$. Taking the limit of each factor in
Eq.~\eqref{eq:small-gap-circular-scaled-kernel} therefore gives
\begin{equation}
 \lim_{\bar b\to\infty}
 \bar b^3\mathcal K_{\bar b}(z)=\frac{2}{z}.
 \label{eq:small-gap-circular-kernel-limit}
\end{equation}

We again establish a uniform bound before taking this limit
inside the moments. The positivity of $G$ proved above and
$|\sin x|\leq x$ imply
\begin{equation}
 0<\frac{G(x)}{x^2}
 \leq1+\frac{|\sin x\cos x|}{x}
 \leq2.
 \label{eq:small-gap-circular-G-bound}
\end{equation}
For $\bar b\geq2$,
$\bar b^4/(\bar b^2-1)^2\leq16/9$.
Using $d_{\bar b}(z)\geq z$ in
Eq.~\eqref{eq:small-gap-circular-scaled-kernel}, we obtain
\begin{equation}
 0<\bar b^3\mathcal K_{\bar b}(z)
 \leq\frac{64}{9z}.
 \label{eq:small-gap-circular-kernel-bound}
\end{equation}
Although the right-hand side is singular at zero, multiplication
by the moment weight $z^n$ gives $z^{n-1}$, which is integrable
there for both $n=1$ and $n=3$. The Gaussian ensures integrability
at infinity. Dominated convergence consequently gives
\begin{equation}
 \lim_{\bar b\to\infty}\bar b^3M_n(\bar b)
 =2\int_0^\infty dz\,z^{n-1}
 e^{-z^2/[4(a\sigma)^2]},
 \qquad n=1,3.
 \label{eq:small-gap-circular-moment-limits}
\end{equation}
The two required Gaussian integrals are
\begin{equation}
 \begin{split}
 \int_0^\infty e^{-z^2/[4(a\sigma)^2]}\,dz
 &=\sqrt\pi(a\sigma),\\
 \int_0^\infty z^2e^{-z^2/[4(a\sigma)^2]}\,dz
 &=2\sqrt\pi(a\sigma)^3.
 \end{split}
 \label{eq:small-gap-circular-gaussian-moments}
\end{equation}
The common $\bar b^{-3}$ scaling of the moments therefore
cancels, leaving
\begin{equation}
 \begin{split}
 \lim_{\bar b\to\infty}\frac{M_1(\bar b)}{M_3(\bar b)}
 &=
 \lim_{\bar b\to\infty}
 \frac{\bar b^3M_1(\bar b)}{\bar b^3M_3(\bar b)}\\
 &=\frac{1}{2(a\sigma)^2}>0.
 \end{split}
 \label{eq:small-gap-circular-ratio-limit}
\end{equation}

It remains to control the ratio between the two endpoints.
The separation series~\eqref{eq:hyperbolic-separation-series}
also represents the circular expression, with alternating
powers of $1-\bar b^2$, and is regular at $\bar b=1$.
For each fixed $z>0$, it therefore makes
$\mathcal K_{\bar b}(z)$ continuous in $\bar b$.
To justify continuity after integration, consider any compact
interval in $(0,\infty)$ and let $B$ be its upper endpoint.
Set $Q=\max\{1,B^2-1\}$, so that $|1-\bar b^2|\leq Q$
throughout this interval. Taking absolute values in the
differentiated series and using $d_{\bar b}(z)\geq z$ gives
\begin{equation}
 \begin{split}
 0<\mathcal K_{\bar b}(z)
 &\leq\frac{4B}{z^3}
 \sum_{k=3}^{\infty}
 \frac{(k-2)Q^{k-3}z^{2k}}{(2k)!}\\
 &=\frac{B}{Q^{3/2}}H(\sqrt Q\,z).
 \end{split}
 \label{eq:small-gap-compact-kernel-bound}
\end{equation}
The bounds on $H$ derived above show that this upper bound is
$O(z^3)$ near zero and grows at most exponentially at infinity.
It thus remains integrable after multiplication by either
Gaussian moment weight. Dominated convergence establishes
continuity of $M_1$ and $M_3$ on the compact interval,
including across $\bar b=1$. Since both moments are strictly
positive, their ratio is continuous and strictly positive there.

The endpoint limits now complete the argument. Sufficiently
near $\bar b=0$, the ratio is larger than $m_1/(2m_3)$;
for sufficiently large $\bar b$, it is larger than
$1/[4(a\sigma)^2]$. On the remaining compact interval it
attains a strictly positive minimum. Taking the smallest
of these three positive bounds proves
\begin{equation}
 \inf_{\bar b\geq0}
 \frac{M_1(\bar b)}{M_3(\bar b)}>0.
 \label{eq:small-gap-positive-global-infimum}
\end{equation}
We may therefore choose
\begin{equation}
 \delta(a\sigma)=\sqrt{6}\,a\sigma
 \left[
 \inf_{\bar b\geq0}\frac{M_1(\bar b)}{M_3(\bar b)}
 \right]^{1/2}>0.
 \label{eq:uniform-small-gap-bound}
\end{equation}
For $0<\Omega\sigma<\delta(a\sigma)$, the sufficient
condition~\eqref{eq:small-gap-monotonicity} holds simultaneously
for every $\bar b>0$, proving
$\partial_{\bar b}\mathcal P_{\bar b}<0$ throughout this range.

\subsection{Boundary and interior thresholds for nonmonotonicity}
\label{app:thresholds}
\label{app:threshold-evaluation}

Starting from the decreasing behavior at small gaps, we now increase
$\Omega\sigma$ at fixed $a\sigma$ and ask when
$\partial_{\bar b}\mathcal P_{\bar b}$ first becomes positive somewhere
in $\bar b\geq0$. This defines the critical gap in
Eq.~\eqref{eq:global-critical-gap}. The first increasing interval can
appear immediately to the right of $\bar b=0$, or around an interior
point $\bar b_c>0$. We derive the conditions for these two cases and
then describe their numerical evaluation.

At the boundary, the first derivative itself cannot identify the
threshold: the response is even in $\bar b$, so
$\partial_{\bar b}\mathcal P_{\bar b}|_0=0$ at every gap.
Instead, expand about zero and differentiate:
\begin{equation}
 \begin{aligned}
 \mathcal P_{\bar b}
 &=\mathcal P_0
 +\frac{\bar b^2}{2}
   \left.\partial_{\bar b}^2\mathcal P_{\bar b}\right|_0
 +O(\bar b^4),\\
 \partial_{\bar b}\mathcal P_{\bar b}
 &=\bar b\left.\partial_{\bar b}^2\mathcal P_{\bar b}\right|_0
 +O(\bar b^3).
 \end{aligned}
 \label{eq:boundary-even-expansion}
\end{equation}
For sufficiently small positive $\bar b$, the sign of the derivative
is therefore the sign of the curvature at zero, provided that this
curvature is nonzero. The boundary threshold
$(\Omega\sigma)_{\mathrm{bdry}}$ is the first gap at which
$\partial_{\bar b}^2\mathcal P_{\bar b}|_0$ changes from negative to
positive. Its zero gives Eq.~\eqref{eq:boundary-threshold-condition};
the sign change must also be checked. From the small-acceleration
result~\eqref{eq:small-acceleration-torsion-derivative}, this threshold
tends to $\sqrt{3/2}$ as $a\sigma\to0^+$.

In the interior case, a local maximum of the first derivative rises
from negative values and first reaches zero at $\bar b_c>0$.
At this point, both the value of the first derivative and its slope
with respect to $\bar b$ vanish:
\begin{equation}
 \left.\partial_{\bar b}\mathcal P_{\bar b}\right|_c=0,
 \qquad
 \left.\partial_{\bar b}^2\mathcal P_{\bar b}\right|_c=0,
 \qquad \bar b_c>0.
 \label{eq:appendix-interior-contact}
\end{equation}
Here the subscript $c$ denotes evaluation at
$(\bar b_c,(\Omega\sigma)_{\mathrm{int}})$ for the chosen acceleration.
When selecting a solution, we check that it is a local maximum of
the first derivative and that this maximum becomes positive as the
gap increases. Just above the threshold, the derivative is positive near
$\bar b_c$ and negative on either side. Its left zero is consequently
a local minimum of the response, and its right zero is a local maximum.
This explains both the interior conditions in
Eq.~\eqref{eq:interior-threshold-condition} and the appearance of the
pair of extrema. Comparing this value $(\Omega\sigma)_{\mathrm{int}}$ with
$(\Omega\sigma)_{\mathrm{bdry}}$ then gives the first loss of decreasing
behavior through Eq.~\eqref{eq:threshold-envelope}.

For their numerical evaluation, it is convenient to use
$q=\bar b^2$ as the differentiation variable. This removes the
factor of $\bar b$ that makes the first derivative vanish identically
at the boundary. Treating the response and separation as functions
of $q$, the chain rule gives
\begin{equation}
 \begin{aligned}
 \partial_{\bar b}\mathcal P
 &=2\bar b\,\partial_q\mathcal P,\\
 \partial_{\bar b}^2\mathcal P
 &=2\partial_q\mathcal P+4q\partial_q^2\mathcal P.
 \end{aligned}
 \label{eq:q-chain-rule}
\end{equation}
Thus the boundary equation becomes
$\partial_q\mathcal P|_{q=0}=0$.
At an interior contact, $q>0$ and the first equation requires
$\partial_q\mathcal P=0$; substituting this into the second then gives
$\partial_q^2\mathcal P=0$.

These derivatives can be evaluated directly from the response
integral, without finite differences between neighboring probability
values. Differentiating Eq.~\eqref{eq:dimensionless-response} gives
\begin{equation}
 \begin{split}
 \partial_q\mathcal P
 ={}&\frac1{4\sqrt\pi}\int_0^\infty dz\,
 e^{-z^2/[4(a\sigma)^2]}\\
 &\times\sin\!\left(\frac{(\Omega\sigma)z}{a\sigma}\right)
 \frac{\partial_q d_{\bar b}^2(z)}{d_{\bar b}^3(z)}.
 \end{split}
 \label{eq:q-response-first-derivative}
\end{equation}
Differentiating the final factor once more yields
\begin{equation}
 \begin{split}
 \partial_q^2\mathcal P
 ={}&\frac1{4\sqrt\pi}\int_0^\infty dz\,
 e^{-z^2/[4(a\sigma)^2]}
 \sin\!\left(\frac{(\Omega\sigma)z}{a\sigma}\right)\\
 &\times\left[
 \frac{\partial_q^2 d_{\bar b}^2(z)}{d_{\bar b}^3(z)}
 -\frac{3[\partial_q d_{\bar b}^2(z)]^2}{2d_{\bar b}^5(z)}
 \right].
 \end{split}
 \label{eq:q-response-second-derivative}
\end{equation}
The separation series~\eqref{eq:hyperbolic-separation-series} shows
that both differentiated kernels are regular at $z=0$, and the
Gaussian controls the large-$z$ tails at fixed finite parameters.
For each acceleration, the boundary equation and the two interior
equations are solved for their respective gaps. The sign checks described above select the thresholds reported in Table~\ref{tab:threshold-checks}.

\begin{table}[tbp]
 \caption{Representative boundary and interior threshold gaps.
 The critical ratio $\bar b_c$ is evaluated at the interior threshold;
 the first row places it in the catenary regime.}
 \label{tab:threshold-checks}
 \centering
 \setlength{\tabcolsep}{5pt}
 \begin{ruledtabular}
 \begin{tabular}{cccc}
 $a\sigma$ & $(\Omega\sigma)_{\mathrm{bdry}}$
 & $(\Omega\sigma)_{\mathrm{int}}$ & $\bar b_c$\\
 1.5 & 1.575504 & 1.554248 & 0.941651\\
 3 & 2.206544 & 1.907728 & 1.173370\\
 15 & 8.693463 & 3.652856 & 1.051047
 \end{tabular}
 \end{ruledtabular}
\end{table}

The point at which the two mechanisms meet can be found using the
same equations. As the acceleration is varied, the numerical
interior solutions approach the boundary continuously,
$q_c\to0$. Both $\partial_q\mathcal P$ and
$\partial_q^2\mathcal P$ vanish along these solutions, so continuity
gives
\begin{equation}
 \left.\partial_q\mathcal P\right|_{q=0}=0,
 \qquad
 \left.\partial_q^2\mathcal P\right|_{q=0}=0
 \label{eq:threshold-switch-conditions}
\end{equation}
at their boundary limit. Here $a\sigma$ and $\Omega\sigma$ are the
two unknowns. Solving these equations and following the interior
solutions to $q=0$ gives
$((a\sigma)_*,(\Omega\sigma)_*)\simeq(1.0759453,1.4287131)$,
in agreement with Eq.~\eqref{eq:threshold-switch-point}.

Finally, the critical ratio locates the first appearance of the
increasing interval, rather than either extremum at a larger gap.
For example, at $a\sigma=15$ the threshold occurs at
$\Omega\sigma\simeq3.652856$ and $\bar b_c\simeq1.051047$.
At gap $4$, solving $\partial_q\mathcal P=0$ instead gives a minimum
at $\bar b\simeq1.046756$ and a maximum at $\bar b\simeq1.078665$.
These values account for the shallow pair of extrema discussed in
Sec.~\ref{subsec:torsion-monotonicity}. By
Appendix~\ref{app:temperature-monotonicity}, the effective temperature
has the same thresholds and extrema.

\bibliography{CorrelationHarvestingRefs}

\end{document}